\documentclass[11pt]{article}

\usepackage[T1]{fontenc}
\usepackage{lmodern}
\usepackage[margin=0.82in]{geometry}
\usepackage{amsmath,amssymb,bm,mathtools}
\usepackage{graphicx}
\usepackage{subcaption}
\usepackage{xcolor}
\usepackage{float}
\usepackage{placeins}
\usepackage{lineno}
\usepackage[numbers,sort&compress]{natbib}
\usepackage{authblk}
\usepackage{microtype}
\usepackage{booktabs,tabularx,array,longtable}
\usepackage[hidelinks]{hyperref}

\allowdisplaybreaks[1]
\newcommand{\dd}{\mathrm{d}}
\newcommand{\ii}{\mathrm{i}}

\newcommand{\Ma}{\mathcal{M}}

\begin{document}

\title{Nonlinear waves and conservative transport in
soluble-surfactant falling films on compliant substrates}

\author[1]{Sanghasri Mukhopadhyay%
\thanks{Corresponding author: sanghasri.m@iiitb.ac.in}}

\author[2]{Asim Mukhopadhyay}

\affil[1]{Centre for Applied Sciences,
International Institute of Information Technology,
Bengaluru 560100, India}

\affil[2]{Department of Mathematics,
Vivekananda Mahavidyalaya,
Burdwan 713103, West Bengal, India}

\date{}

\maketitle

\begin{abstract}
We investigate finite-amplitude waves in a liquid film containing a soluble surfactant as it flows down an inclined compliant substrate. The coupled motion of the liquid and a spring-backed plate is described by a conservative weighted-residual model, which also accounts for surfactant transport in the bulk and at the free surface. The reduction retains viscous terms arising from wall motion, and its pressure closure recovers fluid added mass under uniform translation. Comparison with the coupled Orr-Sommerfeld problem gives a difference of approximately $1\%$ in the peak growth rate of the film mode at $Re=25$. Nonlinear computations are then used to examine travelling states and their dependence on substrate properties and surfactant coupling. Along the prescribed stiffness-damping path, substrate displacement decreases by more than four orders of magnitude, while wave speed changes by less than $1\%$. The extrema of free-surface elevation and liquid thickness vary differently, so surface height alone does not reliably indicate film thinning over a deformable support. In the cases studied, Marangoni coupling reduces surfactant localisation and weakens the wave and substrate response. Moving-frame reconstruction reveals recirculation near the wave crest and distinguishes propagation of the deformation pattern from transverse substrate motion. The load-work budget identifies capillary pressure as the principal net source of substrate work for the two compliant supports considered. A control calculation in which stiffness is varied at fixed damping shows a substantial change in damping loss but a comparatively small change in deformation.
\end{abstract}

\section{Introduction}
\label{sec:introduction}

Gravity-driven liquid films can develop surface waves even when the
underlying flow is laminar. Their stability depends on the balance
between inertia, viscous resistance and capillarity, and their nonlinear
evolution may lead to localized pulses. The classical analyses of
Benjamin \citeyearpar{benjamin1957} and Yih \citeyearpar{yih1963}
established the stability framework for films flowing down rigid
inclined planes. These results remain a starting point for studying
films with more complex boundary conditions. When surfactant is present
or the supporting wall can deform, interfacial transport and substrate
motion must also be included in the description.

The full flow is governed by the Navier-Stokes equations with moving
boundaries, but much of its long-wave behaviour can be described in
terms of a smaller set of depth-integrated variables. In a
weighted-residual formulation, information about the transverse
velocity distribution is retained by projecting the governing
equations onto selected profile functions. Ruyer-Quil and Manneville
\citeyearpar{ruyerquil2000,ruyerquil2002} developed this approach for
falling films and examined its accuracy and convergence as the
transverse resolution was increased. These reductions make
finite-amplitude waves accessible to long-time computation, although
their accuracy remains tied to the profile approximations used in
constructing the model.

For a compliant support, fluid stresses induce wall motion, which in
turn changes the liquid geometry and the hydrodynamic boundary
conditions. Carpenter and Garrad
\citeyearpar{carpenter1985,carpenter1986} developed a framework for
the stability of flow over compliant surfaces that accounts for
substrate inertia, damping, tension, bending and foundation stiffness.
In the falling-film problem, Matar, Craster and Kumar
\citeyearpar{matar2007} derived coupled nonlinear evolution equations
for the liquid and substrate. Sisoev \emph{et al.}
\citeyearpar{sisoev2010} subsequently examined coherent waves and
their selection on flexible walls. Alexander, Kirk and Papageorgiou
\citeyearpar{alexander2020} analysed the coupled Orr-Sommerfeld
problem, distinguishing disturbances of the free surface from
instabilities associated with the substrate.

Surfactants affect the film through their influence on surface
tension. Gradients in interfacial concentration generate Marangoni
stresses, which alter the surface velocity and hence the amplification
of waves and the transport of surfactant. Blyth and Pozrikidis
\citeyearpar{blyth2004} examined the stabilizing influence of
insoluble surfactant on the primary film instability. When the
surfactant is soluble, the surface concentration also depends on
exchange with the liquid beneath it. Adsorption and desorption
introduce additional time scales, while diffusion acts in both the
bulk and the interface. Karapetsas and Bontozoglou
\citeyearpar{karapetsas2013,karapetsas2014} showed how these
processes modify the concentration response and the mechanism of
long-wave instability. The strength of Marangoni stabilization
therefore cannot be inferred from equilibrium coverage alone:
solubility, exchange kinetics and the phase of the concentration
response relative to the hydrodynamic disturbance also matter.

The transport formulations of Stone
\citeyearpar{stone1990} and Huang, Lai and Tseng
\citeyearpar{huang2008} account consistently for changes
in interfacial area. A conservative reduction must
preserve this geometric accounting at the retained
order and use compatible bulk and surface fluxes.
Adsorption and desorption redistribute surfactant
between the two reservoirs and must enter their
balances with equal and opposite contributions.

Weighted-residual models for soluble-surfactant films
were developed by Pascal, D'Alessio and Ellaban
\citeyearpar{pascal2019} and extended to heated films by
D'Alessio \emph{et al.} \citeyearpar{dalessio2020}.
These nonlinear reduced models do not conserve the
combined bulk and surface surfactant inventory.
Their surface-transport reduction contains an erroneous
nonflux contribution, which permits artificial
surfactant-mass drift in a closed periodic domain.

The correction to the nonlinear surface-transport condition was examined in detail by Mukhopadhyay \emph{et al.} \citeyearpar{mukhopadhyay2026pof}. Their analysis showed that the earlier reduced surface equation contains a nonflux contribution that produces an artificial change in surfactant inventory during nonlinear evolution. Correcting its coefficient restores the conservative interfacial flux. The defect thus concerns the governing nonlinear balance, rather than a numerical error that can be removed by refining the computation. Mukhopadhyay \emph{et al.} \citeyearpar{mukhopadhyay2026} then combined Orr-Sommerfeld analysis with nonlinear weighted-residual computations for slipping falling films.

A subsequent analytical and numerical audit by Mukhopadhyay
and Mukhopadhyay \citeyearpar{mukhopadhyay2026audit}
identified a separate inconsistency in the depth-integrated bulk
diffusion of the corrected models. The missing contribution arises from the streamwise concentration gradient in the normal diffusive flux at a sloping interface. Including this contribution gives a local conservation law for the combined reduced inventory. It also leaves the equations linearized about uniform equilibrium unchanged. These findings demonstrate why agreement with linear stability theory, although necessary, is insufficient to establish the consistency of a nonlinear soluble-surfactant model.

The interaction between soluble surfactant and substrate
compliance was examined by Samanta \citeyearpar{samanta2025}
using long-wave analysis and numerical solution of a coupled
Orr-Sommerfeld problem. Linear theory establishes the onset
and initial amplification of disturbances, but does not
determine the amplitude of the ensuing waves, the associated
film thinning or the mechanical work sustaining substrate
motion. These questions require a nonlinear description
with consistent transport and conservation laws. For nonlinear evolution, the conservation of surfactant imposes
a requirement that cannot be established from a linear stability
comparison. 

The present work uses the corrected transport formulation to
investigate finite-amplitude waves over a compliant substrate.
The principal questions concern how a liquid wave sustains
substrate deformation and how this deformation changes the
interpretation of film thinning. These questions require information
beyond the growth rates and phase speeds of infinitesimal
disturbances. We distinguish changes in liquid thickness from
changes in absolute free-surface elevation, examine the signed
work done by the individual contributions to the normal load,
and reconstruct the flow relative to the propagating wave.
A surface-substrate phase diagnostic, independent of the chosen
spatial origin, relates the wave geometry to the mechanical
response. A further calculation at fixed damping separates the
effect of a selected stiffness change from the response along
the prescribed stiffness-damping path.

For this purpose, we extend the corrected conservative formulation
to an inclined compliant substrate. Throughout the derivation,
we distinguish the physical liquid thickness $H=h-\zeta$ from
the absolute free-surface elevation $h=H+\zeta$. The thickness
determines liquid storage and transverse profile moments, while
the elevation describes the position and curvature of the free
surface. The momentum equations are projected in coordinates
fitted to the moving wall, retaining contributions associated
with wall slope and wall-velocity gradients. In constructing
the substrate load, we also retain the leading transverse fluid
acceleration. The resulting closure recovers fluid added mass
in the uniform-translation limit.

The asymptotic analysis, the full linear problem and the nonlinear
simulations use a common normalization and the parameter families
specified in section~\ref{subsec:parameter_regime}. We consider
variations in equilibrium coverage, Marangoni strength and
foundation stiffness. In the stiffness comparison, damping
follows the prescribed relation $\mathcal{D}=\sqrt{\mathcal{K}}$.
The coupled Orr-Sommerfeld problem is solved by Chebyshev
collocation, and its growth rates and marginal stability curves
are compared with those of the linearized weighted-residual
model. Nonlinear simulations then examine wave profiles and
propagation speeds, together with substrate motion, surfactant
redistribution and internal flow. The rigid-wall and
uniform-translation limits provide checks on the fluid-substrate
coupling. Small-amplitude growth, inventory conservation and
travelling-state diagnostics are used to assess the computed
solutions within the stated reduced-model approximation.
\section{Problem statement and formulation}
\label{sec:formulation}

\begin{figure}
    \centering
    \includegraphics[width=0.95\linewidth]
        {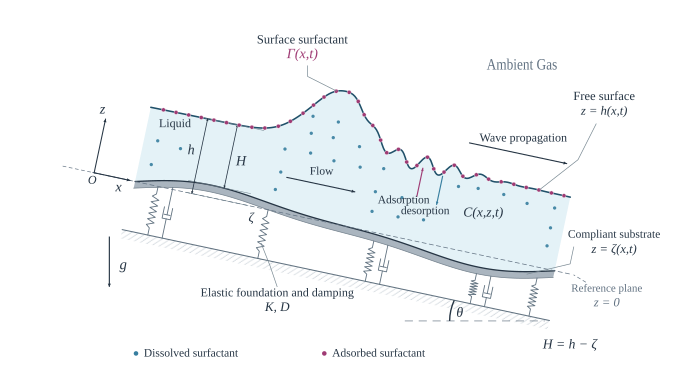}
    \caption{Schematic of a soluble-surfactant-laden liquid film
    flowing down an inclined compliant substrate.
    The free surface and substrate are located at
    $z=h(x,t)$ and $z=\zeta(x,t)$, respectively, giving
    the local liquid thickness $H=h-\zeta$.
    Dissolved surfactant of concentration $C$ exchanges
    with the surface concentration $\Gamma$ through
    adsorption and desorption.
    Springs and dampers represent the elastic foundation
    and substrate damping. The schematic is not to scale.}
    \label{fig:physical_setup}
\end{figure}

We consider a two-dimensional incompressible Newtonian liquid
film flowing under gravity down a compliant substrate inclined
at an angle $\theta$ to the horizontal. The liquid has constant
density $\rho$, dynamic viscosity $\mu$ and kinematic viscosity
$\nu=\mu/\rho$. The coordinate $x$ is directed downslope, while
$z$ is normal to the undeformed substrate. The flow geometry is illustrated in figure~\ref{fig:physical_setup}.

The substrate and free surface are represented by
$z=\zeta(x,t)$ and $z=h(x,t)$, respectively. The physical liquid
thickness is therefore
\begin{equation}
    H(x,t)=h(x,t)-\zeta(x,t)>0.
    \label{eq:H_definition_early}
\end{equation}
This distinction is retained throughout the formulation:
$H$ measures the local liquid depth, whereas $h$ describes
the absolute free-surface elevation relative to the
undeformed substrate.

The substrate undergoes transverse material displacement
and has no streamwise material motion. The liquid satisfies
no slip at the substrate, which is impermeable to both
solvent and surfactant. Adsorption at the liquid-solid
interface is excluded. The surrounding gas exerts a constant
ambient pressure and negligible viscous stress.

\subsection{Scaling and dimensionless parameters}
\label{subsec:lubrication}

Let $\hat h_N$ denote the reference mean liquid thickness
and $\hat\lambda/(2\pi)$ a characteristic streamwise length.
The long-wave aspect ratio is
\begin{equation}
    \varepsilon=\frac{2\pi\hat h_N}{\hat\lambda}\ll1.
\end{equation}
The velocity scale $\hat V$ is initially kept arbitrary.
Dimensional variables are denoted by hats, while material
constants retain their conventional symbols.

The dimensionless variables are defined by
\begin{align}
    x^*&=\frac{\varepsilon\hat x}{\hat h_N},
    &
    z^*&=\frac{\hat z}{\hat h_N},
    &
    t^*&=\frac{\varepsilon\hat V\hat t}{\hat h_N},
    &
    h^*&=\frac{\hat h}{\hat h_N},
    &
    \zeta^*&=\frac{\hat\zeta}{\hat h_N},
    \nonumber\\
    u^*&=\frac{\hat u}{\hat V},
    &
    w^*&=\frac{\hat w}{\varepsilon\hat V},
   &
    p^*&=\frac{\hat p-\hat p_a}{\rho\hat V^2},
    &
    p_s^*&=\frac{\hat p_s-\hat p_a}{\rho\hat V^2},
    &
    C^*&=\frac{\hat h_N\hat C}{\Gamma_\infty},
    &
    \Gamma^*&=\frac{\hat\Gamma}{\Gamma_\infty}.
    \label{eq:dimension}
\end{align}
Here $\hat p_a$ is the ambient gas pressure, $\hat p_s$ is
the backing pressure, and $\Gamma_\infty$ is the maximum
surface packing concentration. The bulk and surface
concentrations are scaled so that their integrated
inventories have compatible units.

After omitting asterisks, we introduce the pressures
\begin{equation}
    \Pi=Re\,p,
    \qquad
    \Pi_s=Re\,p_s,
    \label{eq:pressure_bridge}
\end{equation}
which are based on the viscous-pressure scale
$\mu\hat V/\hat h_N$.

The principal fluid and surfactant parameters are collected
in Table~\ref{tab:notation}. The dimensional diffusivities
$D_b$ and $D_s$ describe transport in the liquid bulk and
along the free surface, respectively. The kinetic constants
$k_1$ and $k_2$ have units of velocity and inverse time and
specify adsorption and desorption in the Langmuir exchange
law. The parameter $\sigma_\Gamma>0$ measures the magnitude
of the surface-tension sensitivity to surfactant concentration.

\begin{table}
    \centering
    \caption{Dimensionless fluid and surfactant parameters.
    Definitions use the scales in Eq.~\eqref{eq:dimension}.}
    \label{tab:notation}
    \small
    \setlength{\tabcolsep}{5pt}
    \renewcommand{\arraystretch}{1.18}
    \begin{tabularx}{\textwidth}
        {@{}ll>{\raggedright\arraybackslash}X@{}}
        \toprule
        Parameter & Definition & Physical interpretation\\
        \midrule
        $Re$
        & $\hat V\hat h_N/\nu$
        & Inertia relative to viscous resistance.\\

        $G$
        & $g\sin\theta\,\hat h_N^2/(\nu\hat V)$
        & Streamwise gravitational forcing relative
          to viscous resistance.\\

        $Ca$
        & $\mu\hat V/\sigma_0$
        & Viscous stresses relative to capillarity.\\

        $Ka$
        & $\sigma_0/
          [\rho\nu^{4/3}(g\sin\theta)^{1/3}]$
        & Capillary-gravity parameter determined
          by the liquid properties and inclination.\\

        $\Ma$
        & $\sigma_\Gamma\Gamma_\infty/(\mu\hat V)$
        & Marangoni coefficient based on viscous stress.\\

        $Pe_b$
        & $\hat V\hat h_N/D_b$
        & Advection relative to bulk diffusion.\\

        $Pe_s$
        & $\hat V\hat h_N/D_s$
        & Advection relative to surface diffusion.\\

        $k_s$
        & $k_2\hat h_N/\hat V$
        & Desorption rate relative to the
          thickness-based advective rate.\\

        $\kappa$
        & $k_1/(k_2\hat h_N)$
        & Scaled adsorption-desorption ratio.\\

        $\Gamma_e$
        & $\hat\Gamma_e/\Gamma_\infty$
        & Reference equilibrium surface coverage.\\
        \bottomrule
    \end{tabularx}
\end{table}

For comparison with formulations based on dynamic pressure,
we define
\begin{equation}
    Fr=\frac{\hat V}
    {\sqrt{g\sin\theta\,\hat h_N}},
    \qquad
    We=\frac{\sigma_0}{\rho\hat V^2\hat h_N},
    \qquad
    Mn=\frac{\sigma_\Gamma\Gamma_\infty}
    {\rho\hat V^2\hat h_N}.
\end{equation}
In this convention,
\begin{equation}
    G=\frac{Re}{Fr^2},
    \qquad
    Ca=\frac{1}{Re\,We},
    \qquad
    \Ma=Re\,Mn.
\end{equation}

We adopt the linear equation of state
\begin{equation}
    \Sigma(\Gamma)
    =\frac{1}{Ca}-\Ma(\Gamma-\Gamma_e),
    \qquad
    \Ma=Re\,Mn
    =\frac{\sigma_\Gamma\Gamma_\infty}{\mu\hat V},
    \label{eq:groups}
\end{equation}
where $\Sigma$ is surface tension scaled by $\mu\hat V$,
$\sigma_0$ is its dimensional value at the reference
equilibrium coverage, and
$\sigma_\Gamma=-\dd\hat\sigma/\dd\hat\Gamma>0$.
The formulation requires positive surface tension and
concentrations within the range of validity of this
equation of state.

The reference equilibrium bulk concentration is
$\phi_e=\Gamma_e/[\kappa(1-\Gamma_e)]$, with
$0\le\Gamma_e<1$. When the actual uniform base coverage
is denoted by $\Gamma_B$, its bulk concentration is
$C_B=\Gamma_B/[\kappa(1-\Gamma_B)]$.

Choosing
$\hat V=g\sin\theta\,\hat h_N^2/(2\nu)$ gives $G=2$
and corresponds to the Nusselt free-surface velocity.
Choosing
$\hat V=g\sin\theta\,\hat h_N^2/(3\nu)$ gives $G=3$
and corresponds to the Nusselt mean velocity.
The present calculations use $G=2$, as specified in
Sec.~III\, \ref{subsec:parameter_regime}. A Reynolds number
based on the Nusselt mean velocity would be
$Re_N=(G/3)Re$. Changes of velocity normalization require
the consistent transformation of all dimensionless groups.

The volumetric flux per unit width is
$q=\int_\zeta^h u\,\dd z$. For the subsequent reduction,
we define the interfacial bulk concentration
$\phi=C(x,h(x,t),t)$ and the integrated concentration
departure
$\chi=\int_\zeta^h(C-\phi)\,\dd z$.
Thus $H\phi+\chi$ is the bulk surfactant inventory per
unit projected streamwise length. The wall-fitted coordinate
used in the weighted-residual derivation is
$y=(z-\zeta)/H$, with $0\le y\le1$.

Subscripts on fields denote partial derivatives, and
$f(h)$ and $f(\zeta)$ denote evaluation at the free
surface and substrate. In the stability sections, hats
denote complex disturbance amplitudes; bars and tildes
denote base states and perturbations. The thickness-based
coordinates used in the stability analysis and numerical
calculations are introduced in Eq.~\eqref{eq:OS_rescaling}.

\subsection{Governing equations and substrate conditions}
\label{subsec:governing}

The dimensionless continuity, momentum and bulk
surfactant equations are
\begin{equation}
    u_x+w_z=0,
    \label{eq:mass}
\end{equation}
\begin{equation}
    \varepsilon Re
    \left(u_t+u u_x+w u_z\right)
    =
    -\varepsilon\Pi_x+G
    +\varepsilon^2u_{xx}+u_{zz},
    \label{eq:x_momentum}
\end{equation}
\begin{equation}
    \varepsilon^2 Re
    \left(w_t+u w_x+w w_z\right)
    =
    -\Pi_z-G\cot\theta
    +\varepsilon^3w_{xx}+\varepsilon w_{zz},
    \label{eq:z_momentum}
\end{equation}
and
\begin{equation}
    \varepsilon Pe_b
    \left(C_t+u C_x+w C_z\right)
    =
    \varepsilon^2C_{xx}+C_{zz}.
    \label{eq:concentration}
\end{equation}

At the substrate $z=\zeta(x,t)$, no slip, impermeability
and zero normal surfactant flux give
\begin{equation}
    u=0,
    \qquad
    w=\zeta_t,
    \qquad
    C_z-\varepsilon^2\zeta_x C_x=0.
    \label{eq:boundary_bottom}
\end{equation}
The substrate velocity is denoted by $v=\zeta_t$.

The substrate is represented by a spring-backed elastic
plate with mass per unit area $\rho_w b$, where $\rho_w$
and $b$ are its density and thickness. Its dimensional
damping, tension, flexural rigidity and foundation stiffness
are denoted by $\hat D$, $\hat T$, $\hat B$ and $\hat K$.
Their dimensionless counterparts and physical roles are
given in Table~\ref{tab:compliant_wall_coefficients}.

\begin{table}
    \centering
    \caption{Compliant-wall coefficients and their
    physical interpretation.}
    \label{tab:compliant_wall_coefficients}
    \small
    \setlength{\tabcolsep}{5pt}
    \renewcommand{\arraystretch}{1.18}
    \begin{tabularx}{\textwidth}
        {@{}ll>{\raggedright\arraybackslash}X@{}}
        \toprule
        Coefficient & Definition & Physical interpretation\\
        \midrule
        $\mathcal I$
        & $\rho_w b\hat V/\mu$
        & Substrate inertia associated with wall mass
          per unit area.\\

        $\mathcal D$
        & $\hat D\hat h_N/\mu$
        & Linear damping opposing substrate velocity.\\

        $\mathcal T$
        & $\hat T/(\mu\hat V)$
        & Prescribed in-plane tension.\\

        $\mathcal B$
        & $\hat B/(\mu\hat V\hat h_N^2)$
        & Flexural rigidity resisting bending.\\

        $\mathcal K$
        & $\hat K\hat h_N^2/(\mu\hat V)$
        & Foundation stiffness restoring the substrate
          towards its reference position.\\
        \bottomrule
    \end{tabularx}
\end{table}

The dimensional coefficients have units
$[\hat D]=\mathrm{Pa\,s\,m^{-1}}$,
$[\hat T]=\mathrm{N\,m^{-1}}$,
$[\hat B]=\mathrm{N\,m}$ and
$[\hat K]=\mathrm{Pa\,m^{-1}}$.
For a homogeneous isotropic plate,
$\hat B=E_w b^3/[12(1-\nu_w^2)]$, where $E_w$
and $\nu_w$ are Young's modulus and Poisson's ratio.

The normal-traction balance is
\begin{align}
    \varepsilon^2\mathcal I\zeta_{tt}
    +\varepsilon\mathcal D\zeta_t
    -\varepsilon^2\mathcal T\zeta_{xx}
    +\varepsilon^4\mathcal B\zeta_{xxxx}
    +\mathcal K\zeta
    =
    \Pi_s-\Pi\big|_\zeta+\tau_{nn,w},
    \label{eq:wall_dimless}
\end{align}
where the normal viscous traction is
\begin{equation}
    \tau_{nn,w}
    =
    \left.
    \frac{2\varepsilon}
    {1+\varepsilon^2\zeta_x^2}
    \left[
        \varepsilon^2u_x\zeta_x^2
        -(u_z+\varepsilon^2w_x)\zeta_x
        +w_z
    \right]
    \right|_\zeta .
    \label{eq:wall_normal_traction}
\end{equation}

Equation~\eqref{eq:wall_dimless} combines the fluid normal
traction with a linear elastic plate law and is intended
for small substrate slopes and strains. Passive supports
satisfy
\begin{equation}
    \mathcal I\ge0,
    \qquad
    \mathcal D\ge0,
    \qquad
    \mathcal T\ge0,
    \qquad
    \mathcal B\ge0,
    \qquad
    \mathcal K>0.
    \label{eq:wall_sign_conditions}
\end{equation}
These mechanical sign conditions do not determine the
stability of the coupled fluid-substrate system.

The uniform hydrostatic load associated with a unit-thickness
reference film is $G\cot\theta$. We therefore denote the
backing pressure in excess of this reference load by
$P_e=\Pi_s-G\cot\theta$.

\subsection{Free-surface boundary conditions}
\label{subsec:surface_conditions}

At $z=h(x,t)$, the normal-stress condition is
\begin{align}
    -\Pi
    +\frac{2\varepsilon}{1+\varepsilon^2h_x^2}
    \left[
        \varepsilon^2u_xh_x^2
        -(u_z+\varepsilon^2w_x)h_x
        +w_z
    \right]
    =
    \frac{\varepsilon^2\Sigma(\Gamma)h_{xx}}
    {(1+\varepsilon^2h_x^2)^{3/2}}.
    \label{eq:boundary_normal}
\end{align}
The tangential-stress and kinematic conditions are
\begin{equation}
    (u_z+\varepsilon^2w_x)
    (1-\varepsilon^2h_x^2)
    -4\varepsilon^2u_xh_x
    +\varepsilon\Ma\Gamma_x
    \sqrt{1+\varepsilon^2h_x^2}=0,
    \label{eq:boundary_tangential}
\end{equation}
and
\begin{equation}
    w=h_t+u(h)h_x.
    \label{eq:boundary_kinematic}
\end{equation}
All velocity and stress quantities in these conditions
are evaluated at the free surface.

Adsorption and desorption exchange surfactant between the
liquid bulk and the interface at the rate
$k_s[\kappa(1-\Gamma)C(h)-\Gamma]$ per unit actual
interfacial area. A positive value denotes adsorption.
The corresponding normal bulk-flux condition is
\begin{equation}
    \frac{
        \varepsilon^2h_xC_x(h)-C_z(h)
    }{
        Pe_b\sqrt{1+\varepsilon^2h_x^2}
    }
    =
    k_s\left[\kappa(1-\Gamma)C(h)-\Gamma\right].
    \label{eq:ficks}
\end{equation}

The surface concentration is measured per unit actual
interfacial area. Following the material-surface transport
formulation \citep{stone1990,huang2008}, its conservative
balance in graph coordinates is
\begin{align}
    \varepsilon
    \Bigg[
        \frac{\partial}{\partial t}
        \left(\Gamma\sqrt{1+\varepsilon^2h_x^2}\right)
        +
        \frac{\partial}{\partial x}
        \left(
            \Gamma\sqrt{1+\varepsilon^2h_x^2}\,u(h)
        \right)
    \Bigg]
    ={}&
    \frac{\varepsilon^2}{Pe_s}
    \frac{\partial}{\partial x}
    \left(
        \frac{\Gamma_x}{\sqrt{1+\varepsilon^2h_x^2}}
    \right)
    \nonumber\\
    &+
    k_s\sqrt{1+\varepsilon^2h_x^2}
    \left[\kappa(1-\Gamma)C(h)-\Gamma\right].
    \label{eq:boundary_surface_surfactant}
\end{align}

The surface inventory per unit projected streamwise length
is $\Gamma\sqrt{1+\varepsilon^2h_x^2}$.
Its advective flux uses the Cartesian streamwise velocity
$u(h)$, which determines the motion of the endpoints of
a material surface segment. The tangential component of
the physical velocity, scaled by $\hat V$, is
\begin{equation}
    \frac{u(h)+\varepsilon^2h_xw(h)}
    {\sqrt{1+\varepsilon^2h_x^2}}.
\end{equation}
The graph-coordinate balance incorporates interfacial
stretching through the time-dependent surface metric.

In the long-wave limit,
$\sqrt{1+\varepsilon^2h_x^2}=1+O(\varepsilon^2)$.
The projected surface balance is obtained by applying
this approximation consistently to the inventory,
advective flux, diffusive flux and exchange term.
Together with the depth-integrated bulk balance, it
defines the reduced total projected inventory
\begin{equation}
    \mathcal C_{\mathrm{tot}}
    =\chi+H\phi+\Gamma,
    \qquad
    \mathcal M_{\mathrm{tot}}
    =\int_0^L\mathcal C_{\mathrm{tot}}\,\dd x.
\end{equation}
Here $L$ denotes the periodic domain length in the
coordinates used for the reduced calculations. The
transverse profile approximations and conservative
flux closures are developed in the weighted-residual
derivation.

The surface metric has zero first variation about a
flat uniform equilibrium. Its consistent nonlinear
completion therefore leaves the corresponding linearized
surface condition unchanged.
\section{Base state and Orr-Sommerfeld formulation}
\label{subsec:base_OS}

For the linear stability problem, introduce
thickness-based streamwise and temporal coordinates:
\begin{equation}
    x_{\mathrm{iso}}=\frac{x}{\varepsilon},
    \qquad
    t_{\mathrm{iso}}=\frac{t}{\varepsilon},
    \qquad
    w_{\mathrm{iso}}=\varepsilon w.
    \label{eq:OS_rescaling}
\end{equation}
We subsequently omit the subscript.
This transformation removes the explicit long-wave
parameter from the continuum eigenvalue problem.
It is a change of variables, rather than the
substitution $\varepsilon=1$ into an asymptotic expansion.

The uniform base state is
\begin{equation}
    \bar h=1,
    \qquad
    \bar\zeta=0,
    \qquad
    \bar w=0,
    \qquad
    \bar\Gamma=\Gamma_B,
    \qquad
    \bar C=C_B.
    \label{eq:uniform_base_state}
\end{equation}

Adsorption-desorption equilibrium gives
\begin{equation}
    C_B
    =
    \frac{\Gamma_B}{\kappa(1-\Gamma_B)}.
    \label{eq:base_concentration}
\end{equation}

The base pressure satisfies
\begin{equation}
    \bar\Pi(z)=G\cot\theta(1-z),
    \qquad
    \bar\Pi_s=G\cot\theta.
    \label{eq:base_pressure}
\end{equation}
The backing pressure therefore balances the
unperturbed liquid load.

The base velocity is determined by
\begin{equation}
    U''+G=0,
    \qquad
    U(0)=0,
    \qquad
    U'(1)=0,
\end{equation}
and hence
\begin{equation}
    U(z)=G\left(z-\frac{z^2}{2}\right).
    \label{eq:base_velocity}
\end{equation}
In particular,
\begin{equation}
    U_s=U(1)=\frac G2,
    \qquad
    U'(0)=G,
    \qquad
    U'(1)=0,
    \qquad
    U''=-G.
    \label{eq:base_velocity_properties}
\end{equation}

The base surface tension is $Ca^{-1}-\Ma(\Gamma_B-\Gamma_e)$;
for $\Gamma_B=\Gamma_e$, it is $Ca^{-1}$.

The disturbances are written as
\begin{align}
    \{
        \widetilde u,\widetilde w,\widetilde\Pi,
        \widetilde C,\widetilde h,
        \widetilde\zeta,\widetilde\Gamma
    \}
    =
    \{
        \hat u,\hat w,\hat\Pi,
        \hat C,\hat h,\hat\zeta,\hat\Gamma
    \}
    e^{\ii k(x-\alpha t)}
    +\mathrm{c.c.},
    \label{eq:OS_normal_modes}
\end{align}
where $k>0$ and
\begin{equation}
    \alpha=\alpha_r+\ii \alpha_i.
\end{equation}

The temporal growth rate is $k\alpha_i$.

We introduce the disturbance streamfunction $\psi(z)$:
\begin{equation}
    \hat u=D\psi,
    \qquad
    \hat w=-\ii k\psi,
    \qquad
    D=\frac{\dd}{\dd z}.
    \label{eq:OS_streamfunction}
\end{equation}

The pressure reconstruction is detailed in
appendix~\ref{app:pressure_integrals}.

The Orr-Sommerfeld equation is
\begin{equation}
    (D^2-k^2)^2\psi
    -
    \ii kRe
    \left[
        (U-\alpha)(D^2-k^2)\psi-U''\psi
    \right]
    =0,
    \qquad 0<z<1.
    \label{eq:OS}
\end{equation}

The bulk concentration perturbation satisfies
\begin{equation}
    (D^2-k^2)\hat C
    -
    \ii kPe_b(U-\alpha)\hat C=0,
    \qquad 0<z<1.
    \label{eq:bulk_eigen}
\end{equation}

The surface-surfactant and normal bulk-flux
conditions are
\begin{equation}
    \ii k
    \left[
        (U_s-\alpha)\hat\Gamma
        +\Gamma_BD\psi(1)
    \right]
    +\frac{k^2}{Pe_s}\hat\Gamma
    -k_s\left[\kappa(1-\Gamma_B)\hat C(1)-(1+\kappa C_B)\hat\Gamma\right]=0,
    \label{eq:surface_linear}
\end{equation}
and
\begin{equation}
    \frac1{Pe_b}D\hat C(1)+k_s\left[\kappa(1-\Gamma_B)\hat C(1)-(1+\kappa C_B)\hat\Gamma\right]=0.
    \label{eq:fick_linear}
\end{equation}

Equivalently,
\begin{equation}
    D\hat C(1)
    +
    \ii kPe_b
    \left[
        (U_s-\alpha)\hat\Gamma+\Gamma_BD\psi(1)
    \right]
    +
    \frac{Pe_b}{Pe_s}k^2\hat\Gamma=0.
    \label{eq:combined_surfactant}
\end{equation}

The linearized tangential-stress condition is
\begin{equation}
    (D^2+k^2)\psi(1)
    +U''(1)\hat h
    +\ii k\Ma\hat\Gamma=0.
    \label{eq:tangential_OS}
\end{equation}

The normal-stress condition is
\begin{equation}
    (D^3-3k^2D)\psi(1)
    -
    \ii kRe(U_s-\alpha)D\psi(1)
    -
    \ii k
    \left(
        G\cot\theta+k^2\left[Ca^{-1}-\Ma(\Gamma_B-\Gamma_e)\right]
    \right)\hat h=0.
    \label{eq:normal_OS}
\end{equation}

The free-surface kinematic condition is
\begin{equation}
    (U_s-\alpha)\hat h+\psi(1)=0.
    \label{eq:kinematic_OS}
\end{equation}

At $z=0$, the no-slip, kinematic and no-flux
conditions become
\begin{equation}
    D\psi(0)+G\hat\zeta=0,
    \qquad
    \psi(0)-\alpha\hat\zeta=0,
    \qquad
    D\hat C(0)=0.
    \label{eq:wall_kinematic_OS}
\end{equation}

The wall-pressure elimination and normal-traction cancellation
are detailed in appendix~\ref{app:pressure_integrals}.

The pressure-eliminated wall condition is therefore
\begin{align}
    &(D^3-k^2D)\psi(0)
    -\ii \mathcal{I}k^3\alpha^2\hat\zeta
    +\mathcal{D}k^2\alpha\hat\zeta
    +\ii \mathcal{T}k^3\hat\zeta
    +\ii \mathcal{B}k^5\hat\zeta
    \nonumber\\
    &\hspace{35mm}
    +\ii k
    \left(\mathcal{K}-G\cot\theta\right)\hat\zeta=0.
    \label{eq:wall_dynamic_OS}
\end{align}

The wall inertia makes the eigenvalue problem
quadratic in $\alpha$. Introducing
\begin{equation}
    \xi=\alpha\hat\zeta
\end{equation}
gives
\begin{equation}
    \psi(0)=\xi,
    \qquad
    \xi=\alpha\hat\zeta,
\end{equation}
and
\begin{align}
    &(D^3-k^2D)\psi(0)
    +\mathcal{D}k^2\xi
    +\ii \mathcal{T}k^3\hat\zeta
    +\ii \mathcal{B}k^5\hat\zeta
    +\ii k(\mathcal{K}-G\cot\theta)\hat\zeta
    \nonumber\\
    &\hspace{35mm}
    =
    \ii \mathcal{I}k^3\alpha\,\xi.
    \label{eq:wall_dynamic_linearized_evp}
\end{align}
This form is linear in the eigenvalue.

The rigid-wall problem is obtained by removing
$\hat\zeta$ and the wall dynamic equation.
Its lower boundary conditions are
\begin{equation}
    \psi(0)=0,
    \qquad
    D\psi(0)=0,
    \qquad
    D\hat C(0)=0.
    \label{eq:rigid_wall_OS}
\end{equation}

For modes with finite temporal frequency as $k\to0$,
it is preferable to use
$e^{\lambda t+\ii kx}$ with $\lambda=-\ii k\alpha$.
The primitive pressure equations should then be retained.
The factor $\mathcal{K}-G\cot\theta$ in
Eq.~\eqref{eq:wall_dynamic_OS} cannot be interpreted
independently of the fluid pressure perturbation as
the uniform foundation stiffness.

\subsection{Parameter choices and continuation paths}
\label{subsec:parameter_regime}

The linear and nonlinear analyses use a common dimensionless
parameter family, with $G=2$ corresponding to the Nusselt
free-surface velocity normalization. The Kapitza number is
defined by
\begin{equation}
    Ka
    =
    \frac{\sigma_0}
    {\rho\nu^{4/3}(g\sin\theta)^{1/3}},
\end{equation}
which gives
\begin{equation}
    Ca(Re)
    =
    \frac{Re^{2/3}}{G^{1/3}Ka}.
    \label{eq:reference_capillary_number}
\end{equation}
The inclination and Kapitza number are held fixed at
$\theta=30^\circ$ and $Ka=3905$ throughout the present study.

The common reference parameters for the linear dispersion
calculations and nonlinear simulations are
\begin{equation}
\begin{gathered}
    Re=25,
    \qquad
    G=2,
    \qquad
    \theta=30^\circ,
    \qquad
    Ka=3905,
    \qquad
    \mathcal{M}=6,
    \\
    Pe_b=Pe_s=1,
    \qquad
    k_s=0.1,
    \qquad
    \kappa=10,
    \qquad
    \Gamma_e=0.1,
    \\
    P_e=0,
    \qquad
    L=250.
\end{gathered}
\label{eq:PRF_comparison_parameters}
\end{equation}
The corresponding reference wall parameters are
\begin{equation}
    \mathcal{I}=1,
    \qquad
    \mathcal{T}=500,
    \qquad
    \mathcal{B}=10,
    \qquad
    \mathcal{K}=100,
    \qquad
    \mathcal{D}=10.
    \label{eq:reference_wall_parameters}
\end{equation}
At this reference point,
\begin{equation}
    Mn=\frac{\mathcal{M}}{Re}=0.24,
    \qquad
    Ca\simeq1.7377834\times10^{-3},
    \qquad
    \phi_e
    =
    \frac{\Gamma_e}{\kappa(1-\Gamma_e)}
    =
    \frac{1}{90}.
    \label{eq:reference_derived_parameters}
\end{equation}
The periodic domain has fundamental wavenumber
$k_1=2\pi/L\simeq0.0251327$, which is also used for the
linear fundamental-mode comparison.

Three parameter sweeps are considered in the linear analysis:
\begin{equation}
\begin{aligned}
    \Gamma_e
    &\in\{0.01,0.1,0.3,0.5,0.9\},
    &&\mathcal{M}=6,\quad \mathcal{K}=100,
    \\
    \mathcal{M}
    &\in\{0,1.5,3,6\},
    &&\Gamma_e=0.1,\quad \mathcal{K}=100,
    \\
    \mathcal{K}
    &\in\{100,500,5\times10^6\},
    &&\Gamma_e=0.1,\quad \mathcal{M}=6.
\end{aligned}
\label{eq:study_parameter_sweeps}
\end{equation}
The nonlinear calculations use the same Marangoni and
stiffness sets, with the coverage sweep restricted to
$\Gamma_e\in\{0.01,0.1,0.3\}$.
All nonlinear simulations are performed at $Re=25$.
For each coverage, the equilibrium bulk concentration is
updated using
$\phi_e=\Gamma_e/[\kappa(1-\Gamma_e)]$.
This preserves adsorption equilibrium in the initial state,
while allowing the total surfactant loading to vary with
$\Gamma_e$.

Along the stiffness sweep, the damping coefficient is chosen as
\begin{equation}
    \mathcal{D}
    =
    2r_d\sqrt{\mathcal{I}\mathcal{K}}
    =
    \sqrt{\mathcal{K}},
    \qquad
    r_d=\frac12,
    \qquad
    \mathcal{I}=1.
    \label{eq:stiffness_damping_path}
\end{equation}
Here, $r_d$ is the nominal damping ratio of the bare foundation
oscillator. The damping of the coupled fluid-membrane modes
is determined by their spectrum.
The three stiffnesses represent compliant, stiffer and nearly
rigid substrates within the present parameter regime.
Because $\mathcal{D}$ changes with $\mathcal{K}$, this sweep
follows a prescribed stiffness-damping path.

The Reynolds-number continuation in the linear analysis is
performed over $0.1\le Re\le100$, holding the viscosity-scaled
Marangoni coefficient $\mathcal{M}=Re\,Mn$ fixed for each curve.
Consequently,
\begin{equation}
    Mn(Re)=\frac{\mathcal{M}}{Re},
    \qquad
    Ca(Re)=\frac{Re^{2/3}}{G^{1/3}Ka}.
    \label{eq:Re_dependent_parameters}
\end{equation}
The coverage sweep also retains a fixed reference capillary
number in the equation of state
\[
    \Sigma(\Gamma)
    =
    Ca^{-1}-\mathcal M(\Gamma-\Gamma_e).
\]
Consequently, its clean-surface intercept is
$\Sigma(0)=Ca^{-1}+\mathcal M\Gamma_e$ and changes
with $\Gamma_e$.
The prescribed sweep is therefore a dimensionless
comparison family; it does not represent changing the
coverage of a single liquid-surfactant system with an
otherwise fixed dimensional equation of state.
An experimental realization requires compatible
surface-tension, diffusion, kinetic and substrate properties.

The parameters $G$, $\theta$, $Ka$, $Pe_b$, $Pe_s$, $k_s$,
$\kappa$, $\Gamma_e$ and the wall coefficients are otherwise
held at the values prescribed for that curve.
A continuation at fixed $Mn$ would instead require
$\mathcal{M}(Re)=Re\,Mn$ and would define a different
parameter path. Numerical evaluation uses $Re>0$, since
the fixed-$Ka$ relation gives $Ca\to0$ as $Re\to0$.

The dimensional interpretation of the transport parameters
must satisfy
\begin{equation}
    \frac{Pe_b}{Re}=\frac{\nu}{D_b},
    \qquad
    \frac{Pe_s}{Re}=\frac{\nu}{D_s}.
    \label{eq:Schmidt_constraints}
\end{equation}
At the reference point, both ratios are $1/25=0.04$,
which implies $D_b=D_s=25\nu$.
These values specify the transport regime of the dimensionless
comparison problem. Assigning the calculation to a particular
liquid-surfactant system requires compatible dimensional
diffusivities, kinetic constants and wall properties.
Furthermore, holding $Pe_b=Pe_s=1$ during the Reynolds-number
sweep gives $\nu/D_b=\nu/D_s=1/Re$.
For fixed dimensional viscosity and diffusivities, the
Péclet numbers would instead vary proportionally to $Re$.

The selected wall coefficients satisfy mechanical passivity:
$\mathcal{I}>0$, $\mathcal{D}>0$, $\mathcal{T}>0$,
$\mathcal{B}>0$ and $\mathcal{K}>0$.
The elastic-plate approximation additionally requires small
substrate slopes and strains, together with the pointwise
conditions in Eq.~\eqref{eq:WR_admissibility}.
No universal mechanical lower bound
$\mathcal{K}>G\cot\theta$ is imposed.
For the present choices, however,
$G\cot\theta=2\sqrt{3}\simeq3.464$, below all three prescribed
foundation stiffnesses.

Finally, a change of velocity normalization requires a
consistent transformation of the dimensionless coefficients.
If $\hat V'=a\hat V$, with the dimensional system and thickness
scale held fixed, then
\begin{align}
    (Re',Pe_b',Pe_s',Ca',\mathcal{I}')
    &=
    a(Re,Pe_b,Pe_s,Ca,\mathcal{I}),
    \nonumber\\
    (G',k_s',\mathcal{M}',\mathcal{T}',\mathcal{B}',\mathcal{K}')
    &=
    a^{-1}
    (G,k_s,\mathcal{M},\mathcal{T},\mathcal{B},\mathcal{K}),
    \nonumber\\
    Mn'&=a^{-2}Mn,
    \qquad
    \mathcal{D}'=\mathcal{D}.
    \label{eq:velocity_scale_conversion}
\end{align}
The quantities $\kappa$, $\Gamma_e$, $Ka$, $\theta$ and $L$
remain unchanged. In particular, replacing $G=2$ by $G=3$
requires transformation of the other coefficients to represent
the same dimensional system. The present linear and nonlinear
calculations retain the $G=2$ normalization throughout.
\subsection{Asymptotic solution of the Orr-Sommerfeld problem}
\label{sec:asymptotic}

Following the long-wave expansion used in
\citet{mukhopadhyay2026}, we consider the limit $k\to0$ with
$G>0$, $0\leq\Gamma_B<1$, and finite positive
$\mathcal{K}$, $\kappa$, $k_s$, $Pe_b$, and $Pe_s$.
The phase speed is denoted by $\alpha$ to distinguish
it from the bulk concentration. Accordingly, the
normal-mode dependence is $e^{\ii k(x-\alpha t)}$,
and the temporal eigenvalue is $\lambda=-\ii k\alpha$.
The Marangoni coefficient appearing in the tangential
stress condition is $\Ma=Re\,Mn$.


The detailed long-wave expansion for the regular film branch,
including the leading eigenfunctions and the first-order phase-speed
correction, is derived in appendix~\ref{app:longwave_film_branch}.

The temporal growth rate of the film branch is
$\sigma_f(k)=\operatorname{Re}\lambda
=k\operatorname{Im}\alpha$. Hence,
\begin{equation}
\sigma_f(k)
=
\left[
\frac{2ReG^2}{15}
-\frac{G\cot\theta}{3}
\left(1-\frac{G\cot\theta}{\mathcal{K}}\right)
-\frac{3\Ma\Gamma_B\kappa(1-\Gamma_B)^2}
{4+3\kappa(1-\Gamma_B)^2}
\right]k^2
+O(k^4).
\label{eq:growth-LW-explicit}
\end{equation}
For a simple analytic branch, the reality of the
governing equations implies
$\alpha(-k)=\overline{\alpha(k)}$.
The growth rate is therefore even in $k$,
which excludes an order-$k^3$ contribution.

When $\Ma$ is held fixed as $Re$ varies, the
long-wave neutral condition gives
\begin{equation}
Re_c^{\mathrm{LW}}
=
\frac{5\cot\theta}{2G}
\left(1-\frac{G\cot\theta}{\mathcal{K}}\right)
+
\frac{45\Ma\Gamma_B\kappa(1-\Gamma_B)^2}
{2G^2\left[4+3\kappa(1-\Gamma_B)^2\right]}.
\label{eq:Rec-explicit}
\end{equation}
Equation~\eqref{eq:growth-LW-explicit} can then
be written as
\begin{equation}
\sigma_f(k)
=
\frac{2G^2}{15}
\left(Re-Re_c^{\mathrm{LW}}\right)k^2
+O(k^4).
\label{eq:growth-LW}
\end{equation}
Here, $Re_c^{\mathrm{LW}}$ is the signed neutral
value for the regular film branch. A negative
value indicates that this branch is already
long-wave unstable for every positive Reynolds
number along the fixed-$\Ma$ parameter path.

The first term in Eq.~\eqref{eq:alpha1-explicit}
represents inertial destabilization, the second
is the restoring contribution of gravity normal
to the substrate, and the third is Marangoni
stabilization. For $\mathcal{K}>G\cot\theta$, substrate
compliance reduces the gravity contribution.
Increasing $\mathcal{K}$ therefore raises the film-mode
threshold towards its rigid-wall value. In the
rigid, surfactant-free limit,
\begin{equation}
Re_c^{\mathrm{LW}}
=\frac{5\cot\theta}{2G}.
\end{equation}
Equivalently, defining the Nusselt mean-flow
Reynolds number by $Re_N=(G/3)Re$ gives
$Re_{N,c}=5\cot\theta/6$.

The parameters $Ca$, $Ka$, $\mathcal{I}$, $\mathcal{D}$, $\mathcal{T}$,
$\mathcal{B}$, $Pe_b$, $Pe_s$, and $k_s$ do not appear
in Eq.~\eqref{eq:Rec-explicit}. This independence
applies only to the leading long-wave threshold
of the regular film branch. Capillarity,
substrate dynamics, and surfactant transport
remain present in the finite-wavenumber
eigenvalue problem. In particular, for finite
positive $k_s$, the wave frequency tends to
zero as $k\to0$, allowing the leading sorption
perturbation to equilibrate. Setting $k_s=0$
before taking the long-wave limit is a different
limiting problem.

The leading threshold depends on the actual
base coverage $\Gamma_B$. If the equilibrium
base state is chosen with $\Gamma_B=\Gamma_e$,
this substitution may be made directly in
Eqs.~\eqref{eq:alpha1-explicit}
and \eqref{eq:Rec-explicit}. If
$\Gamma_B\ne\Gamma_e$, the reference coverage
$\Gamma_e$ also affects the base surface tension,
but that tension first contributes beyond the
order retained in Eq.~\eqref{eq:alpha1-explicit}.

Finally, the identity $\Ma=Re\,Mn$ requires the
parameter held fixed during a Reynolds-number
variation to be specified. Equation~%
\eqref{eq:Rec-explicit} applies at fixed $\Ma$,
as in the present parameter study. If $Mn$ is
held fixed instead, substitution of
$\Ma=Re\,Mn$ into
Eq.~\eqref{eq:growth-LW-explicit} gives
\begin{equation}
Re_{c,Mn}^{\mathrm{LW}}
=
\frac{
\displaystyle
\frac{G\cot\theta}{3}
\left(1-\frac{G\cot\theta}{\mathcal{K}}\right)
}{
\displaystyle
\frac{2G^2}{15}
-
\frac{3Mn\Gamma_B\kappa(1-\Gamma_B)^2}
{4+3\kappa(1-\Gamma_B)^2}
}.
\label{eq:Rec-Mn}
\end{equation}
For $\mathcal{K}>G\cot\theta$, a positive Reynolds-number
onset along this fixed-$Mn$ path requires a
positive denominator in Eq.~\eqref{eq:Rec-Mn}.
If that denominator is nonpositive, this regular
film branch has no finite positive
Reynolds-number destabilization threshold.

We illustrate the consequences of Eq.~\eqref{eq:Rec-explicit}
for $\Gamma_B=\Gamma_e$, $G=2$, $\theta=30^\circ$ and $\kappa=10$.
All comparisons use the fixed-$\Ma$ convention, with
$\Ma=Re\,Mn$. The foundation coefficient is denoted by $\mathcal K$
throughout.
The quantity $\mathcal K_0$ denotes the foundation
coefficient at which the signed long-wave neutral
Reynolds number vanishes.
The plots display nonnegative Reynolds values. A negative signed
neutral value of Eq.~\eqref{eq:Rec-explicit} indicates the absence
of a positive Reynolds-number onset for this film branch, since
Eq.~\eqref{eq:growth-LW} then predicts growth at every $Re>0$
for sufficiently small nonzero wavenumber.

\begin{figure}
\centering
\begin{subfigure}[t]{0.49\textwidth}
  \centering
  \includegraphics[width=\linewidth]{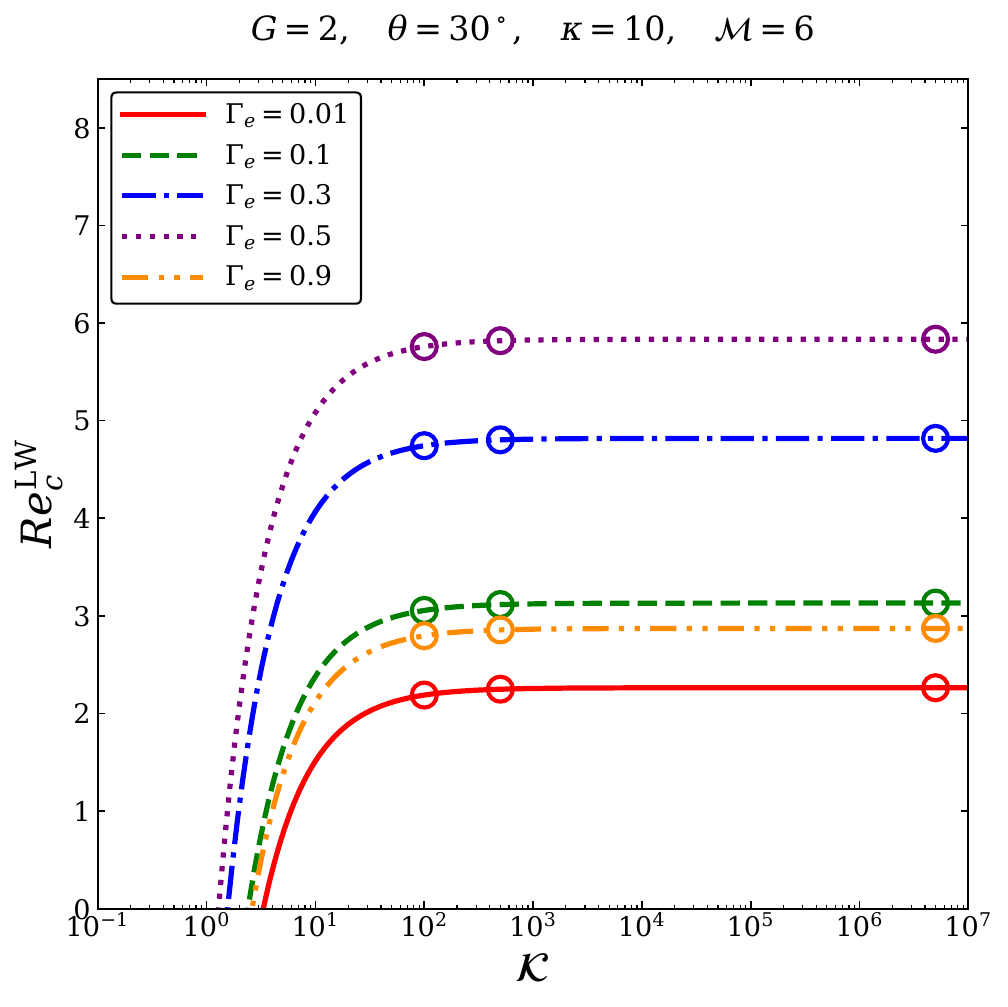}
  \caption{Coverage variation at $\Ma=6$.}
  \label{fig:asymptotic_stiffness_coverage}
\end{subfigure}\hfill
\begin{subfigure}[t]{0.49\textwidth}
  \centering
  \includegraphics[width=\linewidth]{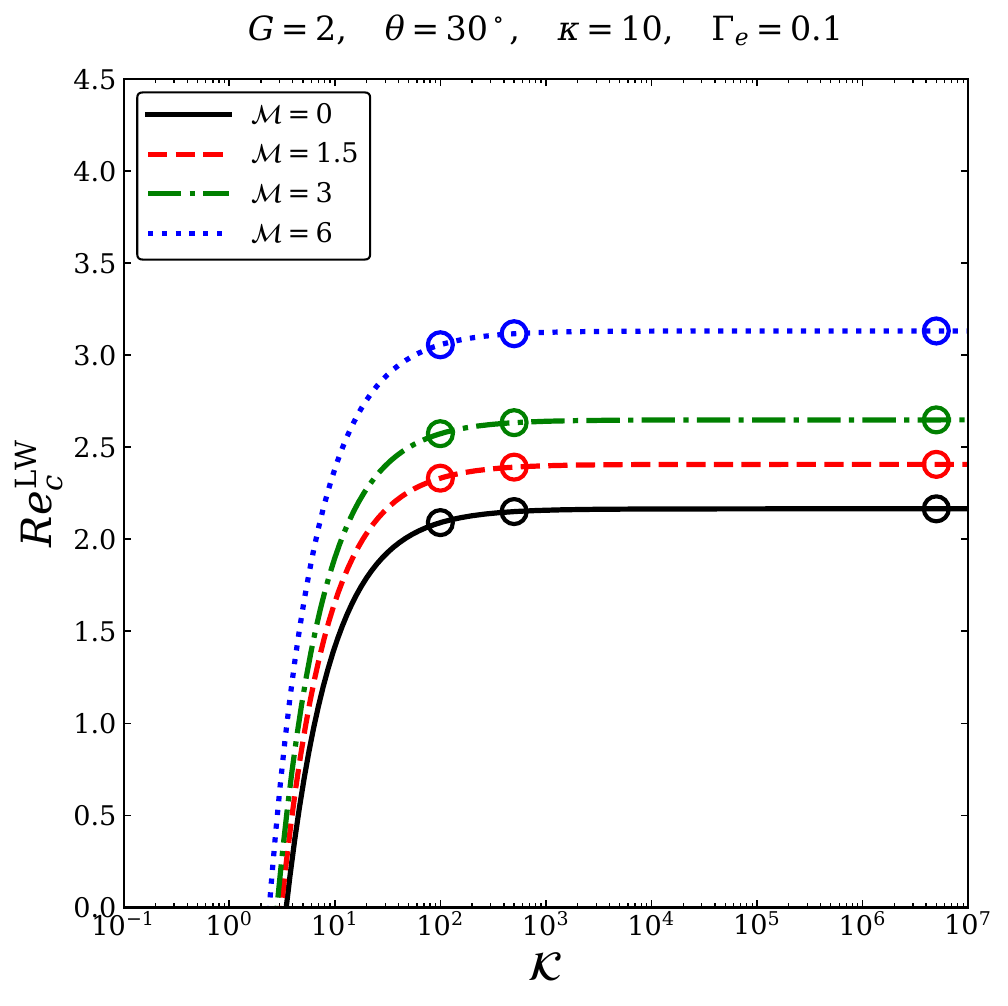}
  \caption{Marangoni variation at $\Gamma_e=0.1$.}
  \label{fig:asymptotic_stiffness_marangoni}
\end{subfigure}
\caption{Long-wave film-mode critical Reynolds number versus
foundation stiffness, with $G=2$, $\theta=30^\circ$, $\kappa=10$
and $\Gamma_B=\Gamma_e$.
(a) $\Gamma_e=0.01,0.1,0.3,0.5,0.9$ at fixed $\Ma=6$.
(b) $\Ma=0,1.5,3,6$ at fixed $\Gamma_e=0.1$.
Open circles mark $\mathcal{K}=100,500,5\times10^6$.
The stiffness axis is logarithmic; only the nonnegative part of
the signed neutral threshold is displayed.}
\label{fig:asymptotic_stiffness_comparisons}
\end{figure}

Figure~\ref{fig:asymptotic_stiffness_comparisons} shows that increasing
foundation stiffness raises the long-wave film threshold towards
its rigid-wall limit. In fact,
\begin{equation}
\frac{\partial Re_c^{\mathrm{LW}}}{\partial\mathcal K}
=\frac{5\cot^2\theta}{2\mathcal K^2}>0,
\qquad
\frac{\partial Re_c^{\mathrm{LW}}}{\partial\Ma}
=\frac{45\Gamma_e\kappa(1-\Gamma_e)^2}
{2G^2[4+3\kappa(1-\Gamma_e)^2]}.
\label{eq:asymptotic_stiffness_marangoni_derivatives}
\end{equation}
The derivative with respect to $\Ma$ is positive for
$0<\Gamma_e<1$.
Thus, stronger Marangoni forcing increases the threshold
linearly at fixed equilibrium coverage, whereas foundation
compliance lowers it by a contribution proportional to
$\mathcal K^{-1}$. All stiffness curves have the same dependence
on $\mathcal K$ and differ by constant vertical offsets.
In panel~(a), $\Gamma_e=0.5$ gives the largest threshold among
the sampled coverages; the smaller threshold at $\Gamma_e=0.9$
already reveals the non-monotone coverage response.

\begin{figure}
\centering
\begin{subfigure}[t]{0.49\textwidth}
  \centering
  \includegraphics[width=\linewidth]{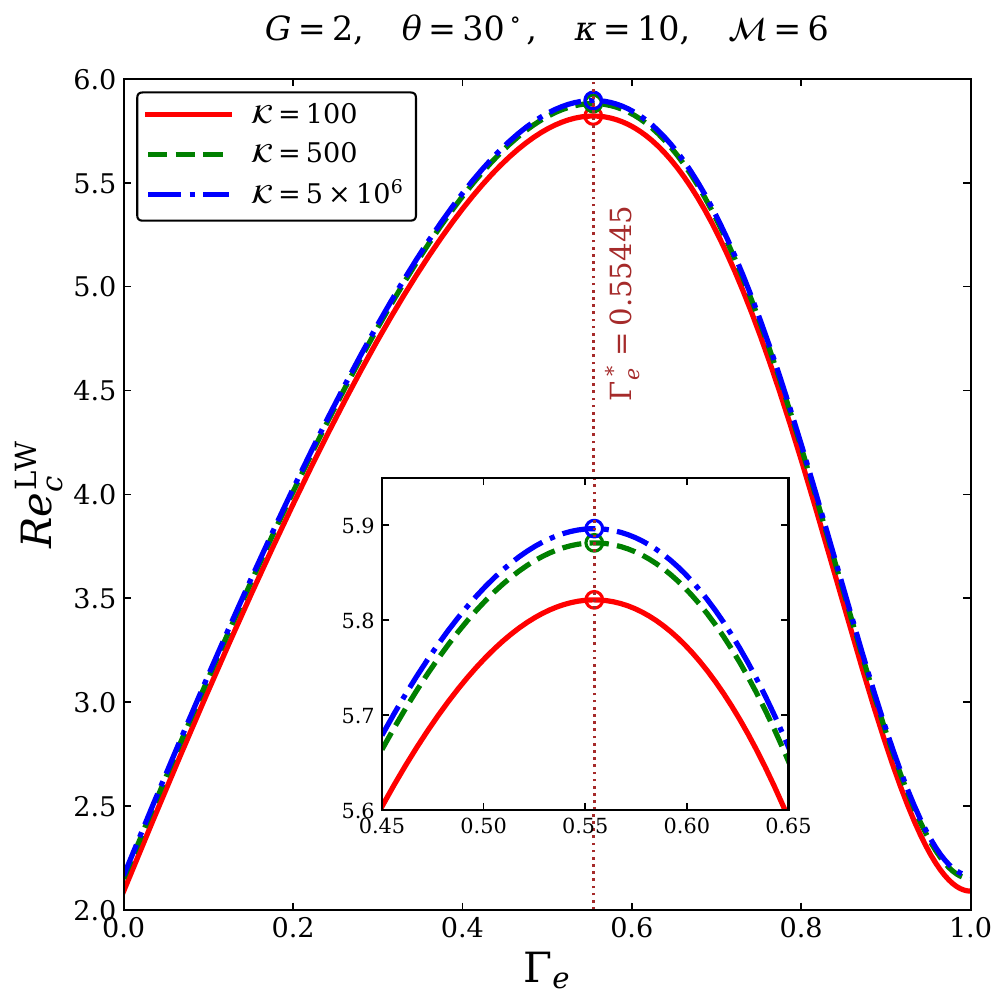}
  \caption{Coverage optimum for three foundations.}
  \label{fig:asymptotic_coverage_optimum}
\end{subfigure}\hfill
\begin{subfigure}[t]{0.49\textwidth}
  \centering
  \includegraphics[width=\linewidth]{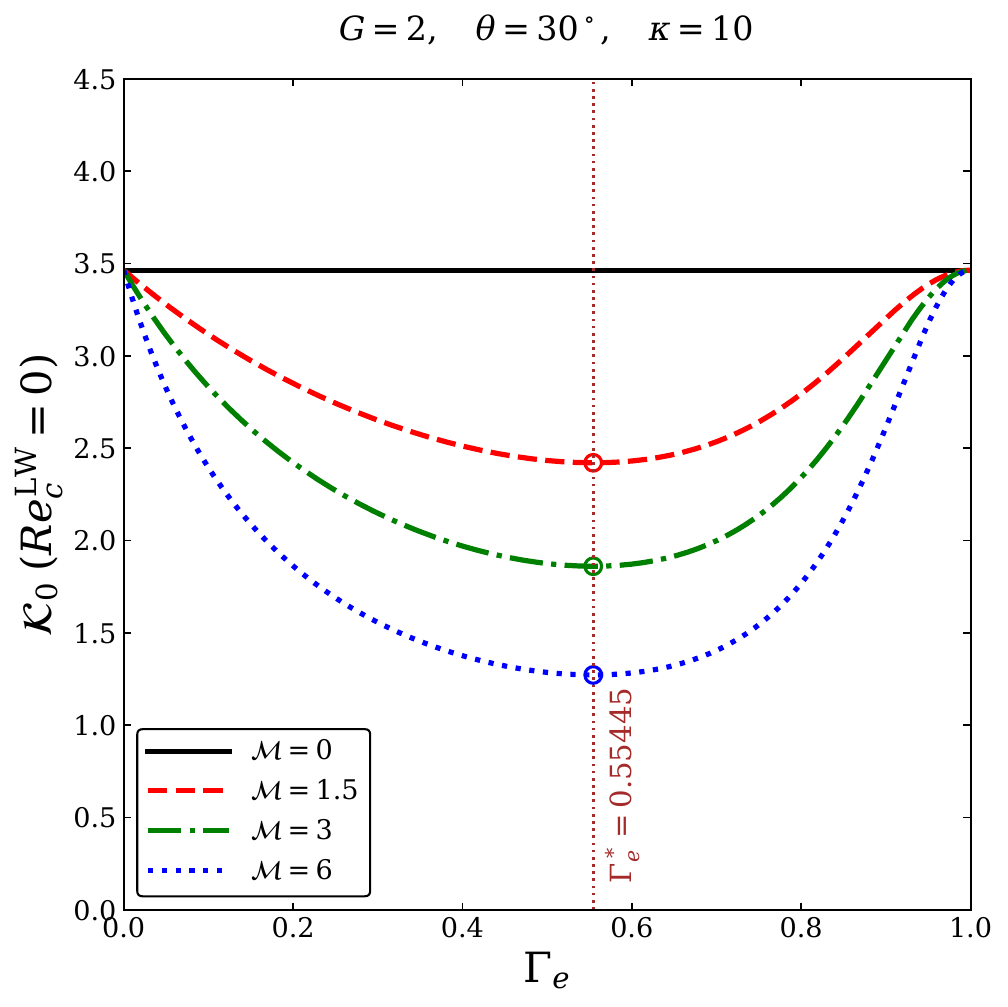}
  \caption{Foundation stiffness at zero signed threshold.}
  \label{fig:asymptotic_zero_threshold}
\end{subfigure}
\caption{Equilibrium-coverage dependence of the long-wave film
threshold for $G=2$, $\theta=30^\circ$ and $\kappa=10$.
(a) $Re_c^{\mathrm{LW}}$ at $\mathcal M=6$ for
$\mathcal K=100,500,5\times10^6$; the inset resolves the peak region
$0.45\leq\Gamma_e\leq0.65$, $5.6\leq Re_c^{\mathrm{LW}}\leq5.95$.
(b) $\mathcal{K}_0$ defined by $Re_c^{\mathrm{LW}}=0$, for
$\Ma=0,1.5,3,6$.
Open circles identify the extrema. The dotted vertical line in
both panels marks $\Gamma_e^*=0.55445$.
The non-monotone coverage dependence is the behaviour reported in
Fig.~3 of \citet{mukhopadhyay2026}, here evaluated for the
compliant-wall threshold.}
\label{fig:asymptotic_coverage_and_zero_threshold}
\end{figure}

The continuous coverage dependence in
Fig.~\ref{fig:asymptotic_coverage_and_zero_threshold}(a)
recovers the dual coverage response reported for soluble
surfactants in Fig.~3 of \citet{mukhopadhyay2026}.
Initially, increasing coverage strengthens the stabilizing
Marangoni contribution; beyond an intermediate maximum,
further loading reduces that contribution. This decrease means
weaker stabilization relative to the optimum: the surfactant
term itself remains nonnegative and tends to zero as
$\Gamma_e\to1^-$. For $\Ma>0$, differentiating
Eq.~\eqref{eq:Rec-explicit} with respect to coverage gives the
condition for the maximum,
\begin{equation}
4(1-3\Gamma_e^*)+3\kappa(1-\Gamma_e^*)^3=0.
\label{eq:asymptotic_coverage_optimum}
\end{equation}
Its left-hand side decreases strictly on $0<\Gamma_e<1$,
so the maximum is unique. At $\kappa=10$,
$\Gamma_e^*=0.55445161$, independently of $\mathcal K$,
$\Ma>0$, $G$ and $\theta$.
Foundation stiffness therefore shifts the height of the maximum
without changing its location. For $\mathcal K=100,500,5\times10^6$, the maximum
long-wave thresholds are approximately
$5.821434$, $5.881434$ and $5.896433$, respectively.
The maximum occurs at
$\Gamma_e^\ast=0.55445161$ for all three foundations.

Figure~\ref{fig:asymptotic_coverage_and_zero_threshold}(b)
expresses the same competition as a foundation-stiffness
boundary. Setting the signed threshold to zero gives
\begin{equation}
\mathcal K_0
=\frac{5\cot^2\theta}{2}
\left[
\frac{5\cot\theta}{2G}
+\frac{45\Ma\Gamma_e\kappa(1-\Gamma_e)^2}
{2G^2[4+3\kappa(1-\Gamma_e)^2]}
\right]^{-1}.
\label{eq:asymptotic_zero_threshold_stiffness}
\end{equation}
For $0<\theta<90^\circ$, $\mathcal K>\mathcal K_0$ gives a
positive long-wave neutral Reynolds value, whereas
$\mathcal K<\mathcal K_0$ gives no positive Reynolds-number
onset for the regular film branch. At $\Ma=0$,
$\mathcal K_0=G\cot\theta=2\sqrt{3}\simeq3.46410$,
independently of coverage. Positive $\Ma$ lowers this boundary,
and its minimum occurs at the same $\Gamma_e^*$ that maximizes
the stabilizing surfactant contribution. Thus optimal loading
permits a softer foundation while retaining a positive
film-mode onset threshold.

\begin{figure}
\centering
\begin{subfigure}[t]{0.49\textwidth}
    \centering
    \includegraphics[width=\linewidth]
    {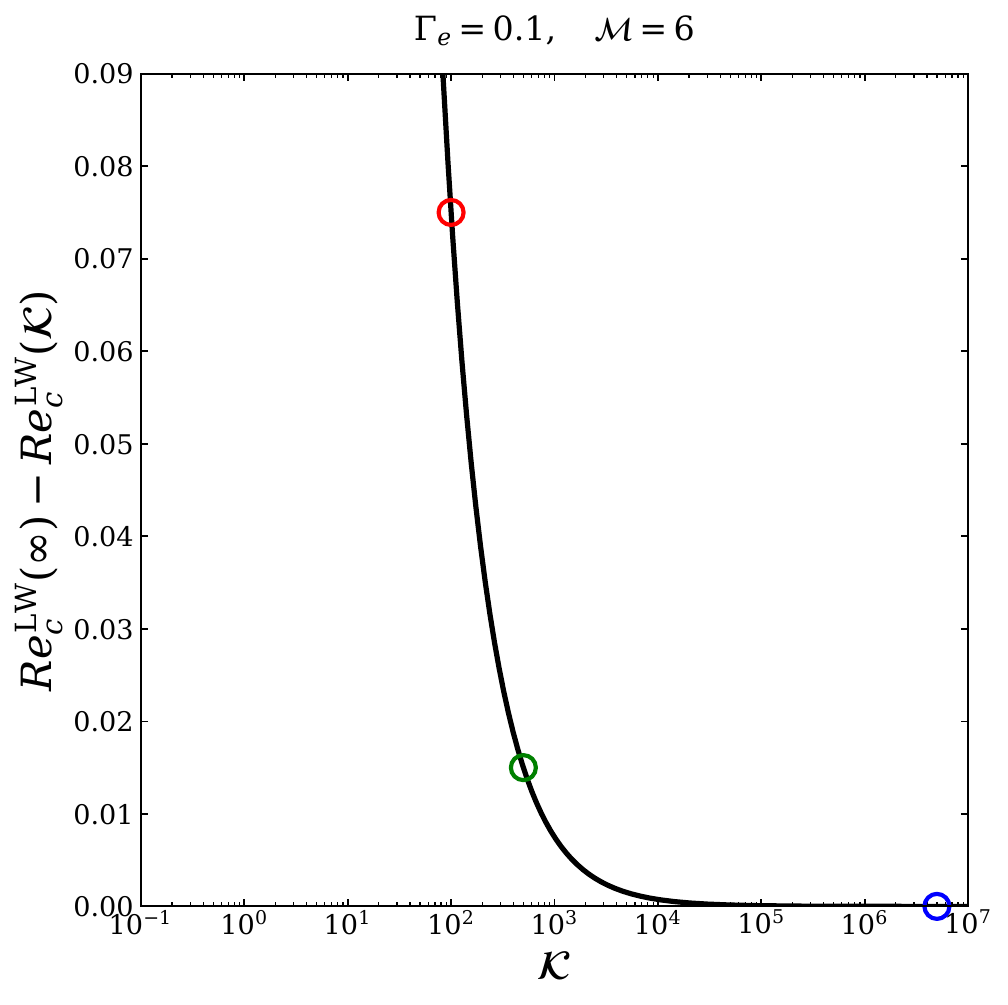}
    \caption{Compliance deficit relative to the rigid wall.}
    \label{fig:asymptotic_compliance_deficit_panel}
\end{subfigure}\hfill
\begin{subfigure}[t]{0.49\textwidth}
    \centering
    \includegraphics[width=\linewidth]
    {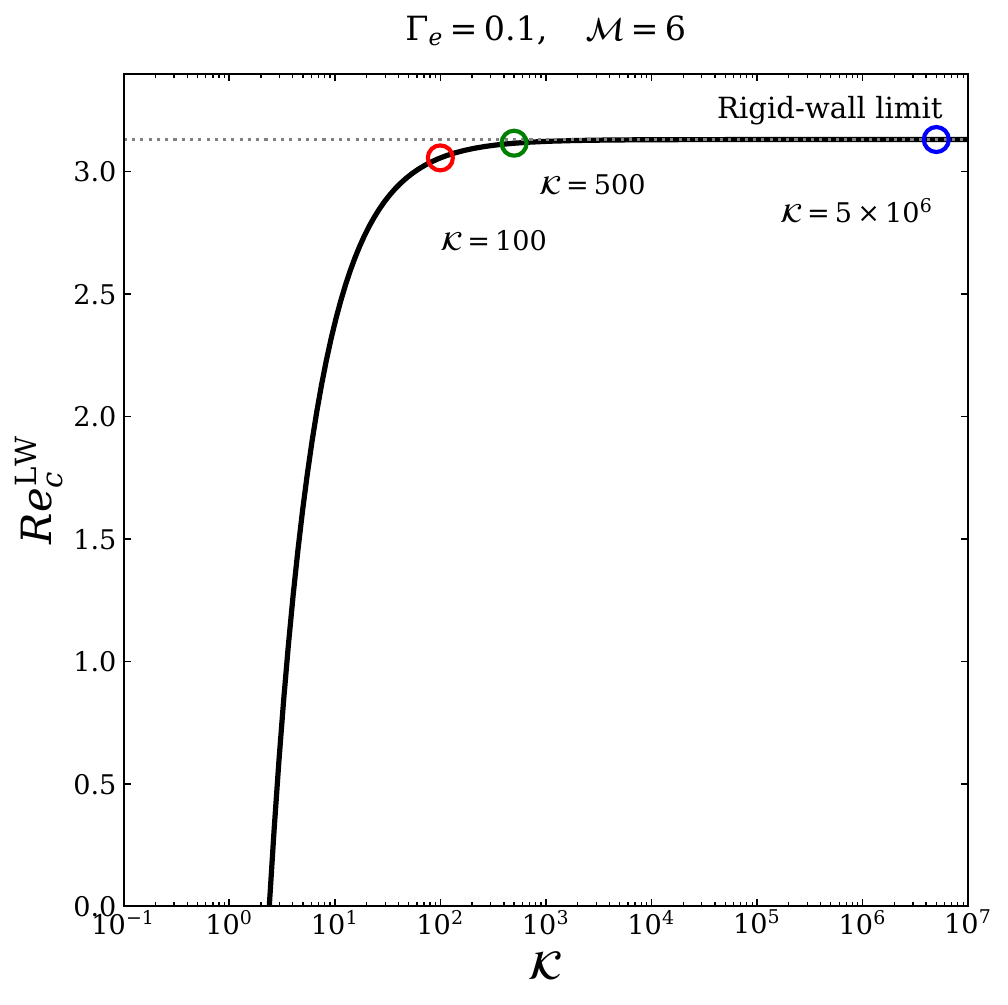}
    \caption{Long-wave threshold versus foundation stiffness.}
    \label{fig:asymptotic_foundation_threshold}
\end{subfigure}
\caption{Foundation-stiffness dependence of the long-wave film
threshold.
(a) Compliance deficit relative to the rigid-wall limit.
At $\theta=30^\circ$,
$Re_c^{\mathrm{LW}}(\infty)
-Re_c^{\mathrm{LW}}(\mathcal{K})=7.5/\mathcal{K}$,
independently of $G$, $\Gamma_e$ and $\Ma$.
(b) Absolute threshold for $G=2$, $\theta=30^\circ$,
$\kappa=10$, $\Gamma_B=\Gamma_e=0.1$ and $\Ma=6$.
The dotted horizontal line marks the rigid-wall limit.
In both panels, open circles mark
$\mathcal{K}=100,500,5\times10^6$.
Only nonnegative signed neutral values are displayed in (b).}
\label{fig:asymptotic_compliance_deficit}
\end{figure}

Finally, Fig.~\ref{fig:asymptotic_compliance_deficit}
presents two complementary measures of the foundation effect.
Panel~(a) isolates the compliance contribution by subtracting
the finite-stiffness threshold from its rigid-wall limit:
\begin{equation}
Re_c^{\mathrm{LW}}(\infty)
-Re_c^{\mathrm{LW}}(\mathcal{K})
=\frac{5\cot^2\theta}{2\mathcal{K}}.
\label{eq:asymptotic_rigid_wall_deficit}
\end{equation}
At $\theta=30^\circ$, this difference is
$7.5/\mathcal{K}$. The deficits at
$\mathcal{K}=100,500,5\times10^6$ are therefore
$0.075$, $0.015$ and $1.5\times10^{-6}$, respectively.
The displayed vertical range resolves the small correction
for the stiffer foundations. Its inverse-stiffness dependence
explains the saturation of the curves in
Fig.~\ref{fig:asymptotic_stiffness_comparisons}.

Panel (b) shows the absolute long-wave threshold for
$\Gamma_e=0.1$ and $\mathcal M=6$.
The threshold increases monotonically with foundation
stiffness, approaching the rigid-wall value
$Re_c^{\mathrm{LW}}(\infty)\simeq3.131053$.
For $\mathcal K=100,500,5\times10^6$, the corresponding
thresholds are approximately
$3.056053$, $3.116053$ and $3.131051$.
The nearly rigid case is visually indistinguishable
from the rigid-wall limit at this scale.
These results isolate the leading compliance correction,
$Re_c^{\mathrm{LW}}(\infty)-Re_c^{\mathrm{LW}}(\mathcal K)
=7.5/\mathcal K$.

The additive separation of compliance and surfactant effects
holds at the order retained in the regular long-wave expansion.
It does not imply their decoupling in the finite-wavenumber
spectrum or in nonlinear travelling waves, nor does a positive
film threshold establish stability of the additional wall and
surfactant branches.

\subsection{Numerical solution of the Orr-Sommerfeld problem}
\label{sec:os-numerical}

The coupled Orr-Sommerfeld (OS) problem is solved using Chebyshev
spectral collocation. The calculation retains fluid inertia,
bulk and surface surfactant transport, adsorption-desorption,
the free-surface stress conditions, and substrate inertia,
damping, tension, bending and foundation stiffness.
The base state is
\begin{equation}
 U(z)=G\left(z-\frac{z^2}{2}\right),\qquad
 \overline{\Gamma}=\Gamma_e,\qquad
 \overline{C}=\frac{\Gamma_e}{\kappa(1-\Gamma_e)},
 \qquad 0\leq z\leq1.
 \label{eq:osn-base}
\end{equation}
All calculations use the thickness-based spatial and temporal
coordinates of the OS formulation.

We consider temporal disturbances proportional to
$e^{ik(x-\alpha t)}=e^{\lambda t+ikx}$, where the wavenumber
$k$ is real and $\alpha=\alpha_r+i\alpha_i$ is the complex
phase speed. Therefore,
\begin{equation}
 \lambda=-ik\alpha,\qquad
 \sigma=\operatorname{Re}\lambda=k\alpha_i.
 \label{eq:osn-growth}
\end{equation}
Positive $\sigma$ corresponds to exponential amplification,
whereas negative $\sigma$ corresponds to decay. Thus,
$\alpha_i$ is the imaginary part of the phase speed and
$k\alpha_i$ is the temporal amplitude-growth rate.

The streamfunction and bulk-concentration amplitudes are
approximated by truncated Chebyshev expansions,
\begin{equation}
 \psi_N(z)=\sum_{m=0}^{N}a_m T_m(2z-1),\qquad
 \widehat C_N(z)=\sum_{m=0}^{N}c_m T_m(2z-1),
 \label{eq:osn-expansion}
\end{equation}
and collocated at the Gauss-Lobatto points
\begin{equation}
 z_j=\frac{1+\cos(j\pi/N)}{2},
 \qquad j=0,\ldots,N.
 \label{eq:osn-nodes}
\end{equation}
Here $T_m$ is the Chebyshev polynomial of the first kind.
The derivative-evaluation matrices are constructed from
\[
 [\mathsf D_r]_{jm}
 =2^r T_m^{(r)}(2z_j-1).
\]
This spectral approach follows the standard treatment of
OS eigenvalue problems \citep{Orszag1971}.

Introducing $\mathcal L_k=\partial_z^2-k^2$, the interior
hydrodynamic and concentration equations become
\begin{align}
 \left[
 \mathcal L_k^2-ikRe
 \left(U\mathcal L_k-U''\right)
 \right]\psi
 &=\lambda Re\,\mathcal L_k\psi,
 \label{eq:osn-fluid}\\
 \left(\mathcal L_k-ikPe_bU\right)\widehat C
 &=\lambda Pe_b\,\widehat C.
 \label{eq:osn-bulk}
\end{align}
Four hydrodynamic rows and two concentration rows are replaced
by the corresponding boundary conditions. The free-surface
kinematic condition, surface-surfactant balance and substrate
conditions complete the discretised system.

An auxiliary substrate-velocity amplitude,
\begin{equation}
 \widehat v=\lambda\widehat\zeta=-ik\psi(0),
 \label{eq:osn-wall-velocity}
\end{equation}
is introduced to express the wall-inertia contribution as a
generalised eigenvalue problem,
\begin{equation}
 \mathsf A(Re,k)\boldsymbol X
 =\lambda\mathsf B(Re,k)\boldsymbol X,
 \label{eq:osn-eigenproblem}
\end{equation}
where
\[
 \boldsymbol X=
 \left(
 a_0,\ldots,a_N,c_0,\ldots,c_N,
 \widehat h,\widehat\Gamma,\widehat\zeta,\widehat v
 \right)^{\mathsf T}.
\]
The matrices have order $2(N+1)+4$. Row and column scaling is
applied before solving the complex generalised eigenvalue
problem. Infinite eigenvalues associated with algebraic
constraints are discarded. Finite eigenpairs are screened
using the normalised residual
\begin{equation}
 r_j=
 \frac{
 \left\|\mathsf A\boldsymbol X_j
 -\lambda_j\mathsf B\boldsymbol X_j\right\|_\infty
 }{
 \left\|\mathsf A\boldsymbol X_j\right\|_\infty
 +|\lambda_j|
 \left\|\mathsf B\boldsymbol X_j\right\|_\infty
 }.
 \label{eq:osn-residual}
\end{equation}
The screening tolerance is $r_j<10^{-6}$, with substantially
smaller residuals for the selected physical modes.

The production calculations use $N=24$, corresponding to
25 transverse collocation points. Selected eigenvalues are
checked using $N=32$ and $N=40$, with differences below
$2\times10^{-8}$ at the verification points. An independently
assembled primitive velocity-pressure formulation provides
an additional check, with agreement within $5\times10^{-10}$
for the selected finite-wavenumber eigenvalues.
Resolution checks are also used to exclude unresolved fast
wall poles in the near-rigid limit.
At the published benchmark parameters, the solver recovers
the analytical long-wave thresholds in table~1 of
Samanta~\citeyearpar{samanta2025} to four decimal places and the
surface-mode phase speed
$\alpha\simeq1.5685+0.1486i$ reported in his figure~6.

The regular film branch is identified from its long-wave
limit $\alpha_f\to G$ and continued in wavenumber by matching
the phase speed to a prediction from the preceding branch
points. Its growth rate is distinguished from the spectral
envelope:
\begin{equation}
 \sigma_f(Re,k)=\operatorname{Re}\lambda_f(Re,k),
 \qquad
 \sigma_{\max}(Re,k)
 =\max_{j\in\mathcal P}\operatorname{Re}\lambda_j(Re,k),
 \label{eq:osn-envelope}
\end{equation}
where $\mathcal P$ contains the resolved physical modes.
The marginal-stability boundary satisfies
$\sigma_{\max}(Re,k)=0$. Sign changes on the sampled
wavenumber grid are bracketed and refined using Brent's
method, with a wavenumber tolerance of $2\times10^{-7}$.
The maxima of the film-growth curves are refined using
bounded scalar optimisation.

The numerical scan covers $0.1\leq Re\leq100$ and
$5\times10^{-4}\leq k\leq4$. The intercept at $k=0$ is
supplied by the analytical long-wave result. Defining
$a_e=\kappa(1-\Gamma_e)^2$, this threshold is
\begin{equation}
 Re_c^{\mathrm{LW}}=
 \frac{5\cot\theta}{2G}
 \left(1-\frac{G\cot\theta}{\mathcal K}\right)
 +\frac{45\mathcal M\Gamma_e a_e}
 {2G^2(4+3a_e)}.
 \label{eq:osn-rec}
\end{equation}
Independent small-wavenumber calculations recover
\[
 \frac{\sigma_f}{k^2}
 \longrightarrow
 \frac{2G^2}{15}\left(Re-Re_c^{\mathrm{LW}}\right).
\]

The common parameters for the following calculations are
\begin{equation}
 \begin{aligned}
 G&=2,\qquad \theta=30^\circ,\qquad Ka=3905,
 \qquad Pe_b=Pe_s=1,\\
 k_s&=0.1,\qquad \kappa=10,\qquad
 \mathcal I=1,\qquad \mathcal T=500,\qquad
 \mathcal B=10.
 \end{aligned}
 \label{eq:osn-parameters}
\end{equation}
The baseline is $\Gamma_e=0.1$, $\mathcal M=6$ and
$\mathcal K=100$. The growth-rate dispersion curves are
evaluated at $Re=25$, whereas the neutral curves are plotted
with $Re$ on the horizontal axis.
During Reynolds-number continuation, $Ka$ and the
viscosity-scaled Marangoni coefficient $\mathcal M=Re\,Mn$
are held fixed, with
\begin{equation}
 Ca(Re)=\frac{Re^{2/3}}{G^{1/3}Ka}.
 \label{eq:osn-ca}
\end{equation}
At $Re=25$, this gives
$Ca\simeq1.7378\times10^{-3}$.
The substrate sweep follows
$\mathcal D=\sqrt{\mathcal K}$ with $\mathcal I=1$.
Consequently, damping varies together with foundation
stiffness, while the nominal damping ratio of the bare
foundation oscillator remains equal to $1/2$.

\begin{figure}
 \centering
 \begin{subfigure}[t]{0.49\linewidth}
  \centering
  \includegraphics[width=\linewidth]
  {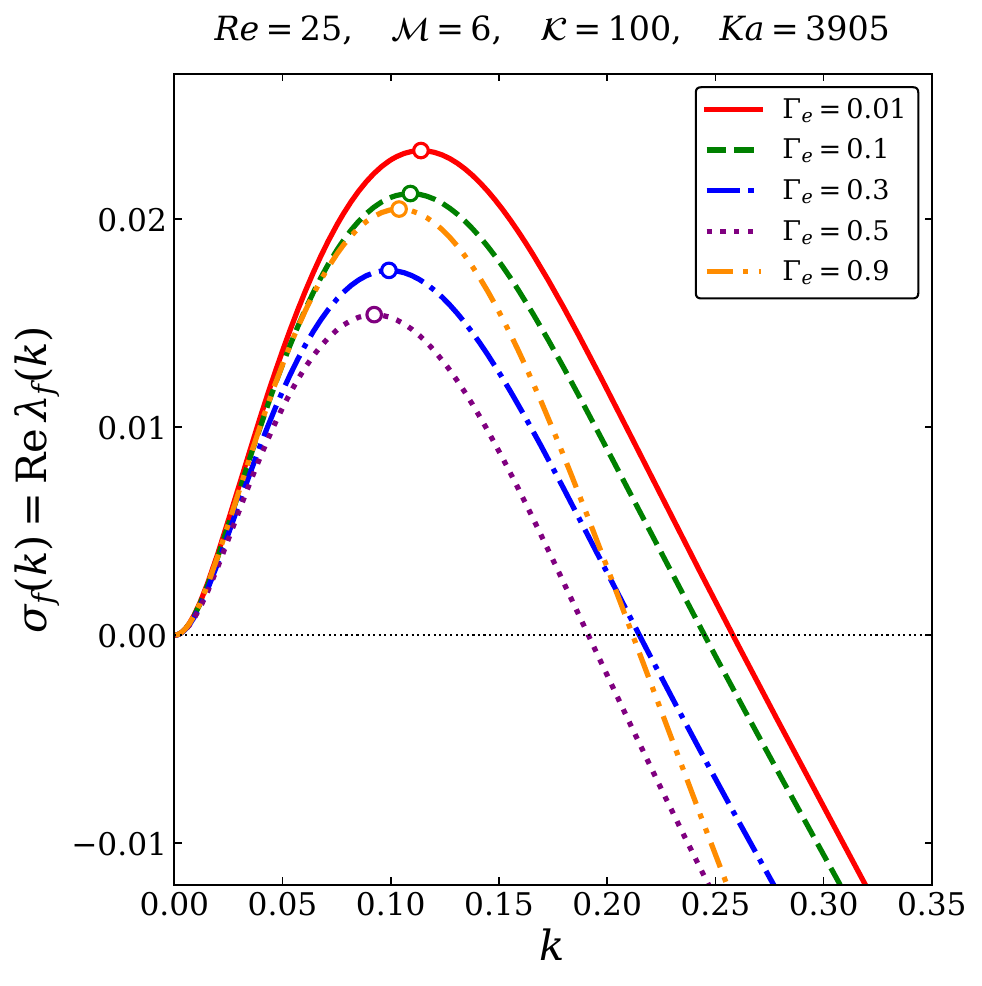}
  \caption{Film-branch growth at $Re=25$.}
 \end{subfigure}\hfill
 \begin{subfigure}[t]{0.49\linewidth}
  \centering
  \includegraphics[width=\linewidth]
  {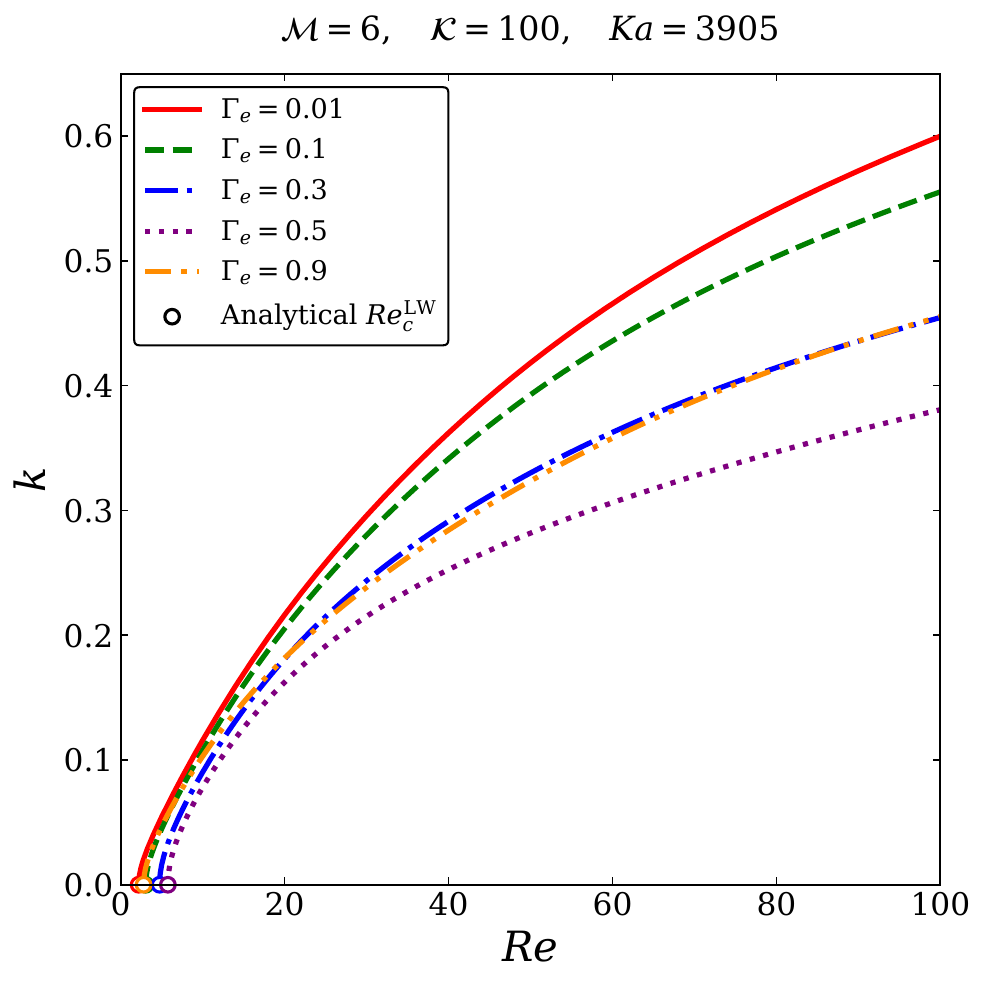}
  \caption{Marginal stability in the $(Re,k)$ plane.}
 \end{subfigure}
 \caption{
 Effect of equilibrium coverage
 $\Gamma_e=0.01,0.1,0.3,0.5,0.9$ at
 $\mathcal M=6$, $\mathcal K=100$ and $Ka=3905$.
 (a) Film-branch temporal growth rate
 $\sigma_f(k)=k\alpha_{i,f}$ at $Re=25$.
 Open circles mark the refined maxima, and the dotted
 horizontal line denotes zero growth.
 (b) Neutral boundaries satisfying $\sigma_{\max}=0$.
 Open circles at $k=0$ mark the analytical long-wave
 thresholds $Re_c^{\mathrm{LW}}$.
 The remaining parameters are given in
 \eqref{eq:osn-parameters}.
 }
 \label{fig:osn-coverage}
\end{figure}

Figure~\ref{fig:osn-coverage} shows the effect of equilibrium
coverage at $\mathcal M=6$ and $\mathcal K=100$.
Increasing $\Gamma_e$ from $0.01$ to $0.5$ reduces the
maximum film-growth rate and narrows the unstable
wavenumber interval at $Re=25$.
For $\Gamma_e=0.01,0.1,0.3,0.5,0.9$, the analytical
long-wave thresholds are approximately
$2.1891,3.0561,4.7431,5.7585,2.7965$, respectively.
Thus, $\Gamma_e=0.5$ gives the strongest stabilisation
among the sampled coverages, while the stabilising
influence weakens at $\Gamma_e=0.9$.

This non-monotone coverage dependence is consistent with
the behaviour reported by Samanta~\citeyearpar{samanta2025}.
The factor $\Gamma_e a_e/(4+3a_e)$ in
\eqref{eq:osn-rec} accounts for the reduction of long-wave
stabilisation as the equilibrium coverage approaches
saturation. The onset threshold does not, however,
determine the entire dispersion curve. Although the
long-wave threshold at $\Gamma_e=0.9$ is lower than that
at $\Gamma_e=0.1$, its peak growth at $Re=25$ is slightly
smaller and its unstable band is narrower.
For $Re>Re_c^{\mathrm{LW}}$, the unstable region lies
below the upper neutral boundary; disturbances above
this boundary decay.

\begin{figure}
 \centering
 \begin{subfigure}[t]{0.49\linewidth}
  \centering
  \includegraphics[width=\linewidth]
  {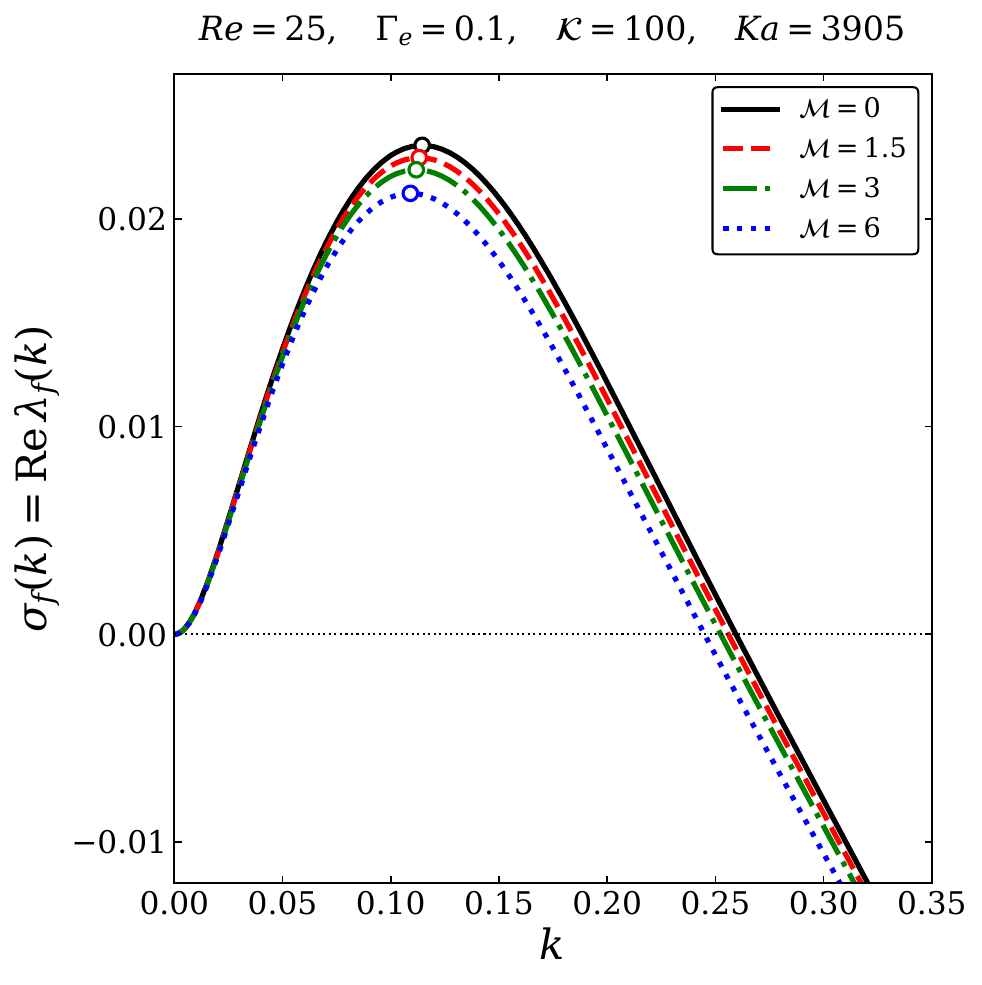}
  \caption{Film-branch growth at $Re=25$.}
 \end{subfigure}\hfill
 \begin{subfigure}[t]{0.49\linewidth}
  \centering
  \includegraphics[width=\linewidth]
  {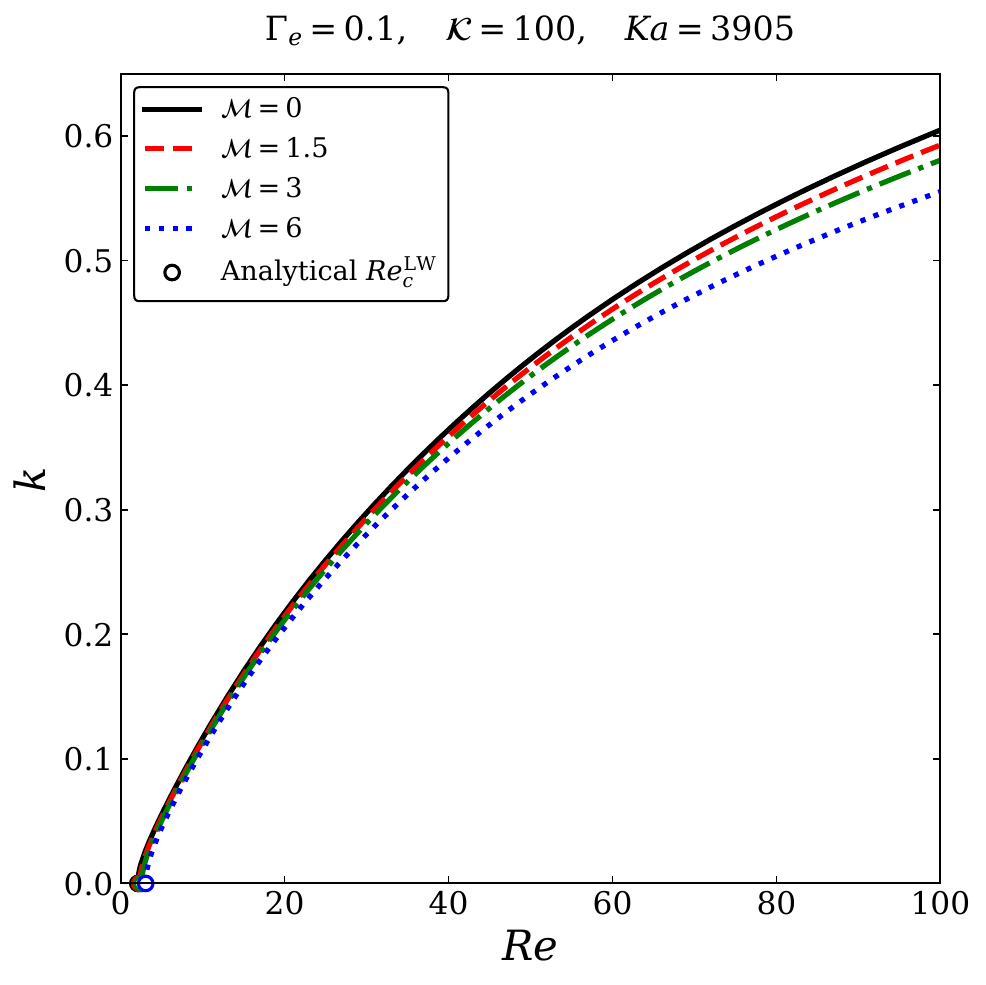}
  \caption{Marginal stability in the $(Re,k)$ plane.}
 \end{subfigure}
 \caption{
 Effect of $\mathcal M=0,1.5,3,6$ at
 $\Gamma_e=0.1$, $\mathcal K=100$ and $Ka=3905$.
 (a) Film-branch temporal growth rate at $Re=25$.
 Open circles mark the refined maxima, and the dotted
 horizontal line denotes $\sigma_f=0$.
 (b) Neutral boundaries, with analytical long-wave
 thresholds indicated by open circles at $k=0$.
 All other parameters are given in
 \eqref{eq:osn-parameters}.
 }
 \label{fig:osn-marangoni}
\end{figure}

Figure~\ref{fig:osn-marangoni} presents the Marangoni
sweep at $\Gamma_e=0.1$ and $\mathcal K=100$.
Increasing $\mathcal M$ from $0$ to $6$ raises the
long-wave threshold from $2.0901$ to $3.0561$.
At $Re=25$, the maximum film-growth rate decreases
from approximately $0.02355$ to $0.02124$, while the
neutral cutoff decreases from $k\simeq0.25948$ to
$0.24489$.
The wavenumber of maximum amplification also decreases,
from approximately $0.11452$ to $0.10906$, corresponding
to a longer preferred linear wavelength.
These results demonstrate the stabilising influence of
Marangoni stresses on the film branch.
The case $\mathcal M=0$ retains surfactant transport
and exchange, with zero Marangoni feedback in the
hydrodynamic stress condition.

\begin{figure}
 \centering
 \begin{subfigure}[t]{0.49\linewidth}
  \centering
  \includegraphics[width=\linewidth]
  {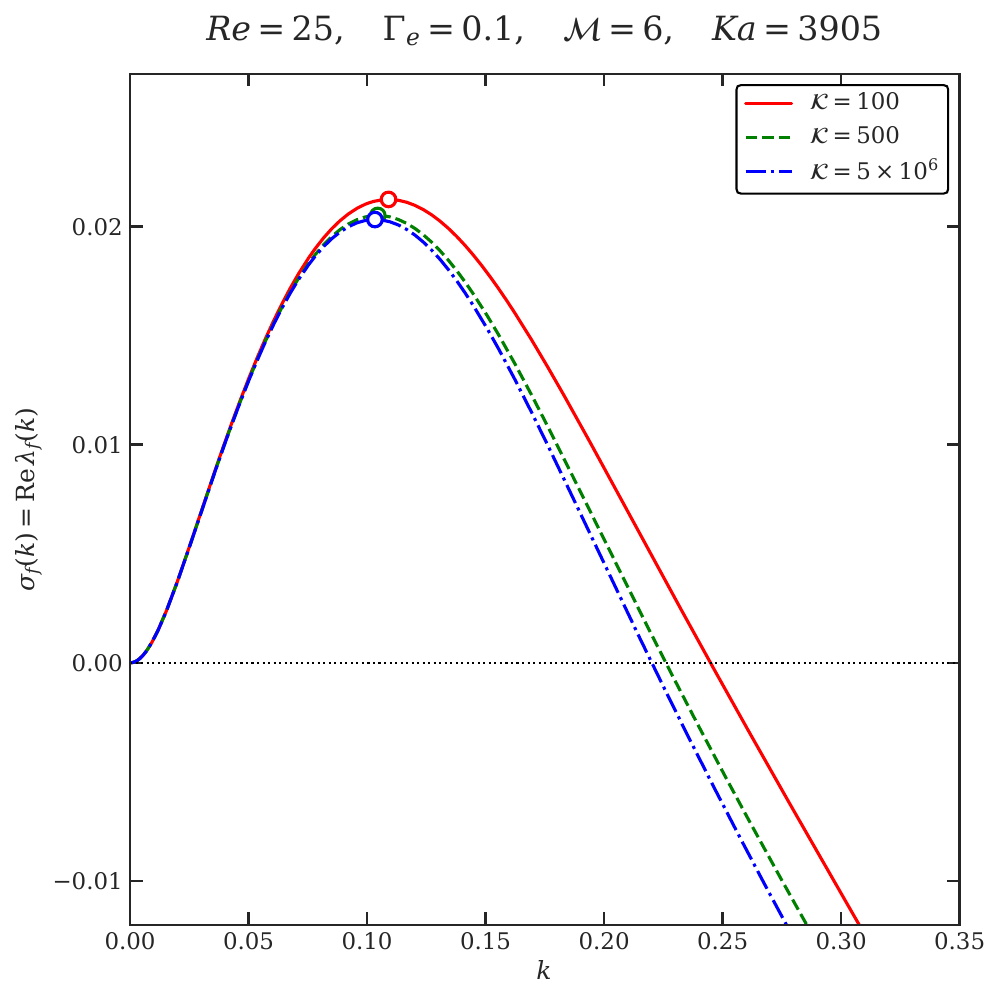}
  \caption{Film-branch growth at $Re=25$.}
 \end{subfigure}\hfill
 \begin{subfigure}[t]{0.49\linewidth}
  \centering
  \includegraphics[width=\linewidth]
  {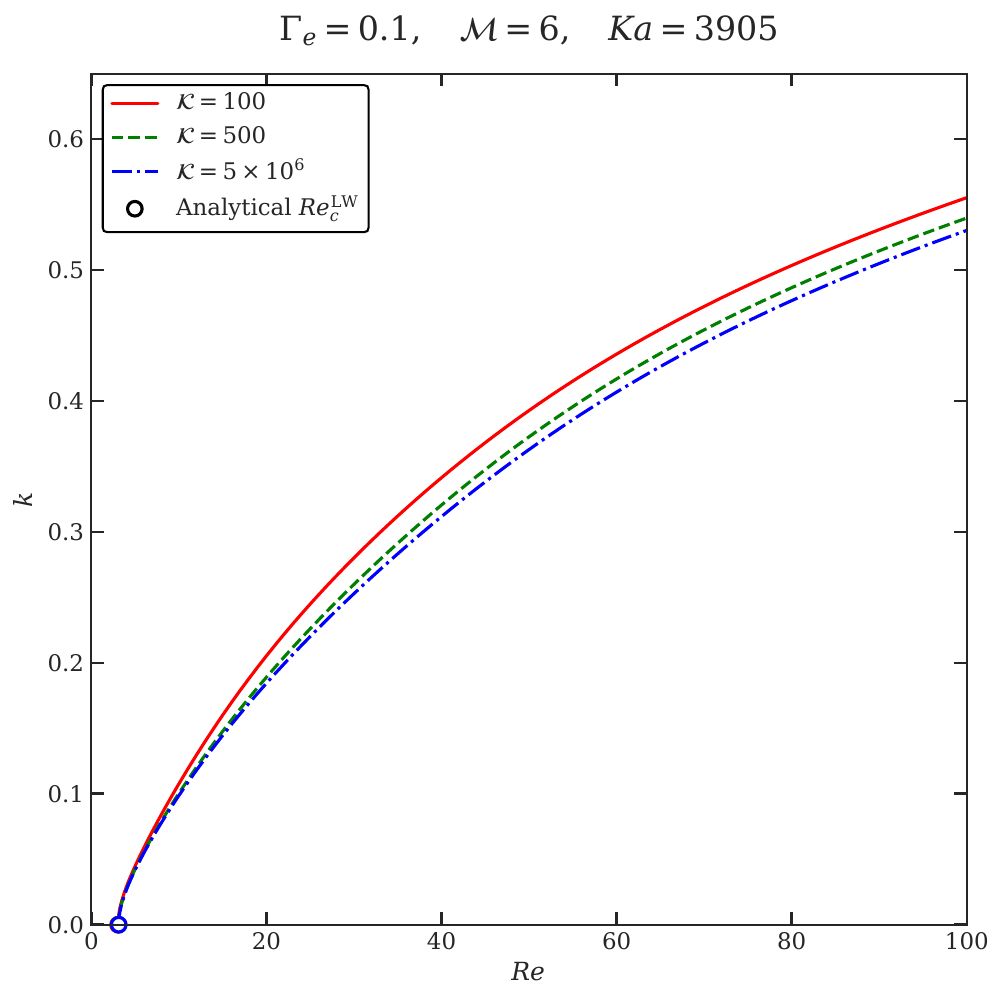}
  \caption{Marginal stability in the $(Re,k)$ plane.}
 \end{subfigure}
 \caption{
 Effect of substrate foundation stiffness
 $\mathcal K=100,500,5\times10^6$ at
 $\Gamma_e=0.1$, $\mathcal M=6$ and $Ka=3905$,
 with $\mathcal D=\sqrt{\mathcal K}$.
 (a) Film-branch temporal growth rate at $Re=25$;
 open circles identify the refined maxima.
 (b) Neutral boundaries and analytical long-wave
 intercepts.
 The remaining parameters are given in
 \eqref{eq:osn-parameters}.
 }
 \label{fig:osn-stiffness}
\end{figure}

Figure~\ref{fig:osn-stiffness} compares
$\mathcal K=100,500,5\times10^6$ along the prescribed
damping path.
The long-wave threshold increases from $3.0561$ to
$3.1161$ and $3.1311$, respectively, approaching the
rigid-wall limit.
At $Re=25$, the maximum film-growth rate decreases
from $0.02124$ to $0.02050$ and $0.02031$.
The corresponding neutral cutoffs decrease from
$0.24489$ to $0.22606$ and $0.22013$.
The growth and neutral curves therefore converge as
the substrate becomes less compliant.
These changes describe the combined
stiffness-damping path
$\mathcal D=\sqrt{\mathcal K}$.

\begin{figure}
 \centering
 \begin{subfigure}[t]{0.49\linewidth}
  \centering
  \includegraphics[width=\linewidth]
  {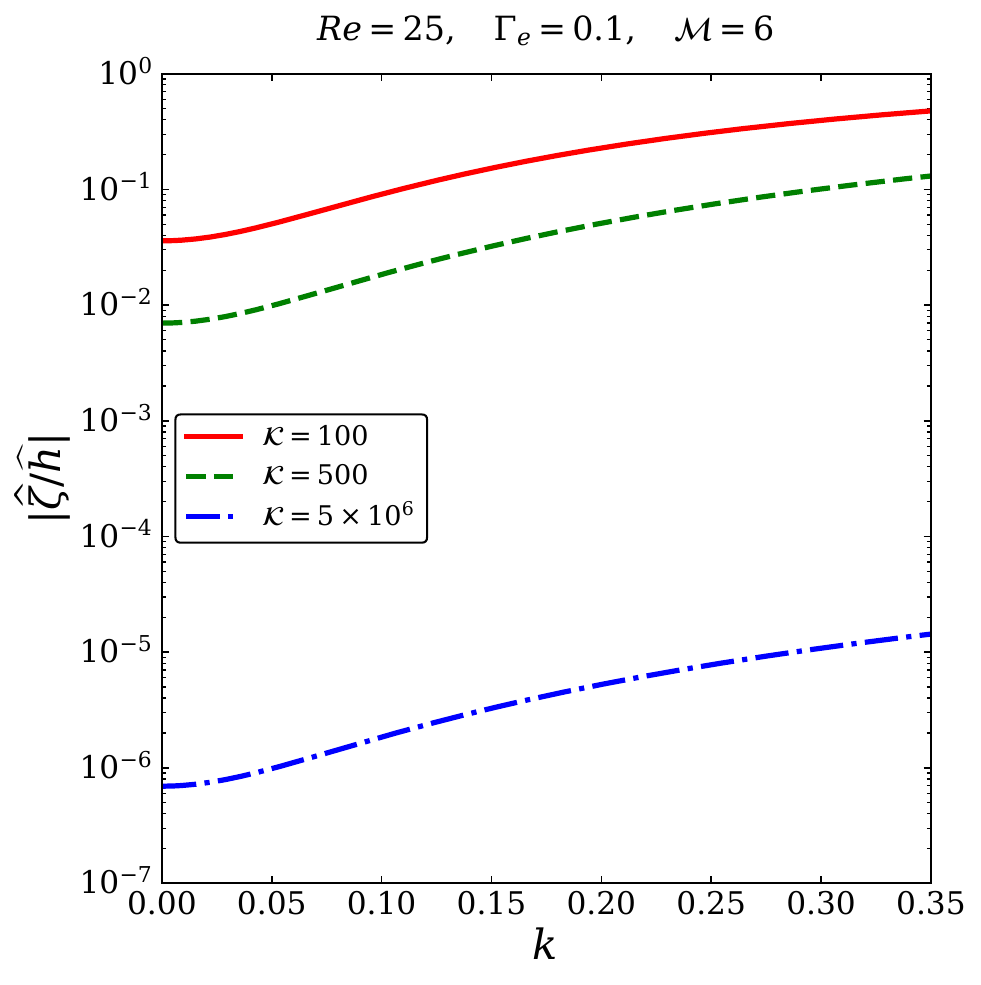}
  \caption{Displacement-ratio magnitude.}
 \end{subfigure}\hfill
 \begin{subfigure}[t]{0.49\linewidth}
  \centering
  \includegraphics[width=\linewidth]
  {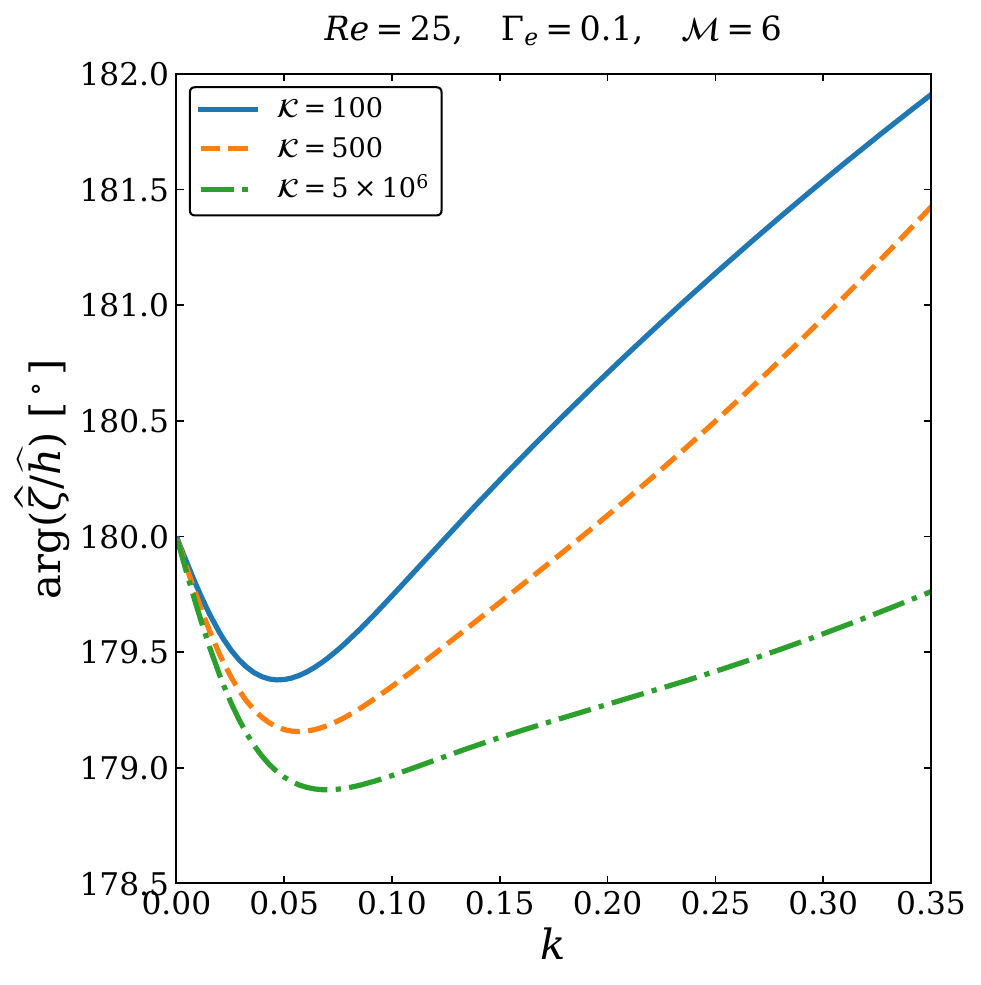}
  \caption{Unwrapped relative phase.}
 \end{subfigure}
 \caption{
 Substrate response of the film branch for
 $\mathcal K=100,500,5\times10^6$ at
 $Re=25$, $\Gamma_e=0.1$, $\mathcal M=6$
 and $Ka=3905$, with
 $\mathcal D=\sqrt{\mathcal K}$.
 (a) Magnitude
 $|\widehat\zeta_f/\widehat h_f|$
 on a logarithmic scale.
 (b) Unwrapped phase
 $\arg(\widehat\zeta_f/\widehat h_f)$
 in degrees.
 The phase remains close to $180^\circ$,
 indicating nearly opposite substrate and
 free-surface displacements.
 }
 \label{fig:osn-response}
\end{figure}

The eigenvectors also determine the substrate response
associated with the film branch. We define
\begin{equation}
 R_f(k)=\frac{\widehat\zeta_f(k)}{\widehat h_f(k)},
 \qquad
 A_f(k)=|R_f(k)|,
 \qquad
 \Delta\phi_f(k)=
 \operatorname{unwrap}\left[\arg R_f(k)\right].
 \label{eq:osn-response}
\end{equation}
The displacement ratio is independent of the arbitrary
eigenvector normalisation.
Figure~\ref{fig:osn-response} shows its magnitude and
phase at $Re=25$.
Over the displayed interval, the substrate displacement
increases relative to the free-surface displacement
with increasing $k$, while increasing $\mathcal K$
strongly suppresses the substrate response.

Near $k=0.1$, the magnitude decreases from approximately
$9.1\times10^{-2}$ at $\mathcal K=100$ to
$1.9\times10^{-6}$ at $\mathcal K=5\times10^6$.
This nearly five-order reduction in substrate response
accompanies a reduction of only about $4.3\%$ in the
maximum film-growth rate.
The long-wave eigenfunctions give
\begin{equation}
 R_f(k)\longrightarrow
 -\frac{G\cot\theta}{\mathcal K-G\cot\theta},
 \qquad k\to0.
 \label{eq:osn-response-lw}
\end{equation}
The predicted long-wave magnitudes are approximately
$0.03588$, $0.006977$ and $6.93\times10^{-7}$ for
the three stiffnesses.
The negative sign agrees with the numerical phase
remaining close to $180^\circ$.
Thus, the substrate and free-surface displacements
are nearly in antiphase, and their relative motion
contributes to the thickness perturbation
$\widehat H=\widehat h-\widehat\zeta$.
The response curves follow the film branch through
both its growing and decaying portions.

\begin{figure}
 \centering
 \includegraphics[width=0.85\linewidth]
 {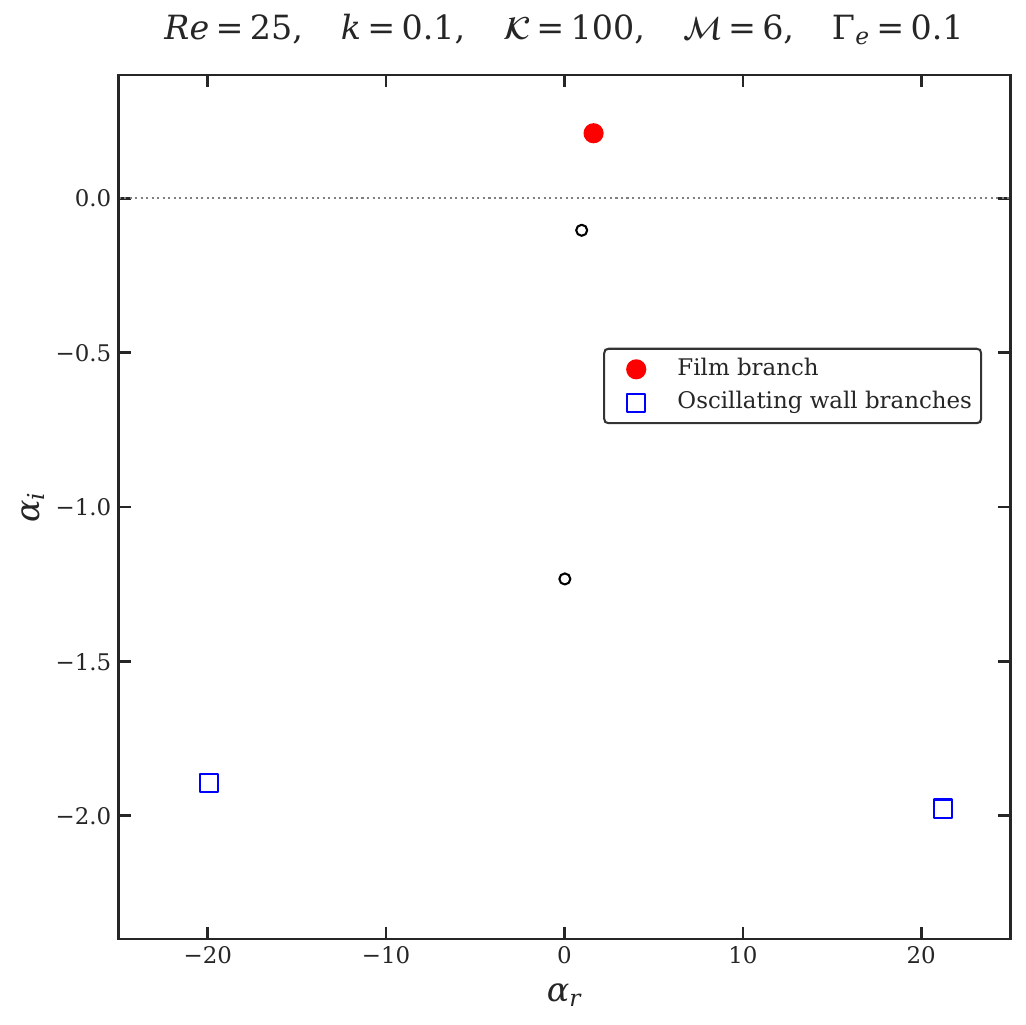}
 \caption{
 Displayed window of the baseline temporal spectrum
 in the complex phase-speed plane
 $(\alpha_r,\alpha_i)$ at
 $Re=25$, $k=0.1$, $\Gamma_e=0.1$,
 $\mathcal M=6$ and $\mathcal K=100$.
 The red filled circle identifies the growing film
 branch, and the blue open squares identify the
 damped oscillating wall branches.
 Black open circles show the other displayed
 eigenvalues.
 The dotted horizontal line denotes $\alpha_i=0$.
 Since $k$ is fixed, the temporal growth rate is
 $\sigma=k\alpha_i=0.1\alpha_i$.
 }
 \label{fig:osn-spectrum}
\end{figure}

Figure~\ref{fig:osn-spectrum} shows a window of the
baseline temporal spectrum at $Re=25$ and the fixed
real wavenumber $k=0.1$.
The spectrum is computed using the primitive
formulation at $N=32$ and displayed in the complex
phase-speed plane using $\alpha=i\lambda/k$.
The growing film mode has
\begin{equation}
 \alpha_f\simeq1.63072+0.21048i,
 \qquad
 \lambda_f\simeq0.021048-0.163072i.
 \label{eq:osn-baseline}
\end{equation}
Its temporal growth rate is therefore
$\sigma_f\simeq0.021048$.

The two highlighted oscillating wall branches have
$\alpha\simeq-19.9214-1.8948i$ and
$21.2018-1.9766i$, corresponding to temporal growth
rates of approximately $-0.18948$ and $-0.19766$.
Both wall branches are damped at the baseline
parameters and remain resolved when the primitive
collocation degree is increased to 40 and 48.
Only one growing mode is found in the baseline
spectrum.

\begin{figure}
 \centering
 \includegraphics[width=0.85\linewidth]
 {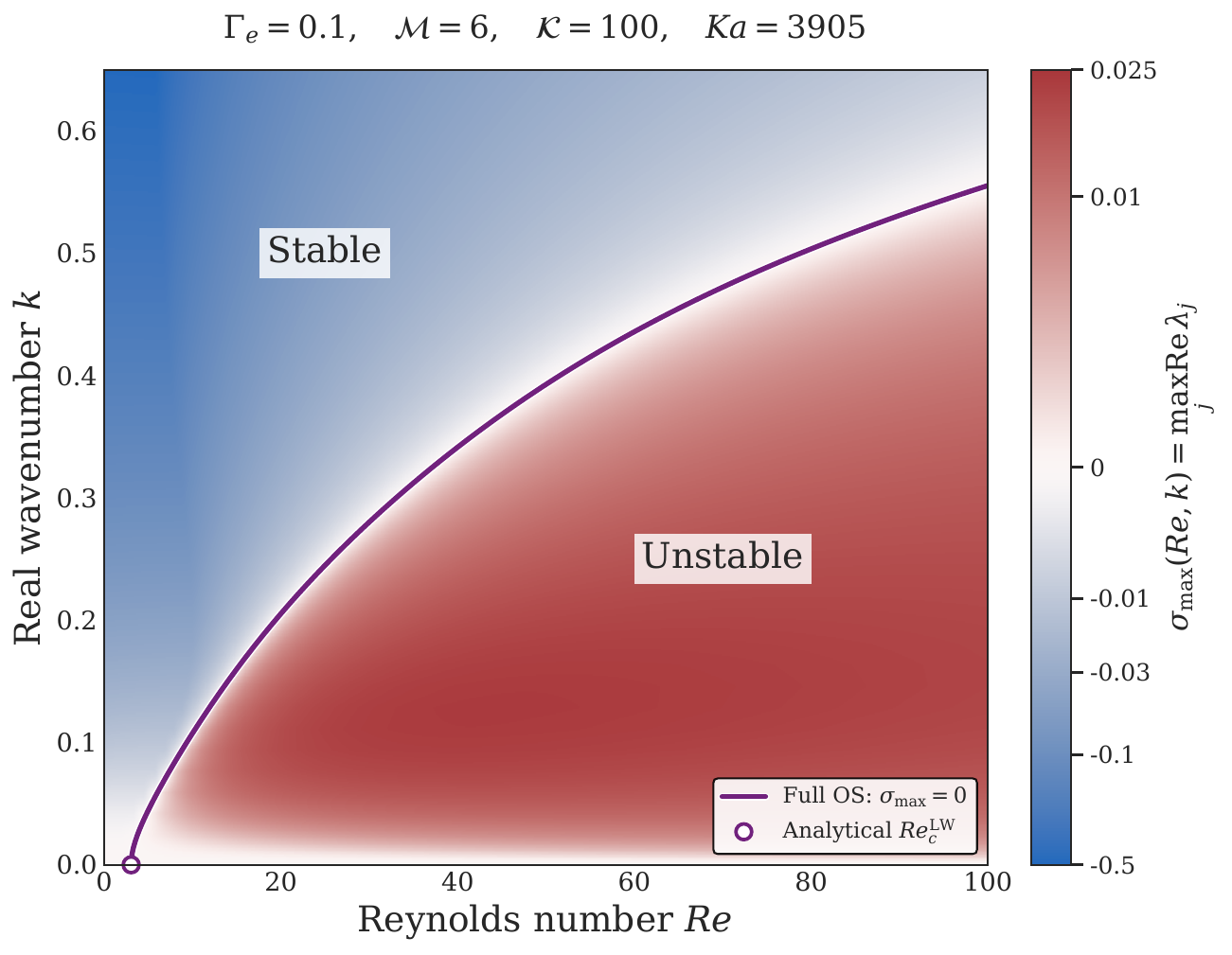}
 \caption{
 Baseline temporal stability map at
 $\Gamma_e=0.1$, $\mathcal M=6$,
 $\mathcal K=100$ and $Ka=3905$.
 Colour denotes the spectral envelope
 $\sigma_{\max}(Re,k)=
 \max_{j\in\mathcal P}\operatorname{Re}\lambda_j$
 at each Reynolds number and real wavenumber.
 Blue indicates decay, red indicates growth,
 and the pale zero-growth level separates them.
 The purple curve is the full-OS neutral boundary.
 The open circle at $k=0$ marks
$Re_c^{\mathrm{LW}}\simeq3.0561$.
 The colourbar gives the actual temporal
 growth-rate values.
 }
 \label{fig:osn-map}
\end{figure}

The stability map in figure~\ref{fig:osn-map} displays
the baseline spectral envelope
$\sigma_{\max}(Re,k)$ at
$\Gamma_e=0.1$, $\mathcal M=6$ and $\mathcal K=100$.
Blue shading denotes decay and red shading denotes
amplification. The purple curve is the directly
refined neutral boundary.
The unstable wavenumber interval broadens as $Re$
increases above the long-wave threshold, while
sufficiently short disturbances remain damped.
The largest amplification occurs at a finite
wavenumber, whereas the growth of the regular film
branch vanishes as $k\to0$.

The colour field is rendered using bilinear
interpolation of the saved spectral envelope onto
a $900\times600$ display mesh.
A zero-centred nonlinear colour scale resolves weak
amplification and stronger decay simultaneously;
the colourbar retains the original growth-rate units.
The neutral curve is determined independently by
eigenvalue-root refinement.
Changes in colour intensity within the stable
region therefore do not define additional neutral
boundaries. Changes in the identity of the
least-damped mode can also produce changes in the
slope of the spectral envelope.

For comparison with a periodic nonlinear calculation
of length $L=250$ in the same coordinates, the
admissible wavenumbers are $k_n=2\pi n/L$.
At the baseline, the continuous maximum occurs at
$k_{\mathrm{peak}}\simeq0.10906$, with
$\sigma_{f,\mathrm{peak}}\simeq0.021236$
and wavelength
$2\pi/k_{\mathrm{peak}}\simeq57.61$.
The most amplified admissible Fourier mode is $n=4$.
The fundamental mode $k_1=2\pi/250$ becomes unstable
at $Re\simeq3.8011$, above the long-wave threshold
$3.0561$.
These results provide the linear onset, initial
amplification rates and preferred wavelengths for
comparison with nonlinear simulations at matching
parameters. The eventual saturation amplitude and
changes in wave number are determined by the
subsequent nonlinear evolution.

\subsection{Integral conservation laws}
\label{subsec:integral_conservation}

Define the volumetric flow rate by
\begin{equation}
    q(x,t)=\int_{\zeta(x,t)}^{h(x,t)}u(x,z,t)\,\dd z.
    \label{eq:flux_definition}
\end{equation}

Integration of continuity gives
\begin{equation}
    q_x-u(h)h_x+u(\zeta)\zeta_x
    +w(h)-w(\zeta)=0.
\end{equation}
Using the boundary conditions,
\begin{equation}
    H_t+q_x=0.
    \label{eq:kinematic_mass_integral}
\end{equation}

Let
\begin{equation}
    \phi(x,t)=C(x,h(x,t),t),
    \qquad
    \chi(x,t)=\int_\zeta^h(C-\phi)\,\dd z.
    \label{eq:chi_definition_early}
\end{equation}
The bulk surfactant inventory is
\begin{equation}
    \int_\zeta^hC\,\dd z
    =\chi+H\phi.
    \label{eq:bulk_inventory_early}
\end{equation}

Depth integration of
Eq.~\eqref{eq:concentration}, including the normal-flux
conditions at both boundaries, yields
\begin{equation}
    \varepsilon
    \left[
        (\chi+H\phi)_t
        +
        \frac{\partial}{\partial x}
        \int_\zeta^h uC\,\dd z
    \right]
    =
    \frac{\varepsilon^2}{Pe_b}
    \frac{\partial}{\partial x}
    \int_\zeta^h C_x\,\dd z
    -k_s\sqrt{1+\varepsilon^2h_x^2}\left[\kappa(1-\Gamma)C(h)-\Gamma\right].
    \label{eq:bulk_integrated_exact}
\end{equation}

Adding Eqs.~\eqref{eq:boundary_surface_surfactant}
and \eqref{eq:bulk_integrated_exact} gives
\begin{align}
    \varepsilon
    \frac{\partial}{\partial t}
    \left(\chi+H\phi+\Gamma \sqrt{1+\varepsilon^2h_x^2}\right)
    +
    \frac{\partial}{\partial x}
    \Bigg[
        &\varepsilon
        \left(
            \int_\zeta^h uC\,\dd z+\Gamma \sqrt{1+\varepsilon^2h_x^2} u(h)
        \right)
        \nonumber\\
        &-\frac{\varepsilon^2}{Pe_b}
        \int_\zeta^hC_x\,\dd z
        -\frac{\varepsilon^2}{Pe_s}
        \frac{\Gamma_x}{\sqrt{1+\varepsilon^2h_x^2}}
    \Bigg]
    =0.
    \label{eq:exact_total_local}
\end{align}

For periodic boundary conditions,
\begin{equation}
    \frac{\dd}{\dd t}
    \int_0^L
    \left(\chi+H\phi+\Gamma \sqrt{1+\varepsilon^2h_x^2}\right)\,\dd x=0.
    \label{eq:exact_total_global}
\end{equation}

The factor multiplying $\phi$ is the liquid thickness
$H$. A common vertical displacement of both boundaries
does not change the bulk inventory.

An exact identity following from Leibniz's rule is
\begin{equation}
    \int_\zeta^hC_x\,\dd z
    =
    \chi_x+H\phi_x
    +\left[C(\zeta)-\phi\right]\zeta_x.
    \label{eq:exact_bulk_diffusive_flux}
\end{equation}

The pressure reduction and depth-integrated momentum identities
are derived in appendix~\ref{app:pressure_integrals}.
The simplified streamwise balance omits the higher-order
transverse-acceleration pressure gradient, while the substrate
load retains the leading acceleration moment because it enters
without an additional streamwise derivative.

\section{Weighted-residual approximation}
\label{subsec:weighted_residual}

We now use the thickness-based coordinates introduced
in Eq.~\eqref{eq:OS_rescaling} and omit their subscripts.
The explicit long-wave parameter is absorbed through
the coordinate transformation. The values of
$\mathcal{I},\mathcal{D},\mathcal{T},\mathcal{B},\mathcal{K}$ are unchanged, and the wall
equation contains
$\mathcal I,\mathcal D,\mathcal T,\mathcal B,\mathcal K$
as defined in Table~\ref{tab:compliant_wall_coefficients}.

We employ a simplified Galerkin approximation in
the convention of \citet{mukhopadhyay2026}.
The velocity profile retains the parabolic contribution
and its leading Marangoni correction. The flow-rate
projection includes the terms linear in the Marangoni
velocity and the second-gradient streamwise viscous
terms. The wall pressure additionally retains the
leading transverse-acceleration moment.
These choices specify the reduced closure; they do
not constitute a complete higher-order expansion
of every transverse field.

The liquid occupies
$\zeta(x,t)\le z\le h(x,t)$.
Introduce the wall-fitted transverse coordinate
\begin{equation}
    H=h-\zeta,
    \qquad
    y=\frac{z-\zeta(x,t)}{H(x,t)},
    \qquad
    0\le y\le1.
    \label{eq:mapped_coordinate}
\end{equation}

The volumetric flow rate becomes
\begin{equation}
    q=H\int_0^1u(x,y,t)\,\dd y,
    \label{eq:mapped_flux}
\end{equation}
and the integrated kinematic condition is
\begin{equation}
    H_t+q_x=0.
    \label{eq:WR_mass_pre}
\end{equation}

The mapped differential operators and material-derivative identities
used below are collected in appendix~\ref{app:wr_closure}.

We introduce the basis functions
\begin{equation}
    f_0(y)=y-\frac{y^2}{2},
    \qquad
    f_1(y)=2y-3y^2.
    \label{eq:basis_functions}
\end{equation}

The velocity approximation is
\begin{equation}
    u(x,y,t)
    =
    u_0+u_1
    =
    \frac{3q}{H}f_0(y)
    +
    \frac{\Ma H}{4}\Gamma_x f_1(y).
    \label{eq:velocity_ansatz}
\end{equation}

Equation~\eqref{eq:velocity_ansatz} satisfies no slip
at the substrate, preserves the prescribed flow rate
and enforces the leading tangential-stress condition
$u_z(h)=-\Ma\Gamma_x$.
The higher-gradient tangential-stress contribution
is included through the weak projection.

The surface velocity is
\begin{equation}
    u(h)
    =
    \frac{3q}{2H}
    -
    \frac{\Ma H}{4}\Gamma_x.
    \label{eq:surface_velocity_WR}
\end{equation}

For the bulk concentration, we use
\begin{equation}
    C(x,y,t)
    =
    \phi(x,t)
    +
    \frac{3\chi(x,t)}{2H(x,t)}(1-y^2),
    \label{eq:concentration_ansatz}
\end{equation}
where
\begin{equation}
    \phi=C(h),
    \qquad
    \chi=\int_\zeta^h(C-\phi)\,\dd z.
    \label{eq:chi_definition}
\end{equation}
The wall concentration is
\begin{equation}
    C(\zeta)=\phi+\frac{3\chi}{2H}.
    \label{eq:bulk_inventory_definition}
\end{equation}

At the leading transverse-diffusion level,
\begin{equation}
    k_s\left[\kappa(1-\Gamma)\phi-\Gamma\right]
    =
    -\frac1{Pe_b}C_z(h)
    =
    \frac{3\chi}{Pe_bH^2}.
    \label{eq:source_closure}
\end{equation}
For $k_s>0$, $\kappa>0$ and $\Gamma<1$,
\begin{equation} 
\phi = \frac{ \displaystyle \Gamma+\frac{3\chi}{Pe_bk_sH^2} }{ \kappa(1-\Gamma) }. \label{eq:phi_closure} 
\end{equation}

The relation involving $C_z(h)$ in
Eq.~\eqref{eq:source_closure} is the leading
normal-flux approximation. The quadratic profile
does not satisfy the complete finite-slope
boundary conditions pointwise.
For example, in the original long-wave variables,
\begin{equation}
    \left[
        C_z-\varepsilon^2\zeta_xC_x
    \right]_\zeta
    =
    -\varepsilon^2\zeta_x\left(\phi+\frac{3\chi}{2H}\right)_x.
    \label{eq:profile_wall_residual}
\end{equation}
The full surface Fick condition also contains
$\varepsilon^2h_xC_x(h)$ and the metric $\sqrt{1+\varepsilon^2h_x^2}$.
Higher-order concentration corrections are needed
if these pointwise terms are to be resolved.
The integral transport equations are therefore
written conservatively before the profile and
exchange approximations are substituted.

To evaluate the streamwise viscous terms, define
the physical streamwise derivative
\begin{equation}
    X
    =
    \left.\partial_x\right|_y
    -
    \frac{\zeta_x+yH_x}{H}\partial_y.
    \label{eq:physical_x_operator}
\end{equation}

The inertial and viscous projection calculations are given in
appendix~\ref{app:wr_closure}. The flow-rate closure retains terms
linear in the Marangoni velocity and the selected second-gradient
viscous contributions; quadratic Marangoni inertia is omitted.

\subsection{Pressure load on the compliant substrate}
\label{subsec:wall_load}

The explicit normal viscous traction has zero linearization and
lies beyond the retained pressure order for slow wall motion.
The traction calculation and the parabolic transverse-acceleration
moment are derived in appendix~\ref{app:wr_closure}.
The acceleration moment is written as
\begin{equation}
    \mathcal A_w
    =
    Hv_t+\widetilde{\mathcal A}_w,
    \label{eq:Aw_split}
\end{equation}
The expanded expression for $\widetilde{\mathcal A}_w$ is given in
equation~\eqref{eq:Aw_explicit} in appendix~\ref{app:wr_closure}.

In thickness-based variables, the selected
pressure closure at the wall is
\begin{align}
    \Pi(\zeta)
    \simeq{}&
    G\cot\theta H
    -
    \frac{\Sigma(\Gamma)h_{xx}}
    {(1+h_x^2)^{3/2}}
    -
    \left[Xu(h)+Xu(\zeta)\right]
    \nonumber\\
    &+2\Ma\Gamma_xh_x
    +Re\,\mathcal A_w.
    \label{eq:pressure_at_wall}
\end{align}

The term $2\Ma\Gamma_xh_x$ follows from
$-2u_z(h)h_x$ in the surface normal stress.
It accompanies the Marangoni contribution to
the velocity-gradient pressure.
The latter is
\begin{align}
    Xu(h)+Xu(\zeta)
    ={}&
    \frac{3}{2H}q_x
    -\frac{3q}{2H^2}H_x
    -\frac{3q}{H^2}\zeta_x
    \nonumber\\
    &+\Ma
    \left[
        \left(\frac34H_x+\frac12\zeta_x\right)\Gamma_x
        -\frac H4\Gamma_{xx}
    \right].
    \label{eq:wall_viscous_load}
\end{align}

The curvature term in
Eq.~\eqref{eq:pressure_at_wall} retains the
surface-tension variation and the geometric
denominator as a compact completion of the
selected pressure approximation.
The leading streamwise capillary force uses
$Ca^{-1}h_{xxx}$, since the corresponding
higher-gradient corrections lie outside the
retained flow-rate truncation.

The acceleration moment
$\mathcal A_w$ is evaluated using the specified
parabolic field, rather than an unknown exact
velocity. Thus Eq.~\eqref{eq:pressure_at_wall}
is an explicit pressure closure.
It recovers the exact fluid added mass for
uniform wall translation, while its accuracy
for finite-wavenumber wall motion requires
comparison with the continuum problem.

\subsection{Conservative surfactant fluxes}
\label{subsec:surfactant_fluxes}

For the concentration profile
\eqref{eq:concentration_ansatz},
Eq.~\eqref{eq:exact_bulk_diffusive_flux} gives
\begin{equation}
    \int_\zeta^h C_x\,\dd z
    =
    H\phi_x+\chi_x+\frac{3\chi}{2H}\zeta_x.
    \label{eq:bulk_diffusive_flux}
\end{equation}

The bulk advective moment is
\begin{equation}
    \int_\zeta^h uC\,\dd z
    =
    \phi q
    +\frac{33}{40}\frac{q\chi}{H}
    +\frac{3\Ma}{80}H\chi\Gamma_x.
    \label{eq:bulk_advective_moment}
\end{equation}

The moving-boundary completion of the diffusion operator is
derived in appendix~\ref{app:wr_closure}.

The diffusion term in the bulk inventory balance
is therefore $Pe_b^{-1}\partial_x\int_\zeta^h C_x\,\dd z$.
The operator $Pe_b^{-1}\int_\zeta^h C_{xx}\,\dd z$ alone omits
the contributions required by the normal-flux
conditions on the moving boundaries.

\section{Conservative weighted-residual model}
\label{sec:complete_model}

The dependent variables are
\begin{equation}
    (H,q,\chi,\Gamma,\zeta,v),
    \qquad
    v=\zeta_t,
    \qquad
    h=H+\zeta.
\end{equation}

The film-mass equation is
\begin{equation}
    H_t+q_x=0.
    \label{eq:model_H}
\end{equation}

The flow-rate equation is
\begin{align}
    q_t
    +\frac{17}{7}\frac qHq_x
    -\frac97\frac{q^2}{H^2}H_x
    ={}&
    \frac5{2Re}
    \left(
        \frac G3H
        -\frac q{H^2}
        -\frac{\Ma}{2}\Gamma_x
    \right)
    \nonumber\\
    &+\frac{5H}{6Re}
    \left(
        \frac1{Ca}h_{xxx}
        -G\cot\theta\,h_x
    \right)
    \nonumber\\
    &+\frac1{Re}
    \Bigg[
        4\frac q{H^2}H_x^2
        -\frac9{2H}q_xH_x
        -6\frac qH H_{xx}
        +\frac92q_{xx}
        \nonumber\\
    &\hspace{13mm}
        -5\frac q{H^2}\zeta_x^2
        -\frac52\frac q{H^2}H_x\zeta_x
        -\frac{15}{4}\frac qH\zeta_{xx}
        -\frac54v_x
    \Bigg]
    \nonumber\\
    &+\Ma
    \Bigg[
        \frac{H^2}{48}\Gamma_{tx}
        +\frac{15}{224}qH\Gamma_{xx}
        +\frac{19}{336}Hq_x\Gamma_x
        +\frac5{112}qH_x\Gamma_x
    \Bigg].
    \label{eq:model_q}
\end{align}

The bulk-surfactant balance is
\begin{align}
    &\frac{\partial}{\partial t}
    \left(\chi+H\phi\right)
    +
    \frac{\partial}{\partial x}
    \left(
        \phi q
        +\frac{33}{40}\frac{q\chi}{H}
        +\frac{3\Ma}{80}H\chi\Gamma_x
    \right)
    \nonumber\\
    &\hspace{15mm}
    =
    \frac1{Pe_b}
    \frac{\partial}{\partial x}
    \left(
        H\phi_x+\chi_x+\frac{3\chi}{2H}\zeta_x
    \right)
    -
    \frac{3\chi}{Pe_bH^2}.
    \label{eq:model_chi}
\end{align}

The surface-surfactant equation is
\begin{equation}
    \Gamma_t
    +
    \frac{\partial}{\partial x}
    \left(\frac{3q\Gamma}{2H}\right)
    =
    \frac1{Pe_s}\Gamma_{xx}
    +
    \frac{3\chi}{Pe_bH^2}
    +
    \frac{\Ma}{4}
    \frac{\partial}{\partial x}
    \left(H\Gamma\Gamma_x\right).
    \label{eq:model_Gamma}
\end{equation}

In particular,
\begin{equation}
    \partial_x(H\Gamma\Gamma_x)
    =
    H\Gamma\Gamma_{xx}
    +H\Gamma_x^2
    +H_x\Gamma\Gamma_x,
\end{equation}
so all three contributions have the common
coefficient $\Ma/4$.

Define the backing pressure relative to the
uniform liquid load by
\begin{equation}
    P_e=\Pi_s-G\cot\theta.
    \label{eq:external_pressure}
\end{equation}

The compliant-wall equations are
\begin{equation}
    \zeta_t=v,
    \label{eq:model_zeta_kinematic}
\end{equation}
and
\begin{align}
    (\mathcal I+ReH)v_t
    +\mathcal Dv
    -\mathcal T\zeta_{xx}
    +\mathcal B\zeta_{xxxx}
    +\mathcal K\zeta
    ={}&
    P_e-G\cot\theta(H-1)
    \nonumber\\
    &+\frac{
        \Sigma(\Gamma)h_{xx}
    }{
        (1+h_x^2)^{3/2}
    }
    \nonumber\\
    &+\frac{3}{2H}q_x
    -\frac{3q}{2H^2}H_x
    -\frac{3q}{H^2}\zeta_x
    \nonumber\\
    &-\Ma
    \left[
        \left(
            \frac54H_x+\frac32\zeta_x
        \right)\Gamma_x
        +\frac H4\Gamma_{xx}
    \right]
    \nonumber\\
    &-Re\,\widetilde{\mathcal A}_w,
    \label{eq:model_zeta}
\end{align}
where $\widetilde{\mathcal A}_w$ is given explicitly in
equation~\eqref{eq:Aw_explicit} in
appendix~\ref{app:wr_closure}.

The structural energy provides a diagnostic of the
fluid-substrate coupling.
Within the retained wall closure, define
\begin{equation}
    \mathcal E_{\mathrm w}
    =
    \frac12\int_0^L
    \left[
    \mathcal I v^2
    +\mathcal T\zeta_x^2
    +\mathcal B\zeta_{xx}^2
    +\mathcal K\zeta^2
    \right]dx.
\end{equation}
Multiplication of the wall equation by $v=\zeta_t$,
followed by integration over the periodic domain, gives
\begin{equation}
    \frac{d\mathcal E_{\mathrm w}}{dt}
    +\mathcal D\int_0^L v^2\,dx
    =
    \int_0^L
    \left[\Pi_s-\Pi_{\mathrm{WR}}(\zeta)\right]v\,dx,
\end{equation}
where $\Pi_{\mathrm{WR}}(\zeta)$ is the pressure closure
in Eq.~\eqref{eq:pressure_at_wall}.
The right-hand side is the work supplied by the liquid
load, and the second term on the left is structural
dissipation.
For a periodic translating state, the integrated structural
energy is constant and the supplied work balances damping.
Decomposition of the load into gravitational, capillary,
viscous, Marangoni and acceleration contributions identifies
the mechanisms transferring energy to the substrate.

The coefficient $ReH$ is the fluid added mass.
It does not redefine the substrate parameter
$\mathcal I$.

The algebraic closure is
\begin{equation}
    h= H + \zeta,
    \qquad
    \phi
    =
    \frac{
        \displaystyle
        \Gamma+\frac{3\chi}{Pe_bk_sH^2}
    }{
        \kappa(1-\Gamma)
    }.
    \label{eq:model_closure}
\end{equation}

The mixed derivative in the flow-rate equation
is determined from
\begin{align}
    \Gamma_{tx}
    ={}&
    -\partial_{xx}
    \left(\frac{3q\Gamma}{2H}\right)
    +\frac1{Pe_s}\Gamma_{xxx}
    +\partial_x
    \left(\frac{3\chi}{Pe_bH^2}\right)
    \nonumber\\
    &+\frac{\Ma}{4}
    \partial_{xx}(H\Gamma\Gamma_x).
    \label{eq:Gamma_tx_closure}
\end{align}

The system can therefore be evaluated in sequence.
Equations~\eqref{eq:model_H} and
\eqref{eq:model_Gamma} first determine $H_t$
and $\Gamma_t$. Equation~\eqref{eq:model_q}
then determines $q_t$, and its streamwise
derivative supplies $q_{tx}$ in
Eq.~\eqref{eq:Aw_explicit}.
Finally, Eq.~\eqref{eq:model_zeta} determines
$v_t$ by division by $\mathcal I+ReH$.
There is no circular dependence of $q_t$ on $v_t$.

The displayed flow-rate equation assumes $Re>0$.
An inertialess limit must be taken in the
undivided momentum balance.

\subsection{Local and global surfactant conservation}
\label{subsec:reduced_conservation}

Define the net bulk and surface fluxes by
\begin{align}
    F_b
    ={}&
    \phi q+\frac{33}{40}\frac{q\chi}{H}
    +\frac{3\Ma}{80}H\chi\Gamma_x
    \nonumber\\
    &-\frac1{Pe_b}
    \left(
        H\phi_x+\chi_x+\frac{3\chi}{2H}\zeta_x
    \right),
    \label{eq:total_bulk_flux}\\
    F_s
    ={}&
    \frac{3q\Gamma}{2H}
    -\frac1{Pe_s}\Gamma_x
    -\frac{\Ma}{4}H\Gamma\Gamma_x.
    \label{eq:total_surface_flux}
\end{align}

The transport equations become
\begin{align}
    (\chi+H\phi)_t+\partial_xF_b
    &=-\frac{3\chi}{Pe_bH^2},\nonumber\\
    \Gamma_t+\partial_xF_s
    &=\frac{3\chi}{Pe_bH^2}.
    \label{eq:bulk_surface_local_conservation}
\end{align}

Adding them yields
\begin{equation}
    \frac{\partial}{\partial t}
    \left(\chi+H\phi+\Gamma\right)
    +
    \frac{\partial}{\partial x}
    \left(F_b+F_s\right)=0.
    \label{eq:total_surfactant_local}
\end{equation}

Hence, for periodic boundary conditions,
\begin{equation}
    \frac{\dd}{\dd t}\mathcal M_{\mathrm{tot}}=0,
    \qquad
    \mathcal M_{\mathrm{tot}}
    =
    \int_0^L
    \left(\chi+H\phi+\Gamma\right)\,\dd x.
    \label{eq:total_surfactant_global}
\end{equation}

This is an exact identity of the reduced system.
The surface contribution uses the projected
inventory approximation.
The exact continuum inventory in
Eq.~\eqref{eq:exact_total_global} contains
$\Gamma \sqrt{1+\varepsilon^2h_x^2}$; its metric factor must be retained
in the transport equations as well if that
inventory is used.

\subsection{Conservative variables and admissibility}
\label{subsec:conservative_variables}

For time integration, an equivalent state is
\begin{equation}
    (H,q,\mathcal C_{\mathrm{tot}},\Gamma,\zeta,v),
\end{equation}
where
\begin{equation}
    \mathcal C_{\mathrm{tot}}
    =
    \chi+H\phi+\Gamma,
    \qquad
    (\mathcal C_{\mathrm{tot}})_t
    =
    -\partial_x(F_b+F_s).
    \label{eq:conservative_state_variable}
\end{equation}

The original concentration variables are recovered from
\begin{equation}
    \phi
    =
    \frac{
        \mathcal C_{\mathrm{tot}}-\Gamma
        +\dfrac{Pe_bk_sH^2}{3}\Gamma
    }{
        H+\dfrac{Pe_bk_sH^2}{3}\kappa(1-\Gamma)
    },
    \qquad
    \chi=\mathcal C_{\mathrm{tot}}-\Gamma-H\phi.
    \label{eq:inventory_reconstruction}
\end{equation}

The denominator is positive when
$H>0$, $k_s>0$, $\kappa>0$ and $\Gamma<1$.

The reconstructed fields must satisfy
\begin{equation}
    H>0,
    \qquad
    0\le\Gamma<1,
    \qquad
    \phi\ge0,
    \qquad
    \phi+\frac{3\chi}{2H}\ge0,
    \qquad
    \Sigma(\Gamma)>0.
    \label{eq:WR_admissibility}
\end{equation}

Because the concentration profile is affine in
$y^2$, non-negative concentrations at $y=0$
and $y=1$ ensure non-negativity throughout the
reconstructed liquid layer.
These conditions are distinct from conservation:
a conservative solution may still leave the
admissible range of the chosen profile.

For the purely insoluble limit, set adsorption-desorption
exchange to zero and remove the bulk concentration subsystem.
Equation~\eqref{eq:model_closure} must not be
evaluated at $k_s=0$.
If an uncoupled passive bulk solute is retained,
it requires a separate transport closure.


\section{Numerical simulations}
\label{sec:nonlinear_simulations}

The nonlinear weighted-residual model is integrated to examine the
finite-amplitude evolution of the film, surfactant and compliant
substrate. These calculations extend the linear stability study by
describing wave formation, propagation and the accompanying membrane
response. All quantities reported below are dimensionless.

\subsection{Effect of substrate stiffness}
\label{sec:nonlinear_stiffness}

We first consider
\[
\mathcal{K}\in\{100,500,5\times10^{6}\},
\qquad
\Gamma_e=0.1,
\qquad
\mathcal{M}=6,
\]
with $Re=25$ and $Ka=3905$. The remaining parameters are
$G=2$, $\theta=30^\circ$, $Pe_b=Pe_s=1$, $k_s=0.1$,
$\kappa=10$, $I=1$, $T=500$, $B=10$ and $P_e=0$.
As in the linear study, the damping coefficient follows
$D=\sqrt{\mathcal{K}}$. The comparison therefore follows a prescribed
stiffness-damping path, along which both membrane stiffness and
damping increase.

The simulations are performed on a periodic domain of length $L=250$,
using the same initial disturbance for all three cases:
\begin{align}
H(x,0)
&=1+0.1\cos\left(\frac{2\pi x}{L}\right), \nonumber\\
q(x,0)
&=\frac{G}{3}H(x,0)^3, \nonumber\\
\Gamma(x,0)
&=\Gamma_e,
\qquad
\zeta(x,0)=v(x,0)=0.
\label{eq:nl_initial_condition}
\end{align}
Here, $H$ is the liquid thickness, $q$ is the liquid flux,
$\zeta$ is the membrane displacement and $v=\partial_t\zeta$.
The initial bulk concentration is set to adsorption equilibrium,
$\phi_e=\Gamma_e/[\kappa(1-\Gamma_e)]$, with the conservative
surfactant density initialized as
$C_{\mathrm{tot}}(x,0)=H(x,0)\phi_e+\Gamma_e$.

Spatial derivatives are evaluated using Fourier multipliers on
$N=512$ grid points. Nonlinear terms are evaluated on a grid padded
by a factor of three before projection onto the retained Fourier
modes. The retained spectral cutoff is $|k_n|\leq4$, where
$k_n=2\pi n/L$; the largest retained discrete wavenumber is
approximately $3.996$. For $\mathcal{K}=100$ and $500$, time
integration uses the classical fourth-order Runge-Kutta method
with $\Delta t=0.0125$. For $\mathcal{K}=5\times10^6$,
fourth-order exponential time differencing with
$\Delta t=0.00625$ is used up to $t=1500$, followed by a
Runge-Kutta continuation with $\Delta t=0.0015625$ up to
$t=1700$. Short continuations with smaller time steps and increased
spatial resolution provide local checks of the late-wave states.
The reported results refer to this retained-band discretization
of the weighted-residual model.

For the full-period profile plots, the displayed coordinate is
\[
    \mathcal X
    =
    \left(x-x_c+\frac{L}{2}\right)\bmod L,
\]
where $x_c$ is the principal free-surface crest position.
The crest therefore appears at $\mathcal X=L/2$.
In the local streamline views, the displayed origin is
shifted by $L/2$ so that the crest appears at zero.
The fixed-origin snapshots retain the original coordinate
$x$.
These translations are applied jointly to all fields and
do not alter their relative phase.

\begin{figure*}
\centering
\begin{minipage}[t]{0.48\textwidth}
    \centering
    \includegraphics[width=\linewidth]
    {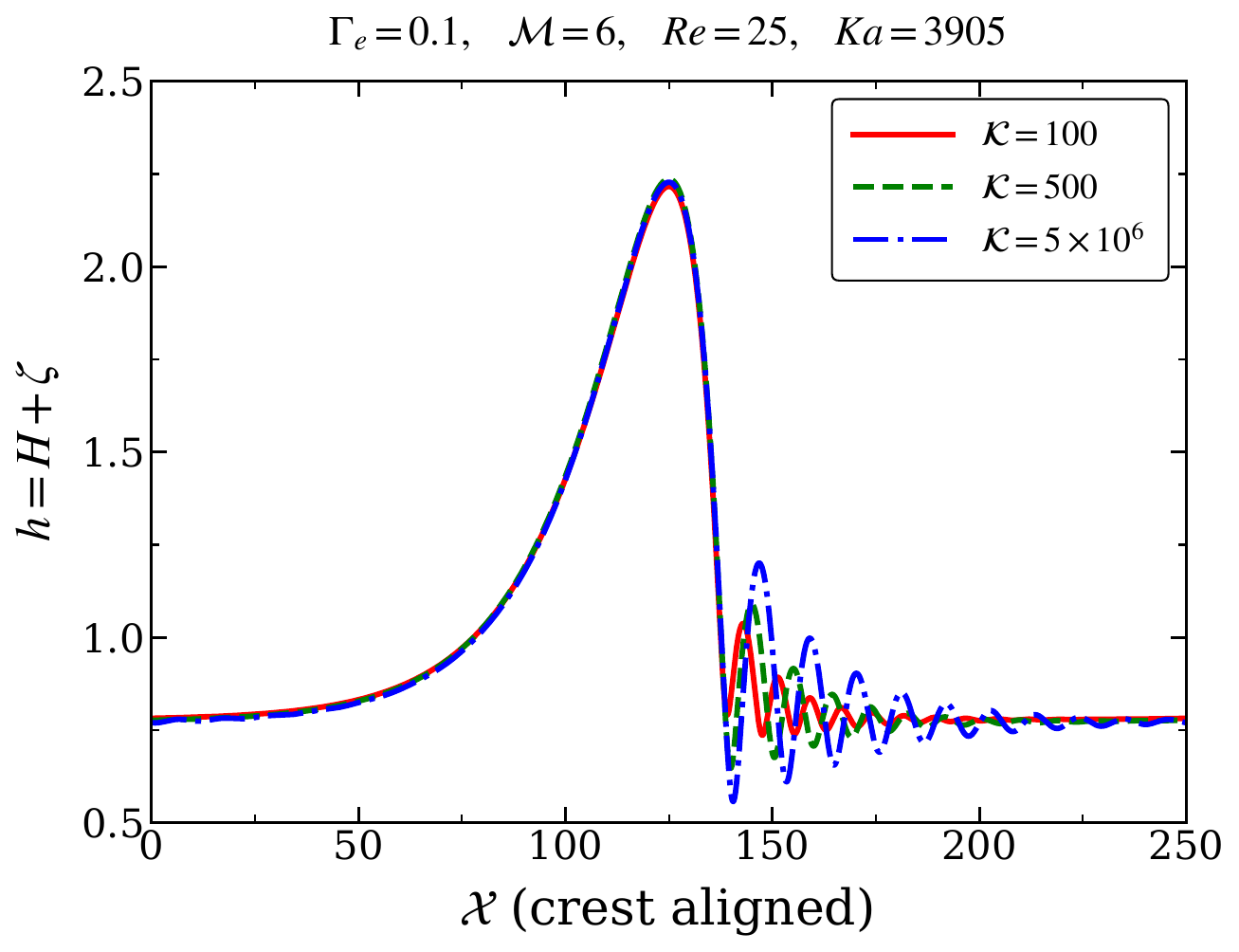}
    \par\smallskip
    {\small (a) Free-surface height $h$.}
\end{minipage}
\hfill
\begin{minipage}[t]{0.48\textwidth}
    \centering
    \includegraphics[width=\linewidth]
    {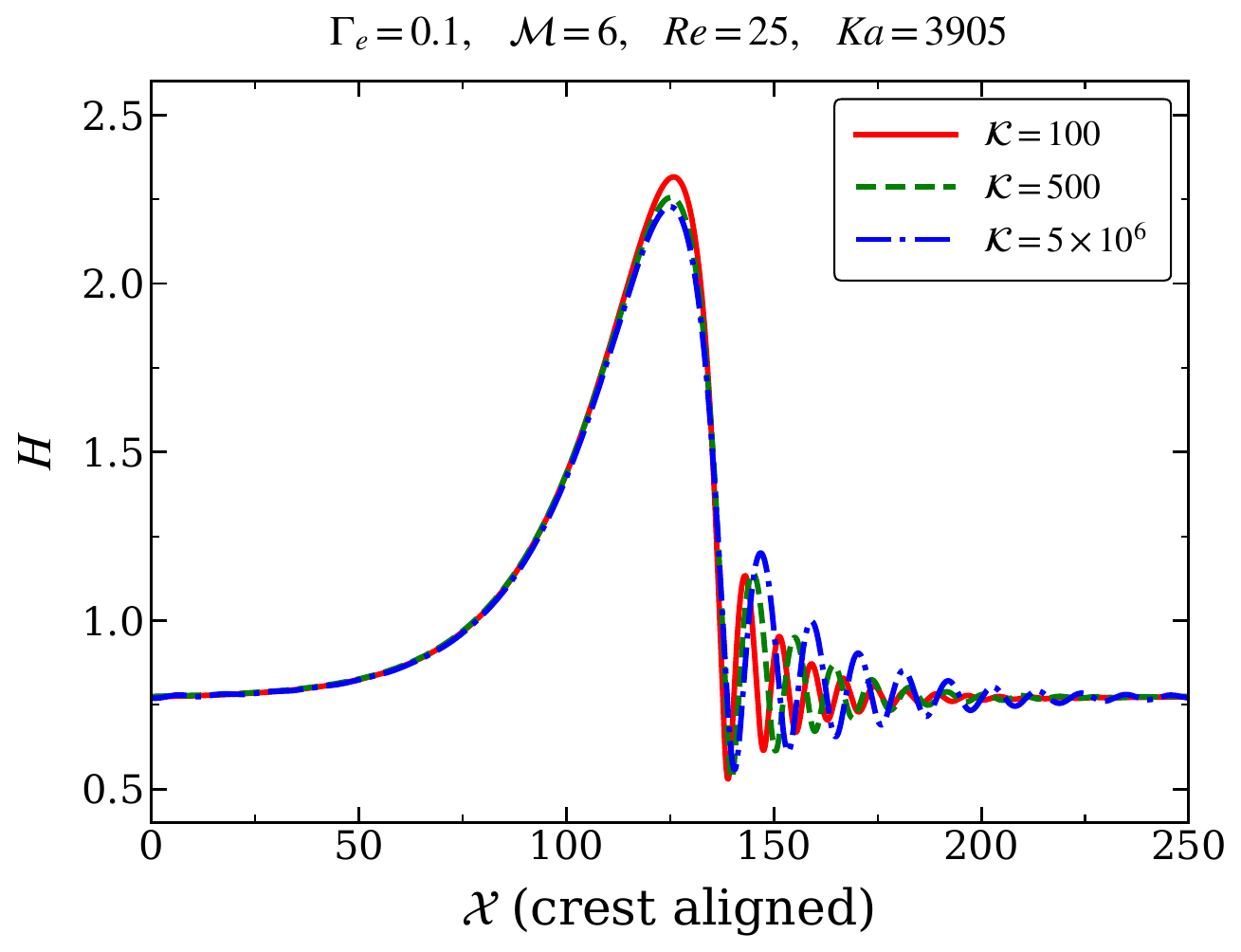}
    \par\smallskip
    {\small (b) Liquid thickness $H$.}
\end{minipage}

\medskip

\begin{minipage}[t]{0.48\textwidth}
    \centering
    \includegraphics[width=\linewidth]
    {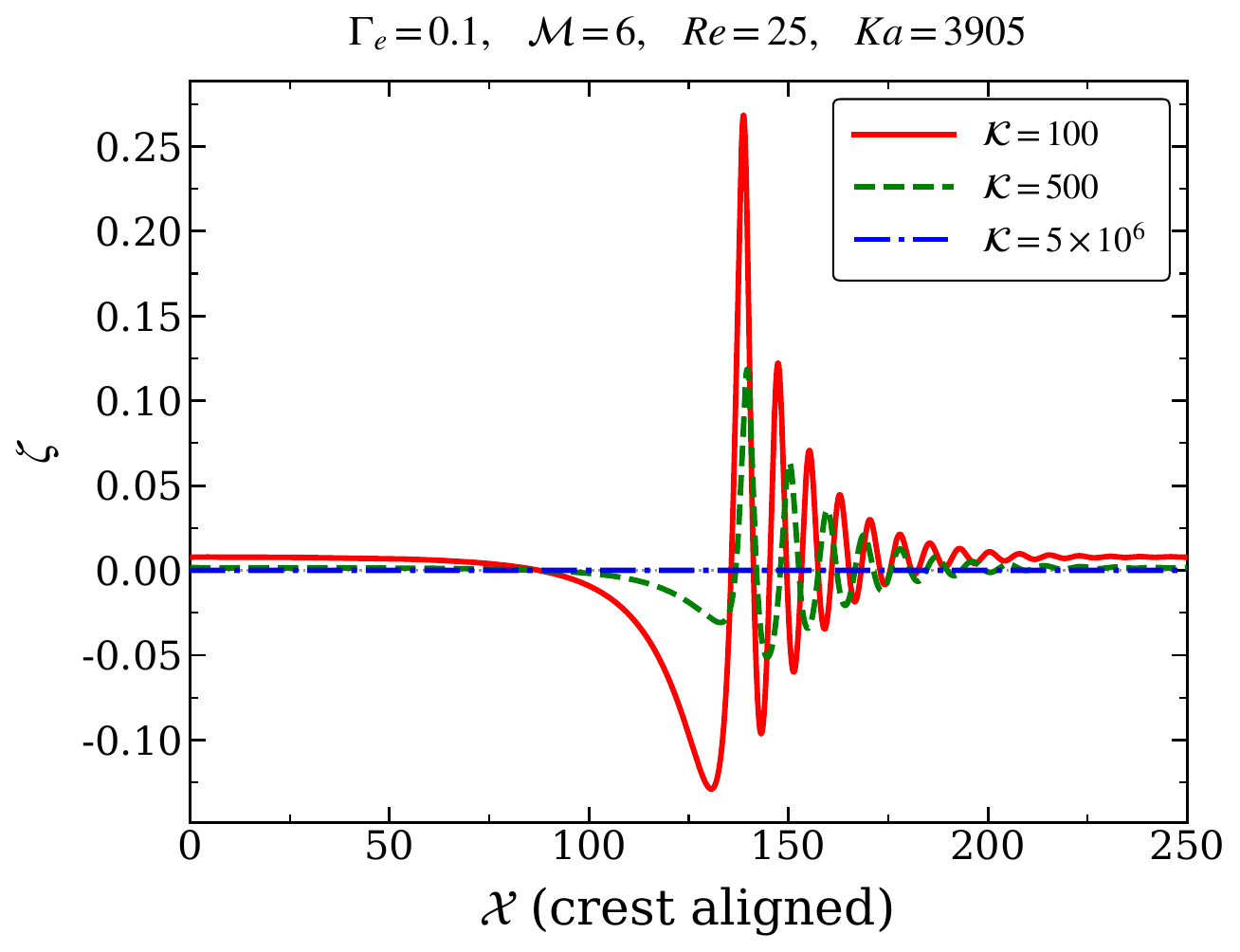}
    \par\smallskip
    {\small (c) Membrane displacement $\zeta$.}
\end{minipage}
\hfill
\begin{minipage}[t]{0.48\textwidth}
    \centering
    \includegraphics[width=\linewidth]
    {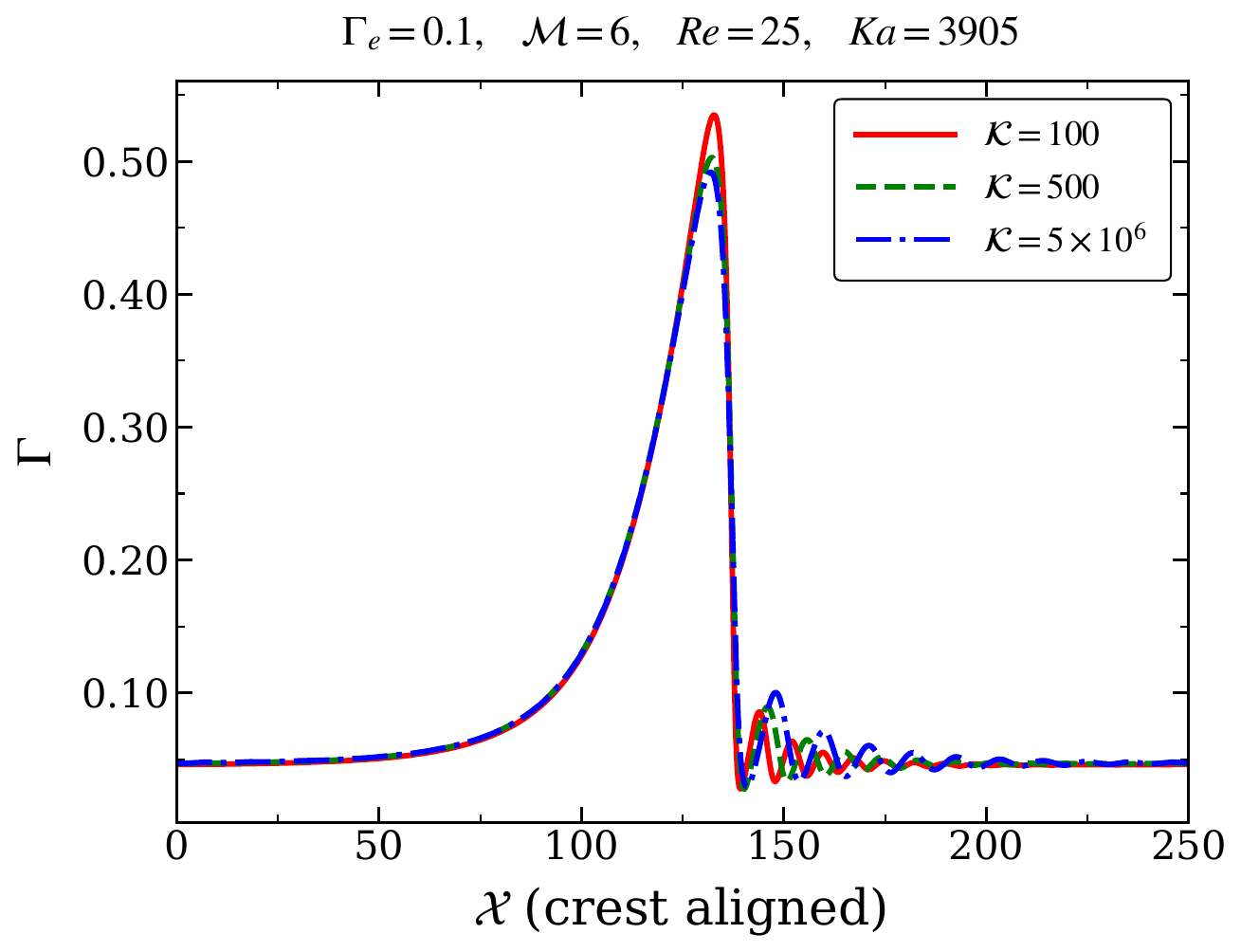}
    \par\smallskip
    {\small (d) Surface surfactant coverage $\Gamma$.}
\end{minipage}
\caption{
Late-wave profiles along the substrate-parameter path
$D=\sqrt{\mathcal{K}}$ for $\Gamma_e=0.1$, $\mathcal{M}=6$,
$Re=25$ and $Ka=3905$.
Red solid, green dashed and blue dash-dotted curves correspond
to $\mathcal{K}=100$, $500$ and $5\times10^6$, respectively.
Profiles are shown at $t=2000$ for $\mathcal{K}=100$ and $500$,
and at $t=1700$ for $\mathcal{K}=5\times10^6$.
Each complete state is translated to place its free-surface crest
at $\mathcal{X}=L/2$, preserving the relative phase of all fields.
}
\label{fig:nl_stiffness_profiles}
\end{figure*}

Figure~\ref{fig:nl_stiffness_profiles} compares the late-wave
profiles. Each case is translated by a single spatial shift,
applied jointly to all fields, so that the principal free-surface
crest is located at $\mathcal{X}=L/2$. This translation preserves
the relative phase between the liquid surface, membrane and
surfactant distributions.

The distinction between the laboratory free-surface height and the
actual liquid thickness is essential:
\begin{equation}
h(x,t)=H(x,t)+\zeta(x,t).
\label{eq:nl_height_thickness}
\end{equation}
The principal free-surface crests remain close to $h_{\max}=2.22$
across the three cases, although their variation is not monotonic.
The steep wave front and its accompanying capillary ripples exhibit
a stronger dependence on the substrate parameters. The laboratory
surface minimum decreases from approximately $0.736$ at
$\mathcal{K}=100$ to $0.557$ at $\mathcal{K}=5\times10^6$.
In contrast, the corresponding minimum liquid thickness increases
slightly, from approximately $0.529$ to $0.557$. Thus, changes in
the laboratory surface trough reflect both liquid thinning and
membrane displacement.

For the most compliant substrate, the membrane develops a broad
downward displacement beneath the principal wave, followed by a
localized upward displacement near the steep front. Increasing
$\mathcal{K}$ strongly reduces this deformation, and the
$\mathcal{K}=5\times10^6$ case approaches the rigid-substrate limit
on the scale of the figure. The surfactant distribution remains
strongly nonuniform, with enrichment on the descending wave front
and smaller oscillations over the capillary-ripple region.
The maximum surface coverage decreases from approximately
$0.535$ to $0.503$ and $0.492$ as $\mathcal{K}$ increases.
These profiles show that a relatively small change in the principal
surface height can accompany appreciable changes in membrane
deformation, liquid thickness and surfactant redistribution.

\begin{figure*}
\centering
\includegraphics[
    width=0.88\textwidth,
    height=0.75\textheight,
    keepaspectratio
]{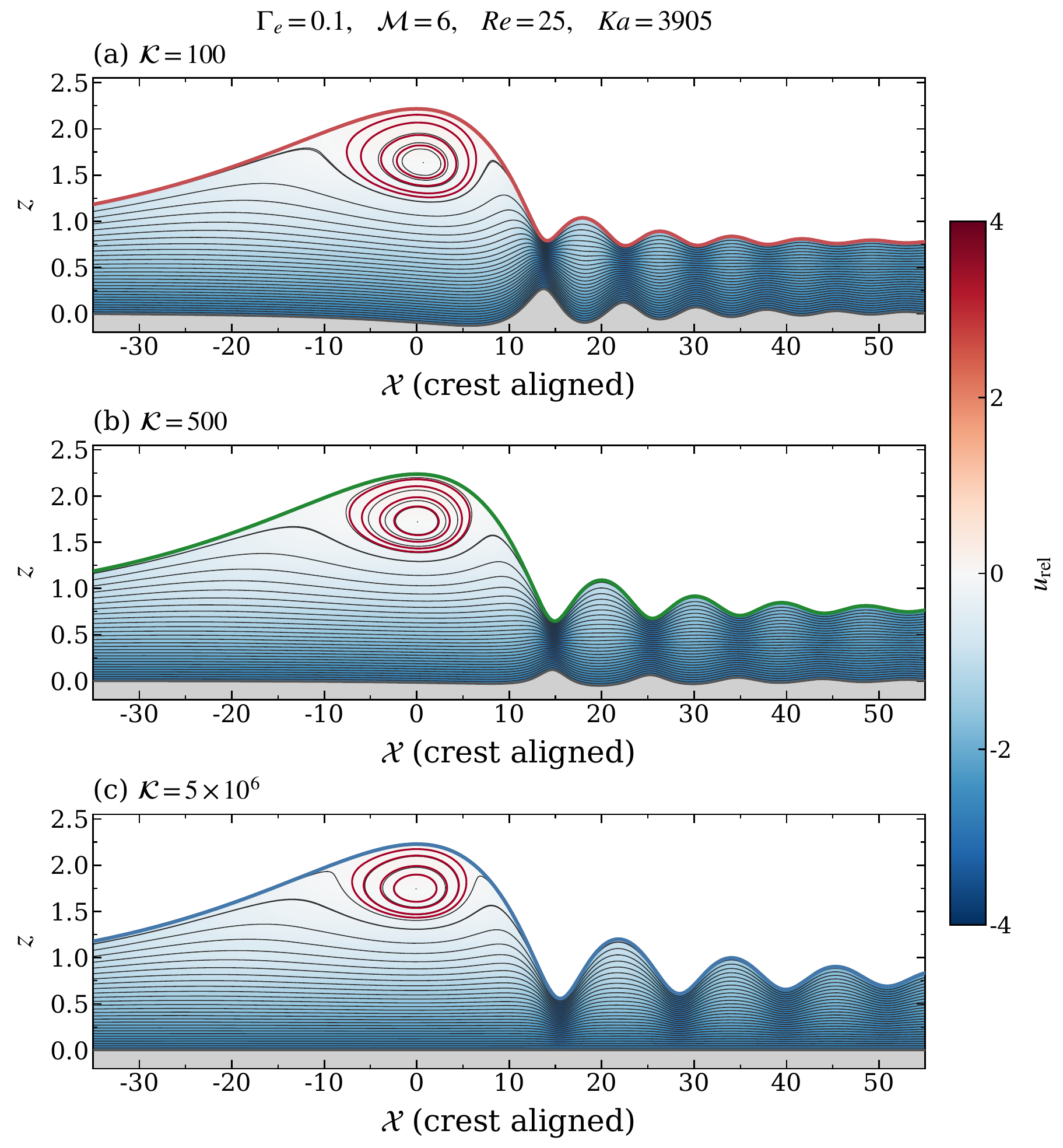}
\caption{
Streamlines in the travelling-wave frame for
(a) $\mathcal{K}=100$,
(b) $\mathcal{K}=500$ and
(c) $\mathcal{K}=5\times10^6$.
The remaining parameters are those of
Fig.~\ref{fig:nl_stiffness_profiles}.
The crest is placed at the origin of the horizontal coordinate
in this local view.
Shading represents the relative streamwise velocity
$u_{\mathrm{rel}}=u-c$, and the burgundy contours highlight
verified closed recirculation regions.
The grey region denotes the substrate.
The vertical scale is exaggerated for visibility;
the coordinate labels retain their dimensionless physical values.
}
\label{fig:nl_stiffness_streamlines}
\end{figure*}

The flow structure associated with these waves is shown in
Fig.~\ref{fig:nl_stiffness_streamlines}. The velocity field is
reconstructed from the weighted-residual velocity profile,
including the actual membrane velocity. If $\psi$ denotes the
laboratory-frame streamfunction, the streamfunction relative
to a wave travelling at speed $c$ is
\begin{equation}
\Psi(x,z,t)=\psi(x,z,t)-cz,
\qquad
\partial_z\Psi=u-c,
\qquad
-\partial_x\Psi=w.
\label{eq:nl_moving_streamfunction}
\end{equation}
Consequently, contours of $\Psi$ describe the instantaneous
streamlines relative to the travelling wave.

A closed recirculation region is present beneath the principal
crest in all three cases. Away from this region, the background
flow is directed predominantly to the left relative to the wave,
while the wave itself propagates to the right in the laboratory
frame. Increasing substrate stiffness flattens the lower boundary
and changes the streamline geometry near the wave front and the
capillary ripples. Nevertheless, crest recirculation persists
into the nearly rigid-substrate case.

\begin{figure*}
\centering
\begin{minipage}[t]{0.48\textwidth}
    \centering
    \includegraphics[width=\linewidth]
    {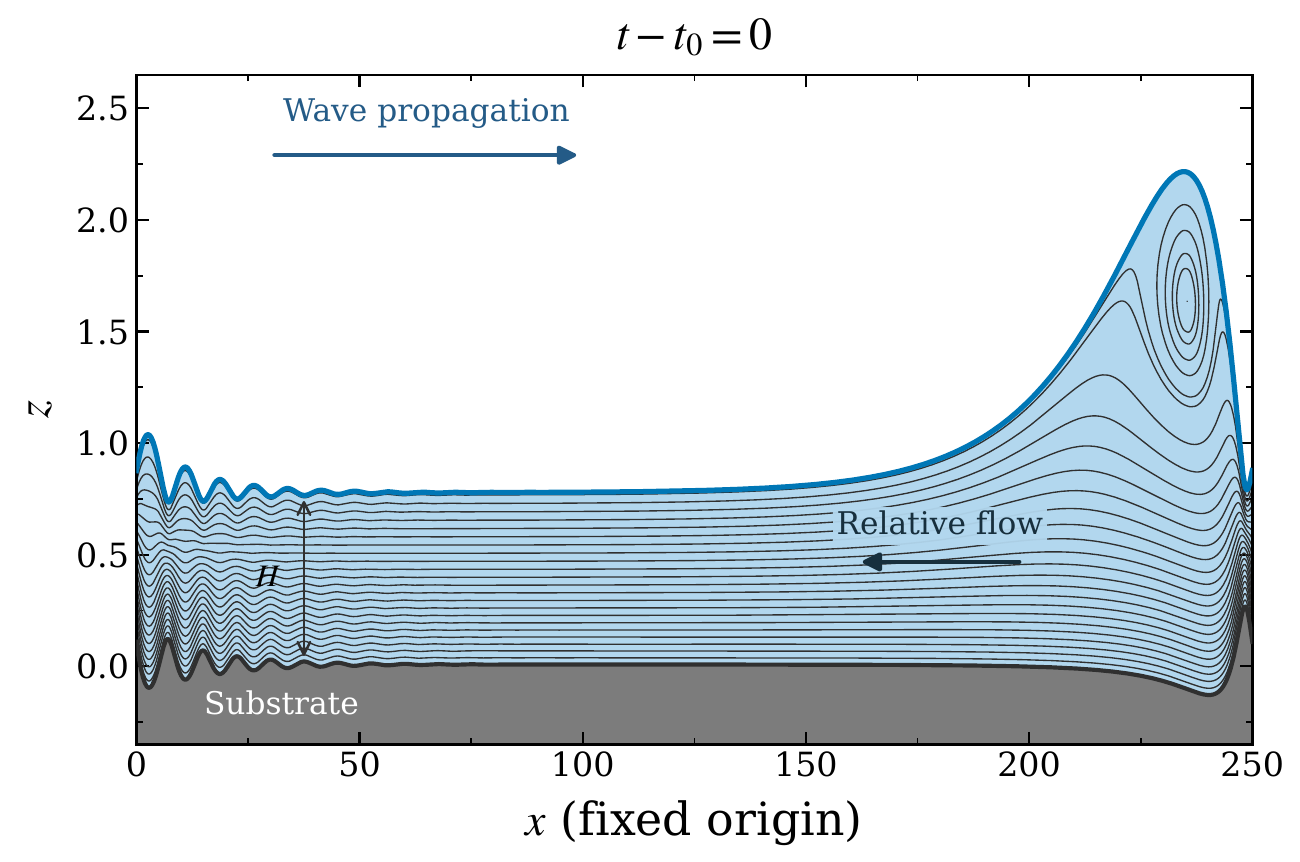}
    \par\smallskip
    {\small (a) $t-t_0=0$.}
\end{minipage}
\hfill
\begin{minipage}[t]{0.48\textwidth}
    \centering
    \includegraphics[width=\linewidth]
    {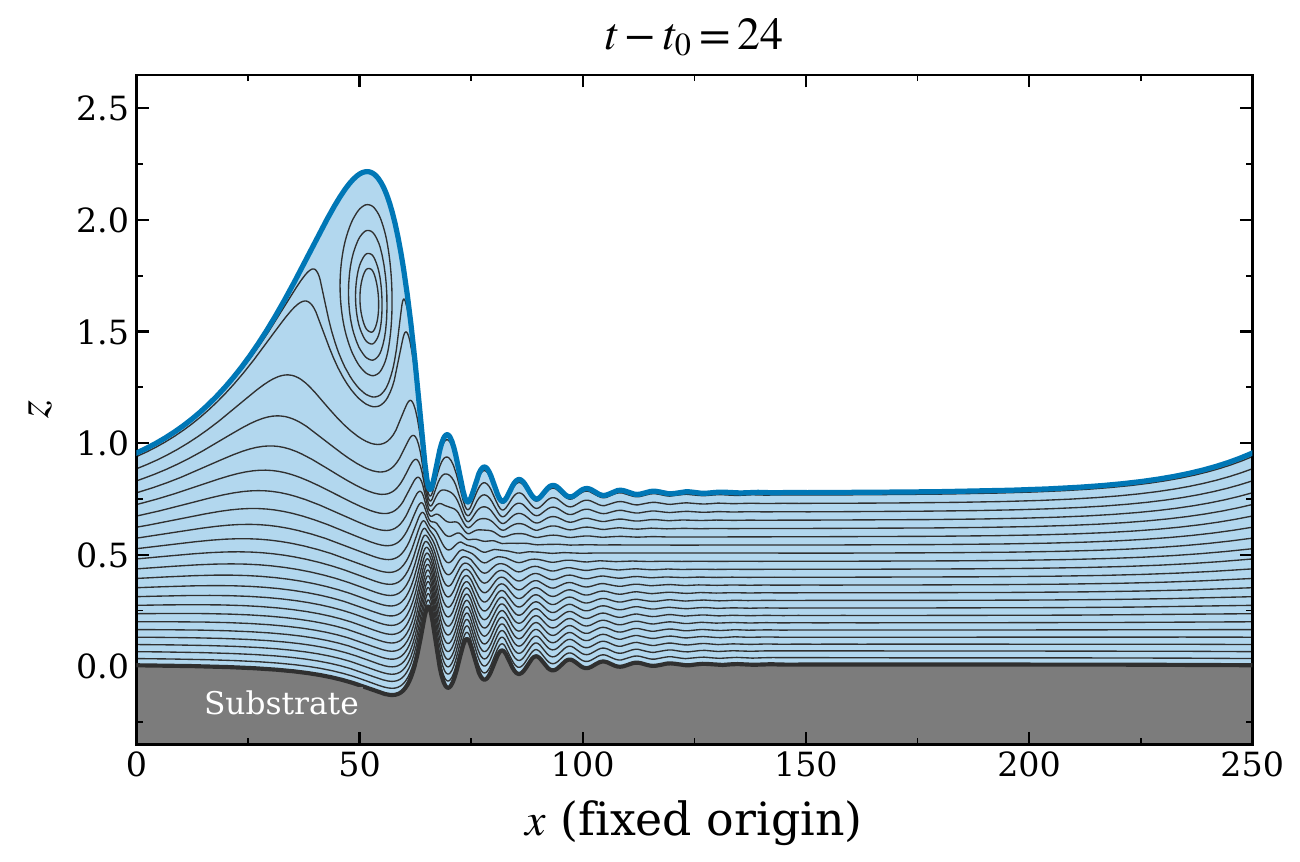}
    \par\smallskip
    {\small (b) $t-t_0=24$.}
\end{minipage}

\medskip

\begin{minipage}[t]{0.48\textwidth}
    \centering
    \includegraphics[width=\linewidth]
    {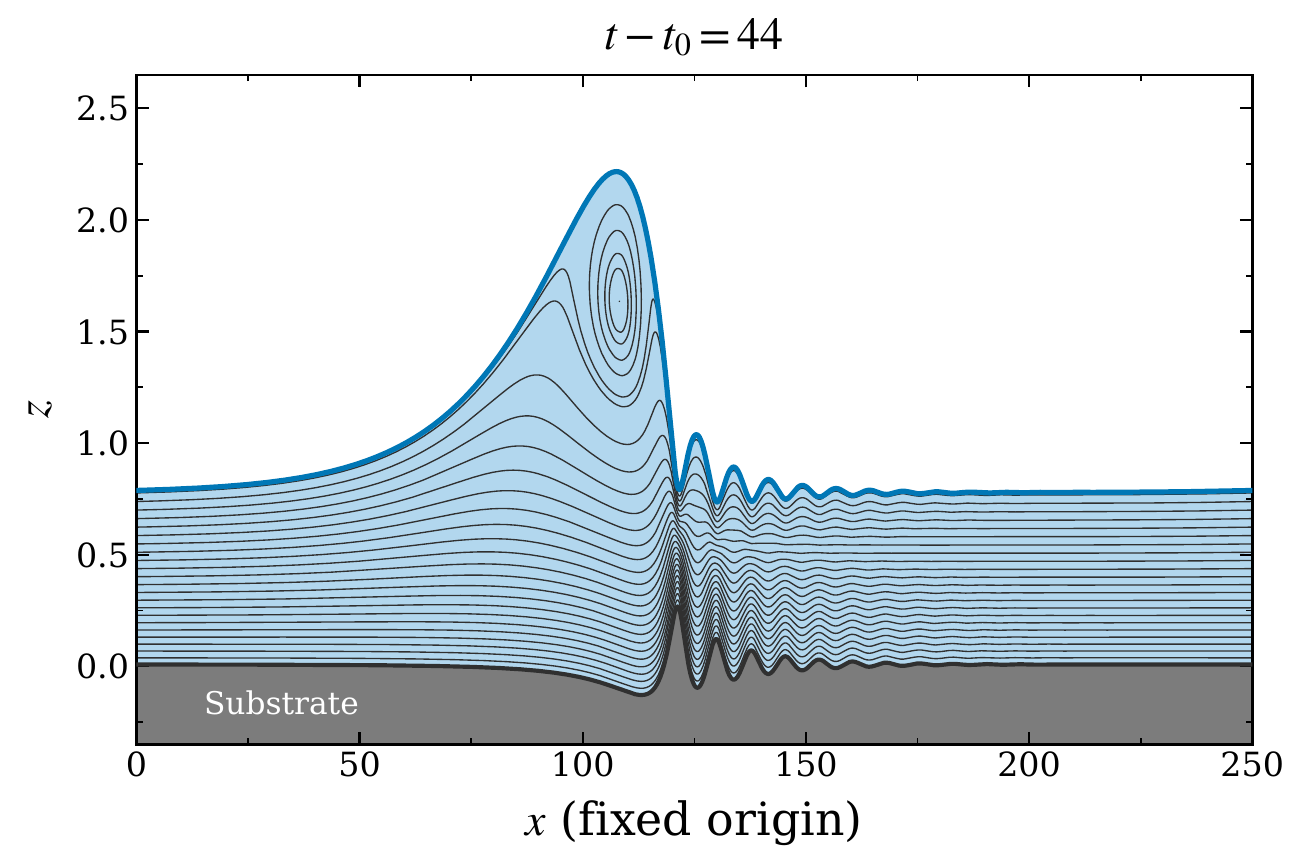}
    \par\smallskip
    {\small (c) $t-t_0=44$.}
\end{minipage}
\hfill
\begin{minipage}[t]{0.48\textwidth}
    \centering
    \includegraphics[width=\linewidth]
    {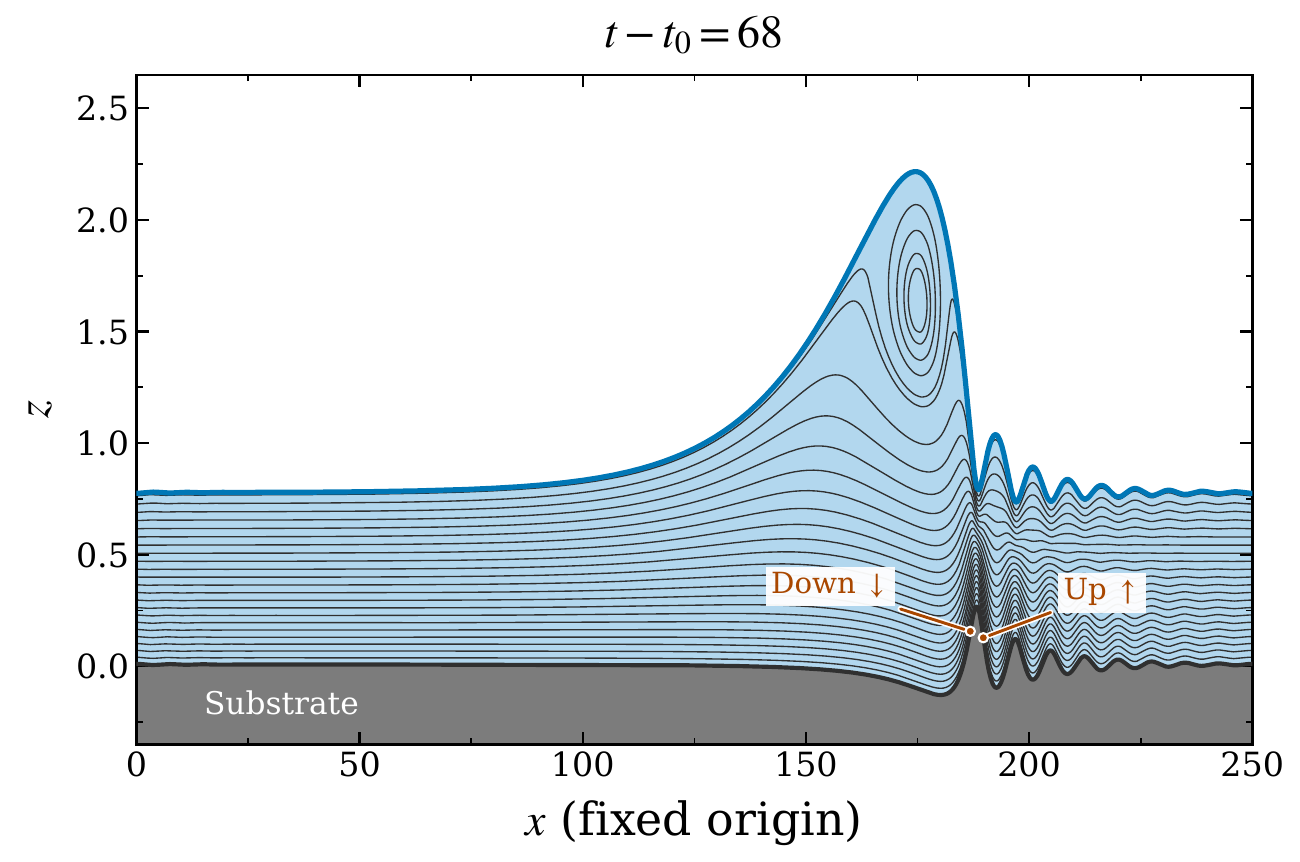}
    \par\smallskip
    {\small (d) $t-t_0=68$.}
\end{minipage}
\caption{
Actual fixed-origin snapshots of the baseline travelling wave,
with $\Gamma_e=0.1$, $\mathcal{M}=6$, $\mathcal{K}=100$,
$Re=25$ and $Ka=3905$.
The reference time is $t_0=1932$.
The blue region denotes the liquid and the grey region the
substrate. Black contours are instantaneous streamlines relative
to the travelling wave.
The first panel indicates rightward wave propagation and
leftward relative background flow.
The final panel identifies local downward and upward membrane
motion. All panels retain the same spatial origin and coordinate
limits; the vertical scale is exaggerated for visibility.
}
\label{fig:nl_fixed_origin_snapshots}
\end{figure*}

Figure~\ref{fig:nl_fixed_origin_snapshots} illustrates the motion
of the baseline wave, with $\mathcal{K}=100$, using four actual
snapshots at a fixed laboratory spatial origin. No crest alignment
is applied to these panels. The principal wave travels to the
right, crosses the periodic boundary, and continues through the
domain while retaining nearly the same shape. The closed
recirculation region and the membrane deformation accompany
this translation.

The horizontal motion of the deformation pattern should be
distinguished from the local vertical response of the membrane
material. For an approximately translating membrane profile,
\begin{equation}
\zeta(x,t)\simeq Z(x-ct)
\quad\Longrightarrow\quad
v(x,t)\simeq-c\,\partial_x\zeta(x,t).
\label{eq:nl_membrane_translation}
\end{equation}
Thus, opposite sides of a membrane feature move vertically in
opposite directions as the deformation pattern passes.
The upward and downward annotations in the final panel indicate
the sign of this local vertical velocity.

Although the horizontal coordinates in
Fig.~\ref{fig:nl_fixed_origin_snapshots} remain fixed in the
laboratory frame, the streamline contours are computed using
$\Psi$. The panels therefore show the physical translation of
the wave together with the flow relative to it.

\begin{figure*}
\centering
\begin{minipage}[t]{0.48\textwidth}
    \centering
    \includegraphics[width=\linewidth]
    {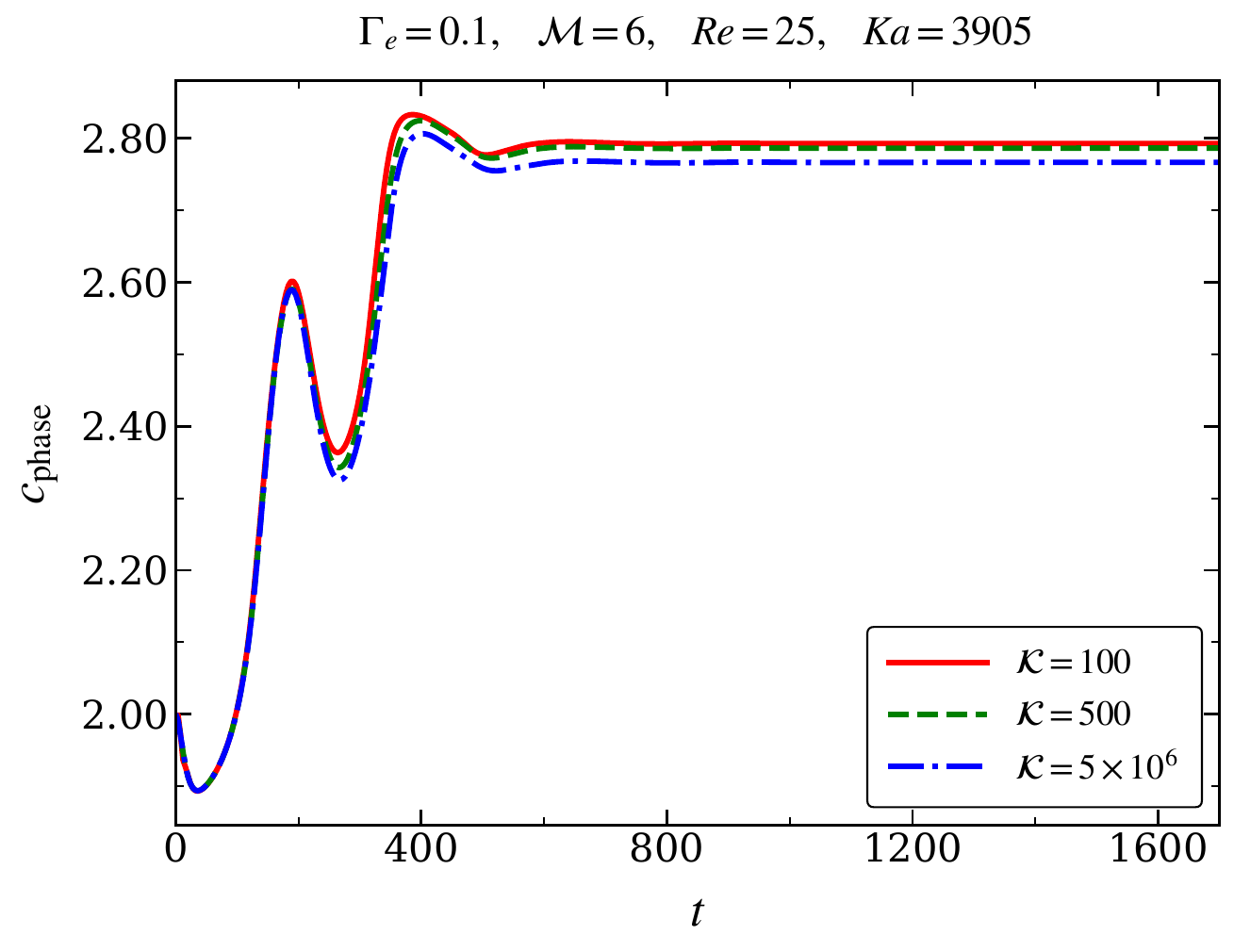}
    \par\smallskip
    {\small (a) Measured phase speed.}
\end{minipage}
\hfill
\begin{minipage}[t]{0.48\textwidth}
    \centering
    \includegraphics[width=\linewidth]
    {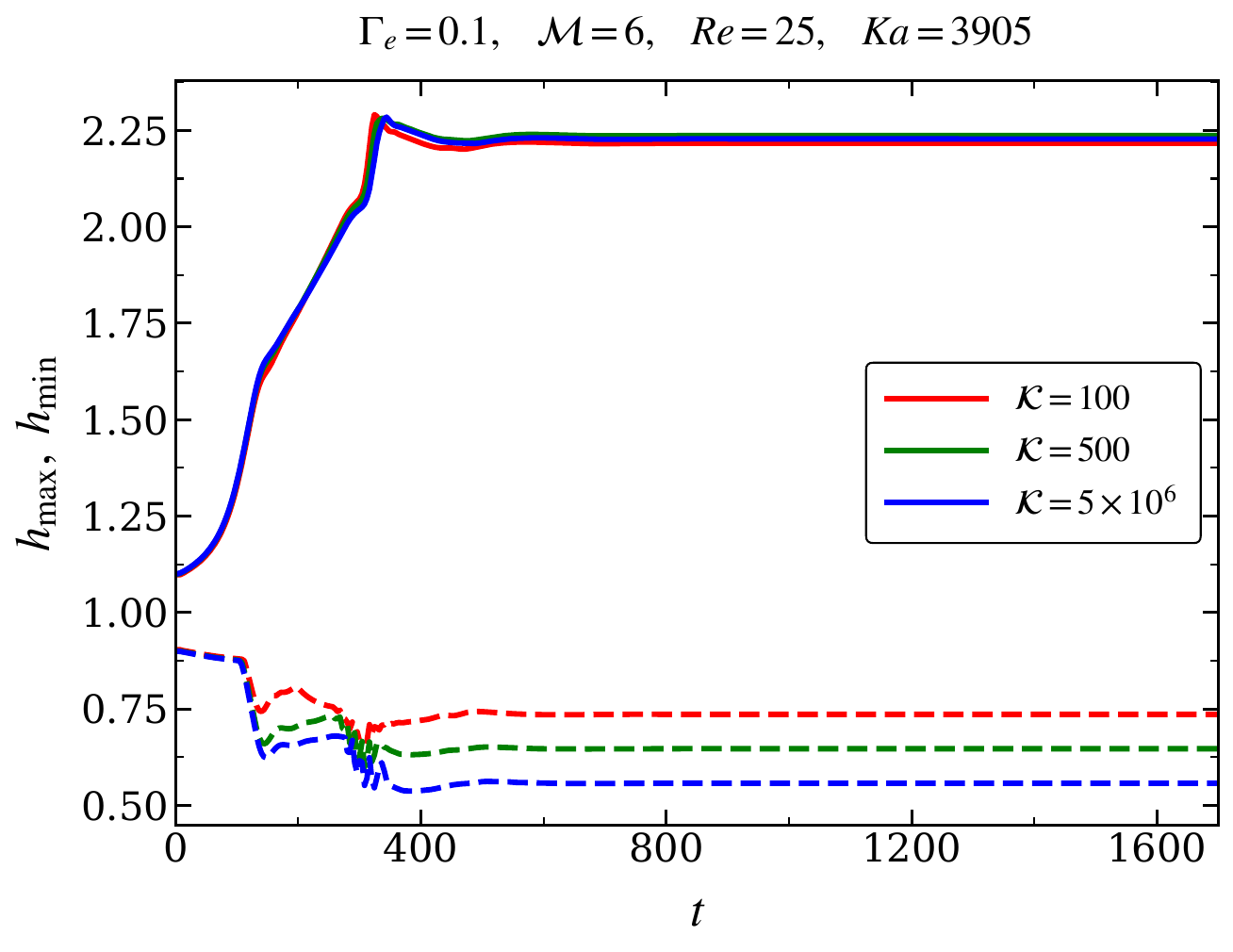}
    \par\smallskip
    {\small (b) Free-surface height extrema.}
\end{minipage}
\caption{
Temporal evolution of
(a) the phase speed defined by
Eq.~\eqref{eq:nl_measured_phase_speed} and
(b) the measured free-surface height extrema.
Colours identify the three substrate stiffnesses as in
Fig.~\ref{fig:nl_stiffness_profiles}.
In panel (b), solid curves denote $\max_x h(x,t)$ and dashed
curves denote $\min_x h(x,t)$.
The common interval $0\leq t\leq1700$ is used for all cases.
}
\label{fig:nl_speed_height}
\end{figure*}

The temporal evolution is quantified in
Fig.~\ref{fig:nl_speed_height}.
The propagation speed is obtained from the unwrapped phase of
the dominant late-time Fourier mode of the free-surface height:
\begin{equation}
\widehat{h}_n(t)
=\frac{1}{L}\int_0^L h(x,t)e^{-ik_nx}\,dx,
\qquad
c_{\mathrm{phase}}(t)
=-\frac{1}{k_n}\frac{d}{dt}\arg\widehat{h}_n(t),
\qquad
k_n=\frac{2\pi n}{L}.
\label{eq:nl_measured_phase_speed}
\end{equation}
The dominant late-time mode is $n=1$ for all three cases.
Intervals in which its amplitude is too small for a reliable
phase measurement are omitted.

Following the formation transient, the measured speeds approach
approximately $2.792$, $2.786$ and $2.766$ for
$\mathcal{K}=100$, $500$ and $5\times10^6$, respectively.
The propagation speed therefore decreases modestly along the
chosen substrate-parameter path. The measured free-surface maximum
and minimum also approach nearly constant values after the initial
growth and overshoot. Together, these plateaus are consistent
with the approximately shape-preserving translation visible in
Fig.~\ref{fig:nl_fixed_origin_snapshots}.

Travelling-wave classification additionally uses the scaled
residual of
$\partial_t\boldsymbol{Y}+c\,\partial_x\boldsymbol{Y}$
for all six evolved fields
$\boldsymbol{Y}=(H,q,C_{\mathrm{tot}},\Gamma,\zeta,v)$,
together with the drift of their Fourier amplitudes.
Both diagnostics satisfy the prescribed $10^{-3}$ tolerance
in the assessed late intervals.
For uniform comparison, the history figures show only the common
saved interval $0\leq t\leq1700$.

\begin{figure*}
\centering
\begin{minipage}[t]{0.48\textwidth}
    \centering
    \includegraphics[width=\linewidth]
    {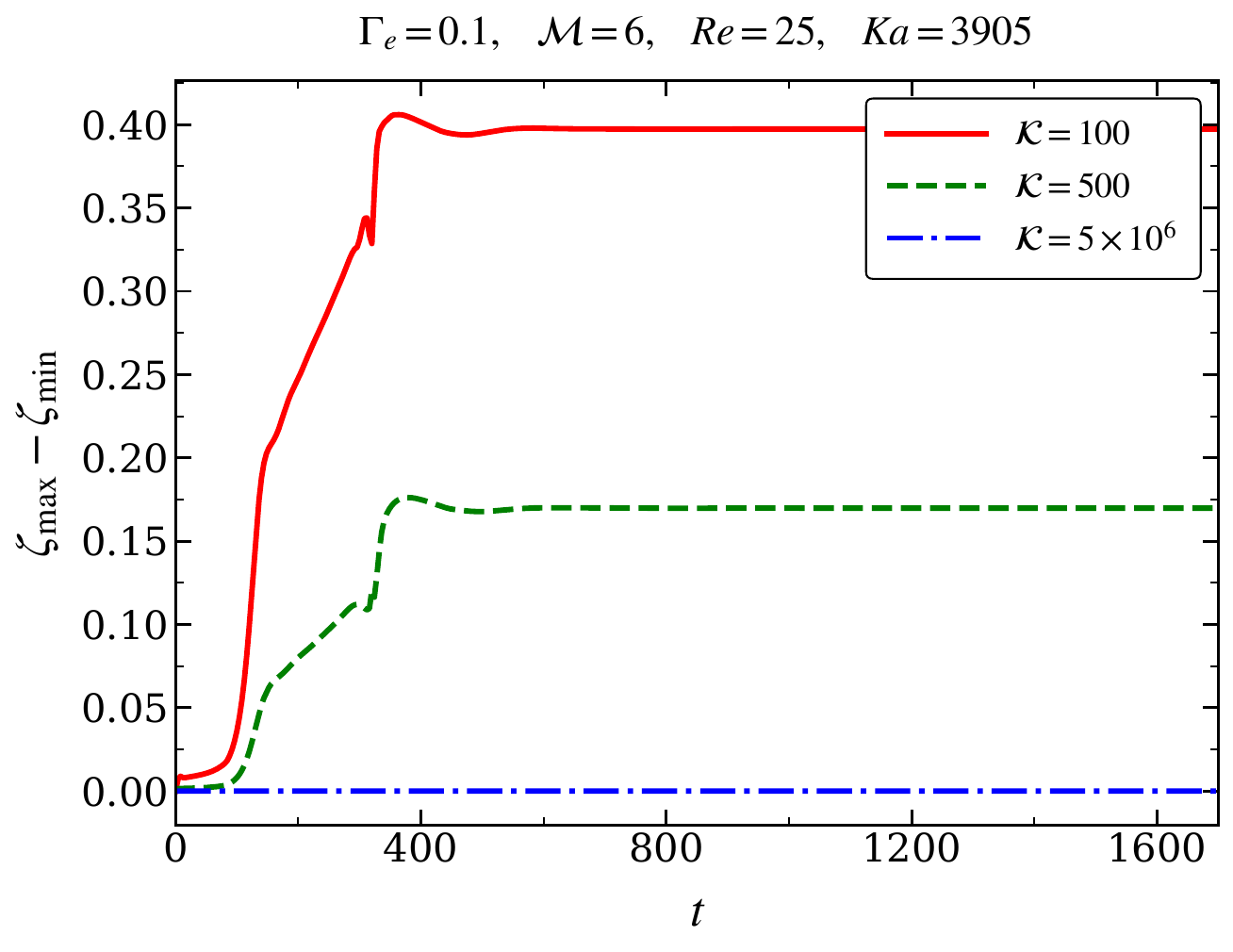}
    \par\smallskip
    {\small (a) Membrane displacement amplitude.}
\end{minipage}
\hfill
\begin{minipage}[t]{0.48\textwidth}
    \centering
    \includegraphics[width=\linewidth]
    {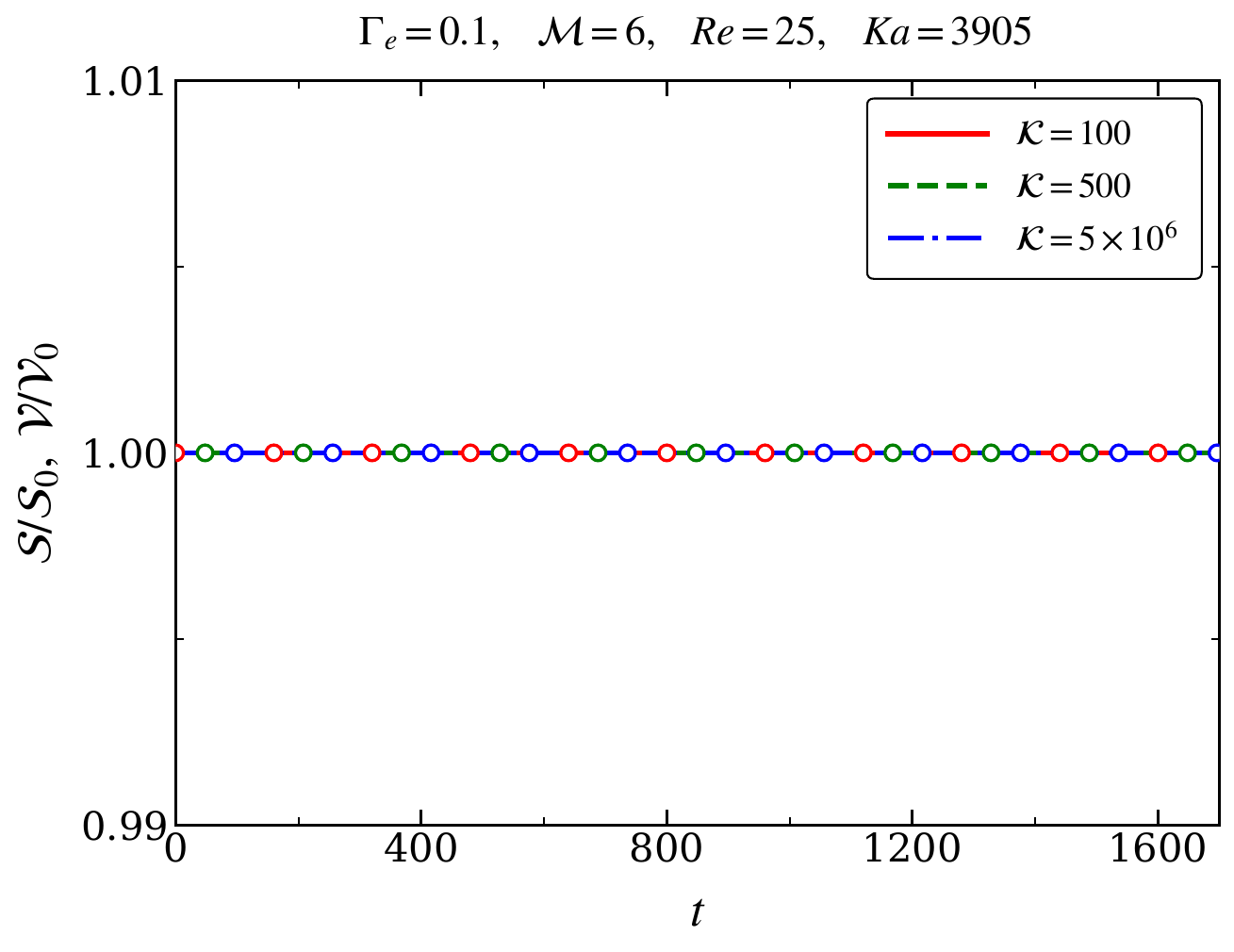}
    \par\smallskip
    {\small (b) Normalized conserved inventories.}
\end{minipage}
\caption{
(a) Membrane peak-to-peak amplitude
$\Delta\zeta=\max_x\zeta-\min_x\zeta$.
(b) Conservation of total surfactant inventory and liquid volume,
normalized by their initial values.
Lines represent $\mathcal{S}(t)/\mathcal{S}(0)$ and open circles
represent $\mathcal{V}(t)/\mathcal{V}(0)$ at selected saved times.
Red, green and blue identify
$\mathcal{K}=100$, $500$ and $5\times10^6$, respectively.
The parameters and common time interval are those of
Fig.~\ref{fig:nl_speed_height}.
}
\label{fig:nl_membrane_conservation}
\end{figure*}

The membrane response and conservation diagnostics are presented
in Fig.~\ref{fig:nl_membrane_conservation}.
The membrane amplitude is measured by its full spatial
peak-to-peak displacement,
\begin{equation}
\Delta\zeta(t)
=\max_x\zeta(x,t)-\min_x\zeta(x,t).
\label{eq:nl_membrane_amplitude}
\end{equation}
Its growth accompanies the development of the liquid wave,
followed by an approximately constant late-time amplitude.
The late values are about $0.397$, $0.170$ and
$2.23\times10^{-5}$ for the three increasing stiffnesses.
The nearly rigid case therefore appears close to zero on the
linear scale of the figure.

The conserved liquid volume and total surfactant inventory of
the weighted-residual model are
\begin{align}
\mathcal{V}(t)
&=
\int_0^L H(x,t)\,dx,
\\[2mm]
\mathcal{S}(t)
&\equiv
\mathcal{M}_{\mathrm{tot}}(t)
=
\int_0^L C_{\mathrm{tot}}(x,t)\,dx,
\\[2mm]
C_{\mathrm{tot}}
&=
H\phi+\chi+\Gamma,
\label{eq:nl_conserved_inventories}
\end{align}

where $\phi$ and $\chi$ are the bulk-concentration variables
defined in the weighted-residual closure.

The liquid volume is determined by the actual thickness $H$,
while $\mathcal{S}$ includes both bulk and surface contributions.
Adsorption and desorption redistribute surfactant between these
contributions, so the integral of $\Gamma$ alone is generally
time dependent.

The normalized inventories
$\mathcal{S}(t)/\mathcal{S}(0)$ and
$\mathcal{V}(t)/\mathcal{V}(0)$ remain at unity throughout the
recorded simulations. Their relative departures are below
$4\times10^{-16}$ in the saved diagnostics.
This conservation follows from the conservative flux formulation
and cancellation of the bulk-surface exchange terms, without
field clipping or inventory renormalization.
It provides a check on the discretized conservation laws,
complementing the time-step and spatial-refinement checks.

\subsubsection{Phase and work transferred to the substrate}

The structural-energy identity given above permits the normal load to be
resolved into signed contributions,
$\mathcal F_{\mathrm w}=\mathcal F_e+\mathcal F_g+
\mathcal F_{\Sigma}+\mathcal F_{\nu}+\mathcal F_M+\mathcal F_a$.
These represent backing pressure, hydrostatic pressure, capillary pressure,
viscous pressure, an explicit Marangoni correction and transverse fluid
acceleration. Their expressions follow from the existing pressure closure
and are listed in appendix~\ref{app:jfm_red_load}.
Define
\begin{equation}
\begin{aligned}
\mathcal P_j(t)&=\int_0^L\mathcal F_jv\,\mathrm dx,
&\mathcal Q_D(t)&=\mathcal D\int_0^L v^2\,\mathrm dx,\\
\overline{\mathcal P}_j&=\frac{1}{t_b-t_a}
\int_{t_a}^{t_b}\mathcal P_j(t)\,\mathrm dt,
&\frac{\mathrm d\mathcal E_{\mathrm w}}{\mathrm dt}
+\mathcal Q_D&=\sum_j\mathcal P_j.
\end{aligned}
\label{eq:jfm_red_work}
\end{equation}
Positive $\mathcal P_j$ supplies work to the substrate; negative
$\mathcal P_j$ removes it. These terms quantify mechanical transfer at the
substrate. They do not measure the total viscous dissipation in the liquid.
For a translating periodic state, the structural energy is stationary and
the net load work balances structural damping.

\begin{figure}
\centering

\begin{subfigure}[t]{0.48\linewidth}
    \centering
    \includegraphics[width=\linewidth]
        {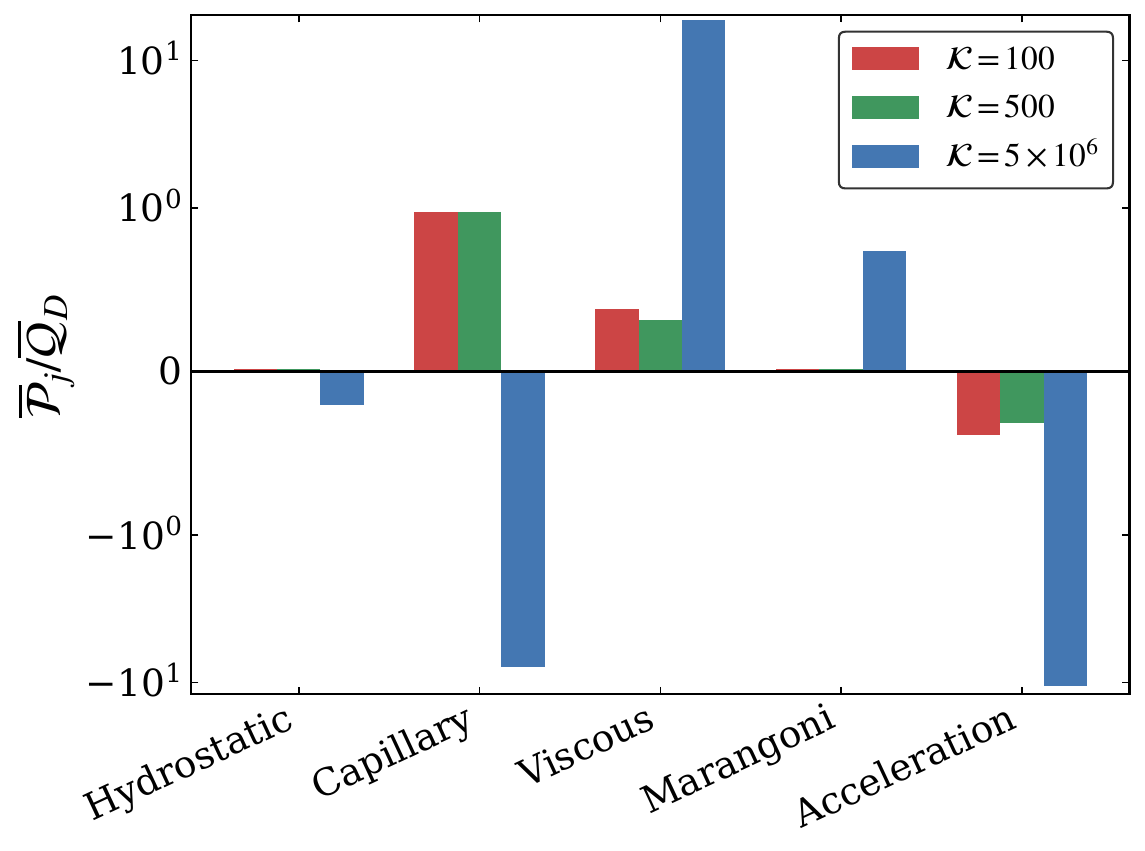}
    \caption{Signed normalised load work.}
    \label{fig:jfm_work_contributions}
\end{subfigure}
\hfill
\begin{subfigure}[t]{0.48\linewidth}
    \centering
    \includegraphics[width=\linewidth]
        {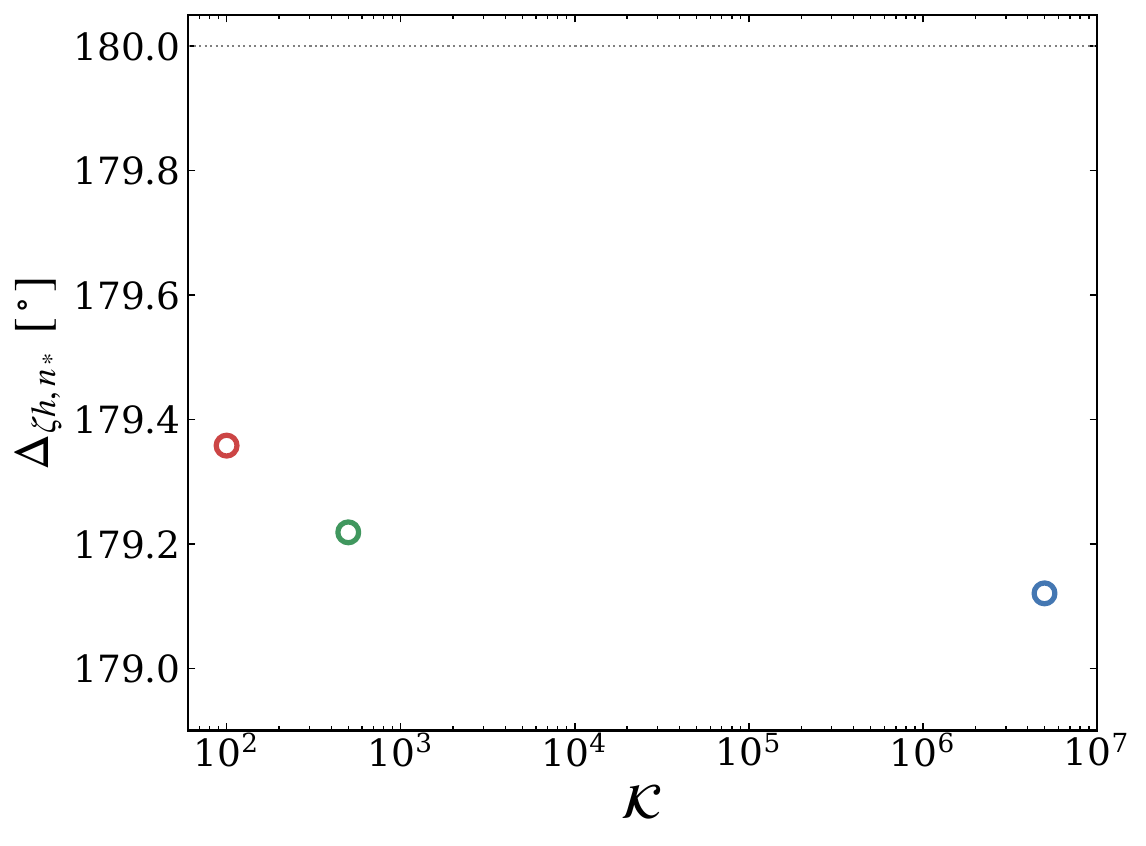}
    \caption{Surface-substrate phase difference.}
    \label{fig:jfm_surface_substrate_phase}
\end{subfigure}

\caption{Mechanical coupling for the support cases in
figure~\ref{fig:nl_stiffness_profiles}.
(a) Signed mean load work normalised by the corresponding mean
substrate damping loss. Positive values supply work to the substrate;
negative values remove it. The ordinate uses a symmetric logarithmic
scale with a linear region between $-1$ and $1$.
(b) Circular mean surface-substrate phase difference for the strongest
late surface mode; the dotted line marks $180^\circ$.
Averages use each case's independently assessed late interval.
For the nearly rigid support,
$\overline{\mathcal Q}_D\simeq8.15\times10^{-6}$,
so large normalised components correspond to small absolute transfers.}
\label{fig:jfm_red_work_phase}

\end{figure}

Figure~\ref{fig:jfm_red_work_phase}(a) resolves this balance. For
$\mathcal K=100$ and $500$ along $\mathcal D=\sqrt{\mathcal K}$,
the mean capillary-pressure work supplies approximately $0.975$ and
$0.973$ of the corresponding mean damping loss. The positive viscous
contributions, approximately $0.379$ and $0.315$, are largely balanced
by negative acceleration contributions, approximately $-0.385$ and
$-0.317$. Thus capillary-pressure work supplies most of the net work
received by these two compliant supports, despite substantial exchange
through the other load components. The capillary contribution includes
the local dependence $\Sigma(\Gamma)$; the explicit Marangoni term
denotes the additional gradient-dependent pressure contribution.

The phase difference is independent of the spatial origin:
\begin{equation}
\Delta_{\zeta h,n_*}
=\arg\!\left(\widehat\zeta_{n_*}\widehat h_{n_*}^{\,*}\right),
\qquad
n_* = \underset{n\geq1}{\operatorname{arg\,max}}
\left\langle|\widehat h_n|^2\right\rangle_t.
\label{eq:jfm_red_phase}
\end{equation}
The circular mean phase differences are approximately
$179.36^\circ$, $179.22^\circ$ and $179.12^\circ$ for the three selected
supports (figure~\ref{fig:jfm_red_work_phase}b).
The nearly opposing surface and substrate displacements explain why
the extrema of $h$ and $H=h-\zeta$ respond differently.
The mean damping loss falls from approximately $8.523$ at
$\mathcal K=100$ to $8.15\times10^{-6}$ in the nearly rigid case,
while the propagating liquid wave remains finite. The nearly rigid
case has different component signs and very small absolute transfers;
the capillary-work balance stated above is specific to the two compliant
supports.
\par

\subsubsection{A control at fixed damping}

\begin{figure}
\centering

\begin{subfigure}[t]{0.48\linewidth}
    \centering
    \includegraphics[width=\linewidth]
        {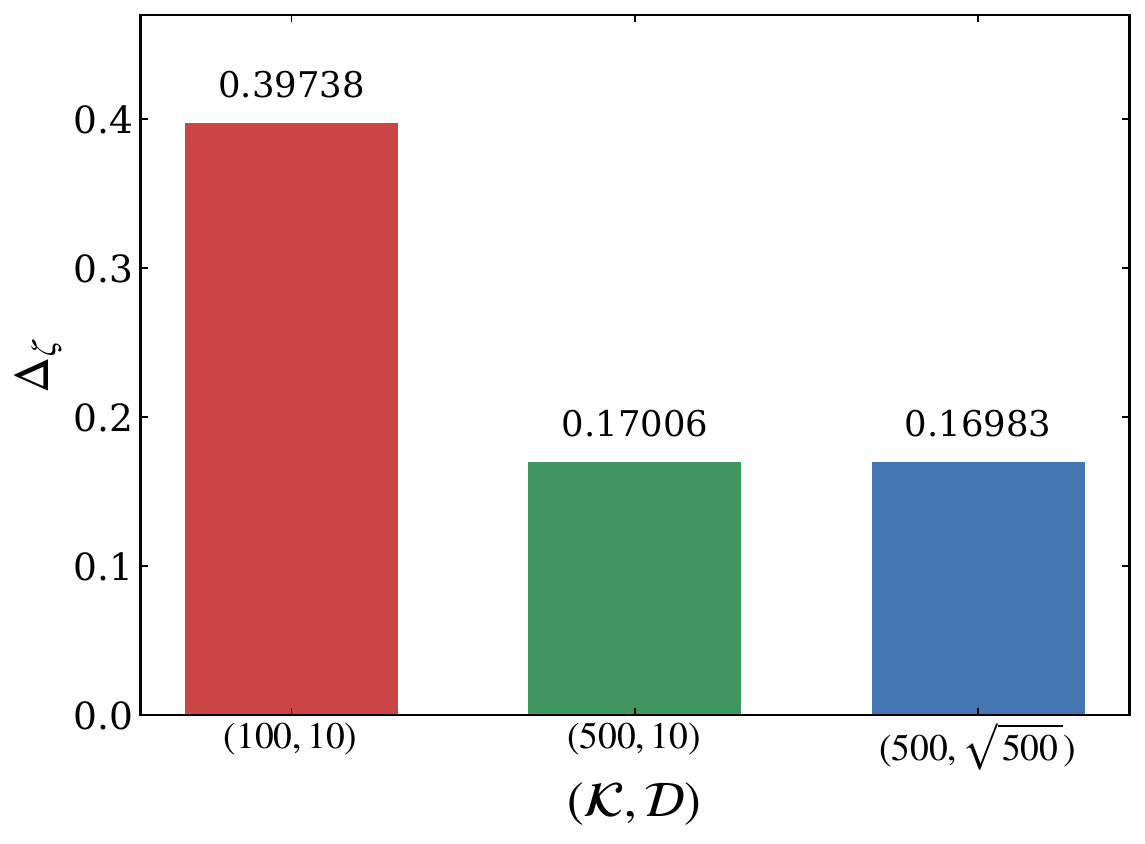}
    \caption{Peak-to-peak substrate displacement.}
    \label{fig:jfm_control_amplitude}
\end{subfigure}
\hfill
\begin{subfigure}[t]{0.48\linewidth}
    \centering
    \includegraphics[width=\linewidth]
        {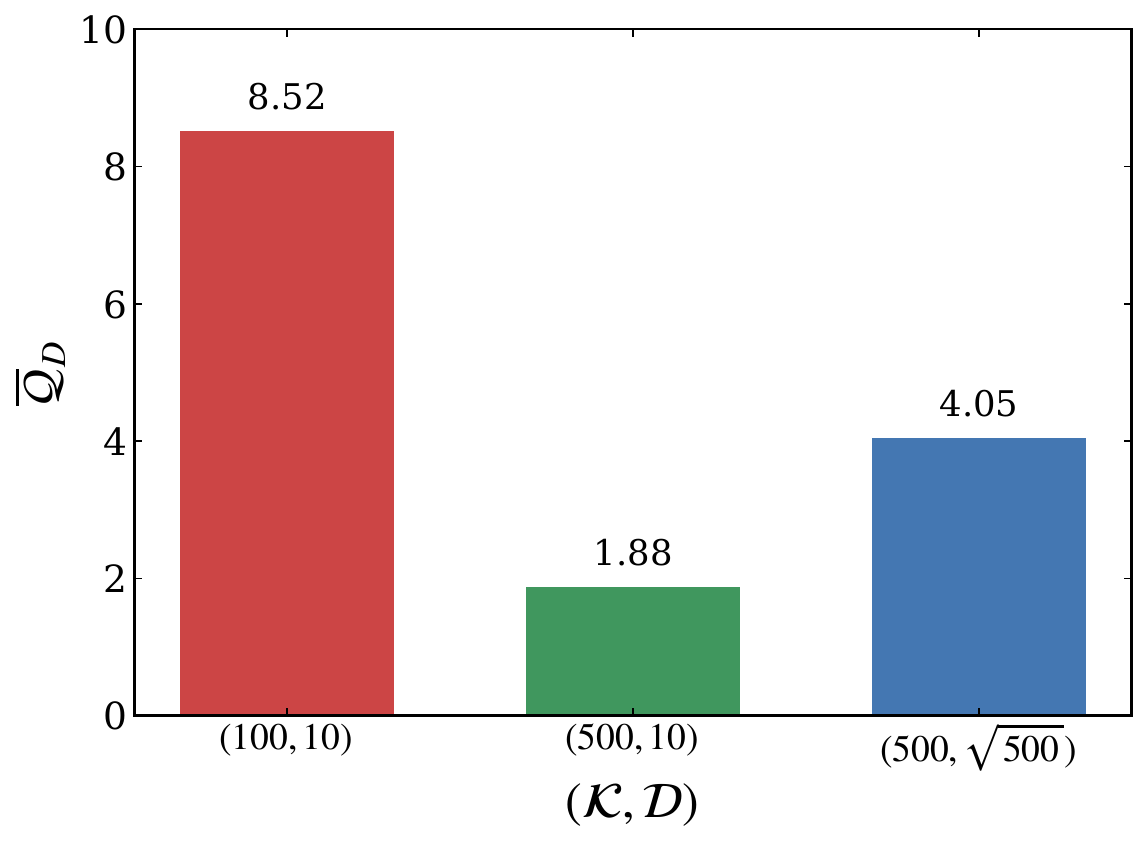}
    \caption{Mean substrate damping loss.}
    \label{fig:jfm_control_damping}
\end{subfigure}

\caption{Finite-time controls at $Re=25$, $\Gamma_e=0.1$,
$\mathcal M=6$ and $Ka=3905$.
The abscissa specifies $(\mathcal K,\mathcal D)$.
(a) Actual peak-to-peak substrate displacement at the common
observation time $t=2000$.
(b) Mean substrate damping loss over $1600\leq t\leq2000$.
All runs use the same initial condition and remaining coefficients.
The $(500,10)$ control is an unsteady propagating state and does
not meet the strict travelling-wave criteria.}
\label{fig:jfm_red_control}

\end{figure}

The original support family changes stiffness and damping together.
A further calculation uses $\mathcal K=500$ and $\mathcal D=10$,
with all other coefficients and the initial condition unchanged.
Figure~\ref{fig:jfm_red_control} compares the actual displacement at
$t=2000$ and the mean damping loss over $1600\leq t\leq2000$.
At fixed $\mathcal D=10$, increasing $\mathcal K$ from $100$ to $500$
reduces the peak-to-peak displacement from approximately $0.3974$ to
$0.1701$. At fixed $\mathcal K=500$, replacing
$\mathcal D=\sqrt{500}$ by $10$ changes this displacement by only
about $0.14\%$, whereas the mean damping loss decreases from
approximately $4.047$ to $1.879$, a reduction of about $53.6\%$.
For these selected states, the control therefore associates the large
deformation change chiefly with stiffness, while damping substantially
changes the work balance.

The additional control is a finite-time propagating state.
Its late full-state travelling residual is approximately $0.0355$ and
its modal amplitude drift is approximately $0.0246$, both above the
$10^{-3}$ acceptance thresholds specified in
section~\ref{sec:verification}\,\ref{subsec:verification_scope}. It is consequently excluded
from the verified travelling-wave family. The comparison isolates the
selected finite-time responses and does not establish equality of their
eventual asymptotic states.
\par

\subsection{Effect of equilibrium surfactant coverage}
\label{sec:nl_coverage}

\begin{figure*}
\centering
\begin{minipage}[t]{0.48\textwidth}
    \centering
    \includegraphics[width=\linewidth]
    {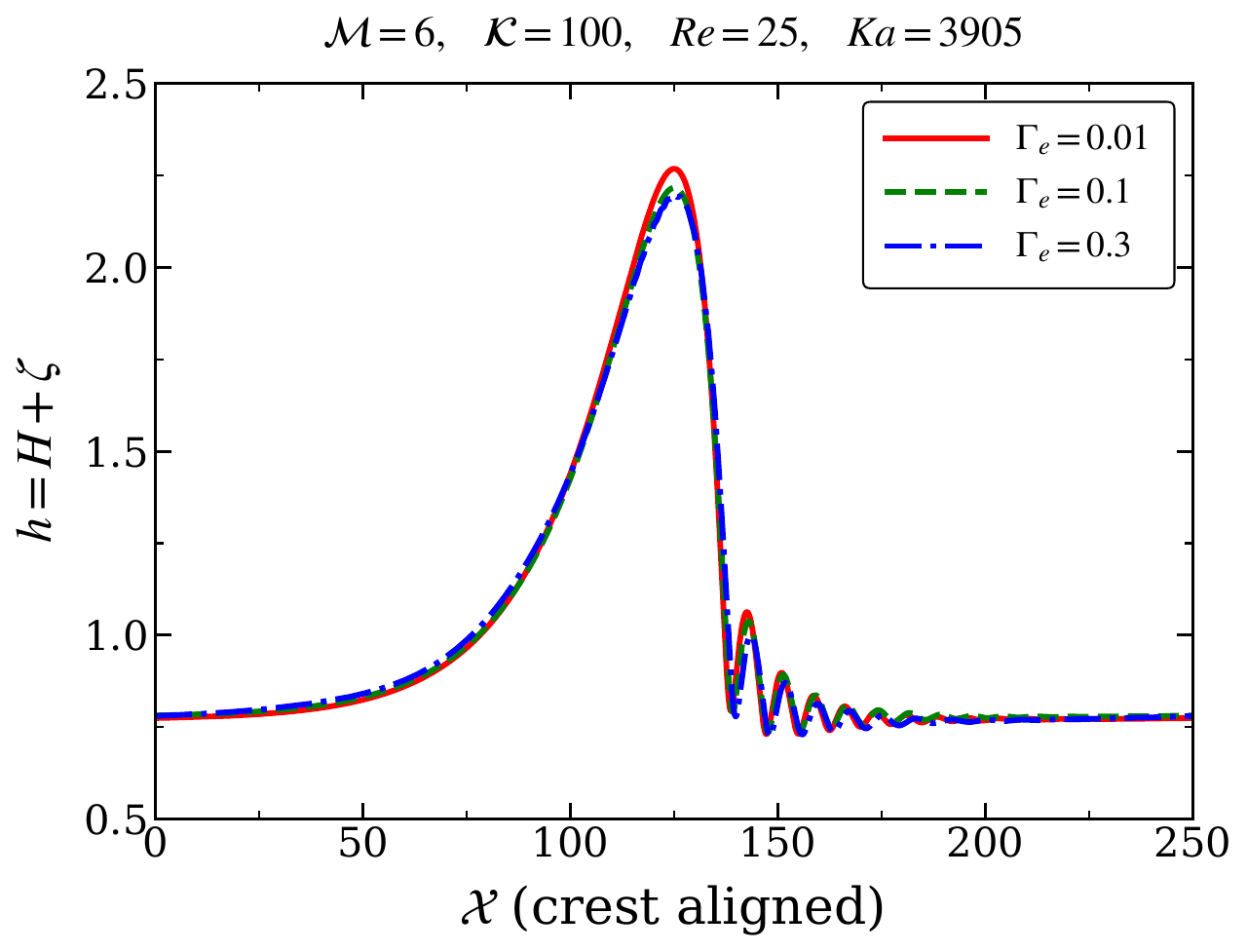}
    \par\smallskip
    {\small (a) Free-surface height $h$.}
\end{minipage}
\hfill
\begin{minipage}[t]{0.48\textwidth}
    \centering
    \includegraphics[width=\linewidth]
    {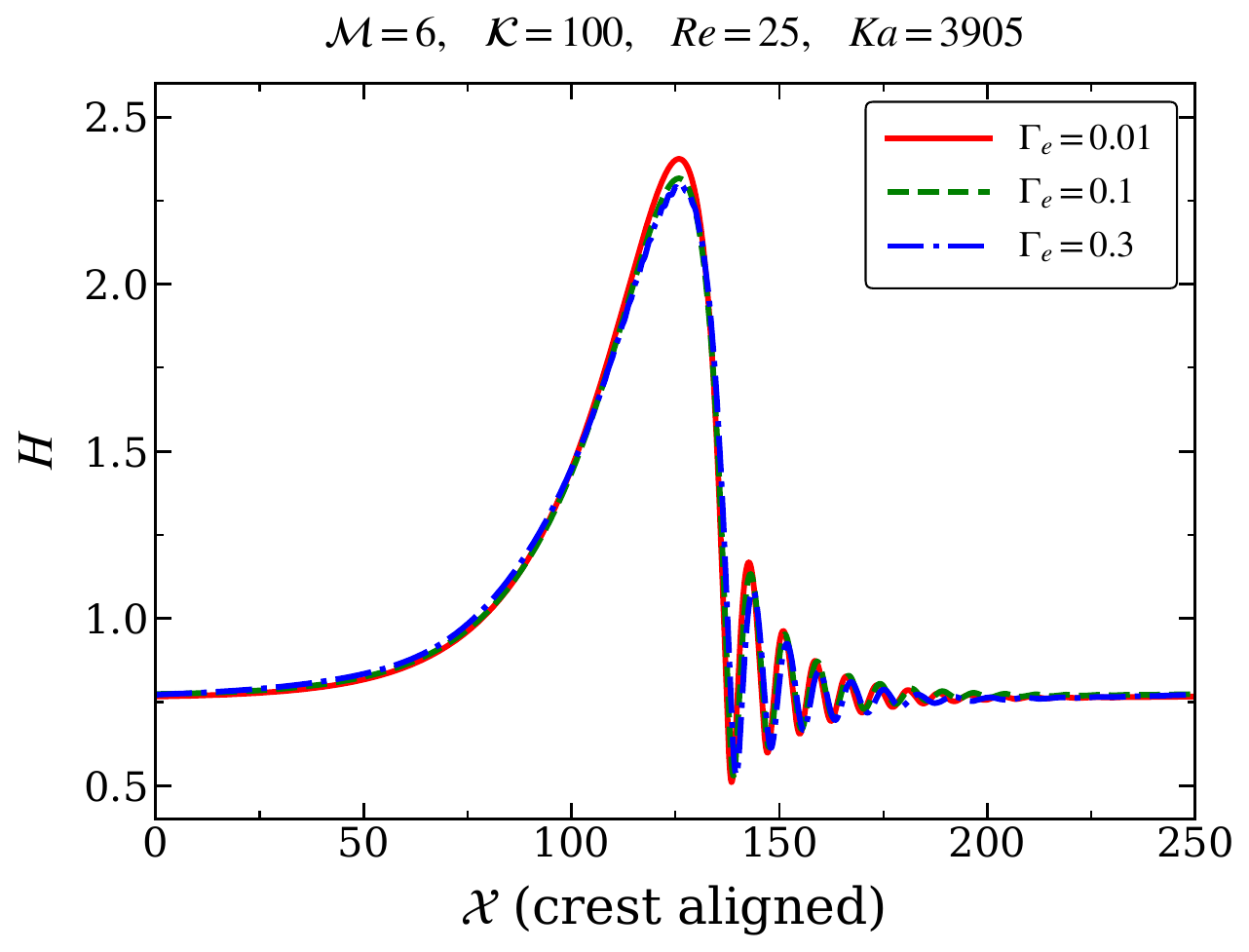}
    \par\smallskip
    {\small (b) Liquid thickness $H$.}
\end{minipage}

\medskip

\begin{minipage}[t]{0.48\textwidth}
    \centering
    \includegraphics[width=\linewidth]
    {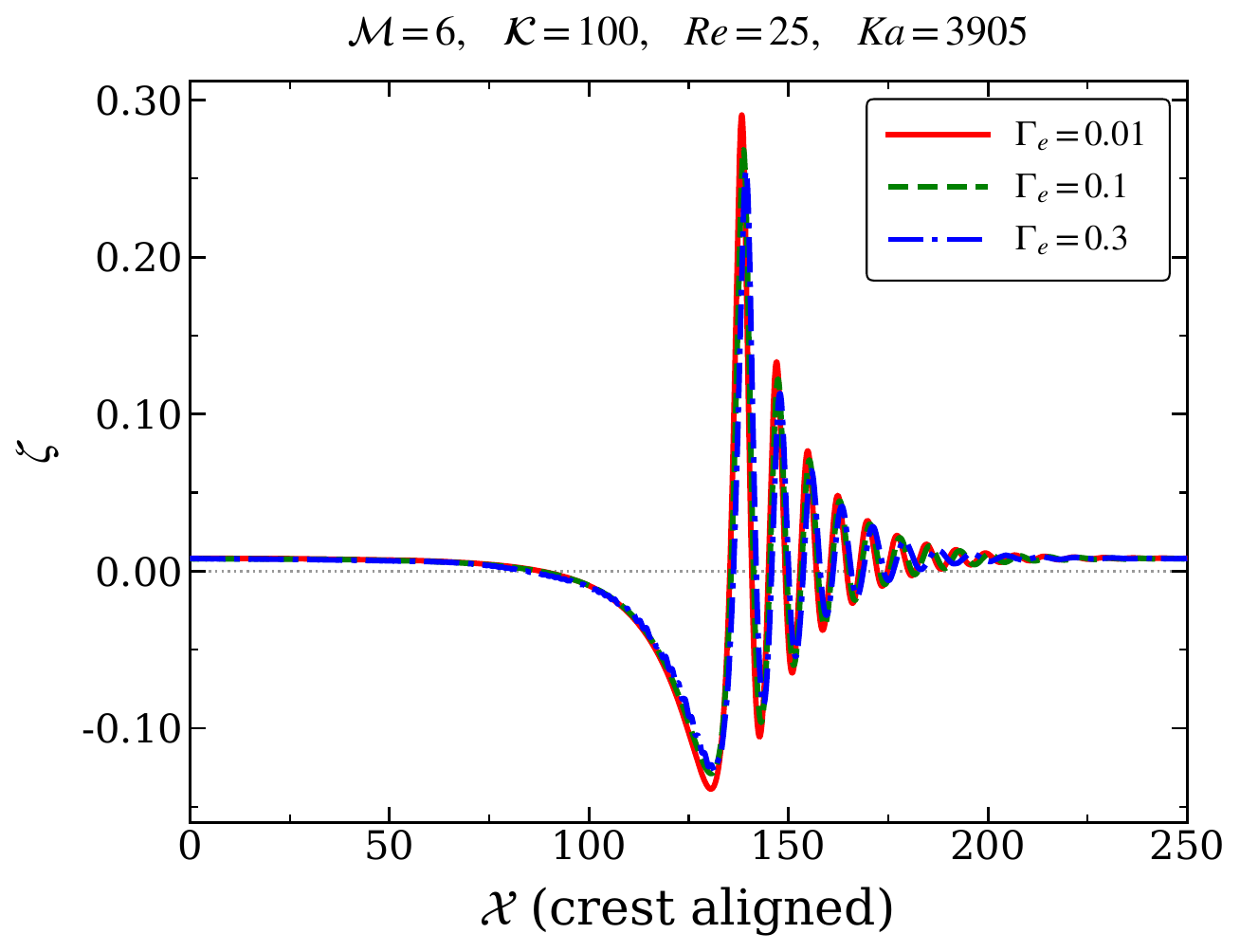}
    \par\smallskip
    {\small (c) Membrane displacement $\zeta$.}
\end{minipage}
\hfill
\begin{minipage}[t]{0.48\textwidth}
    \centering
    \includegraphics[width=\linewidth]
    {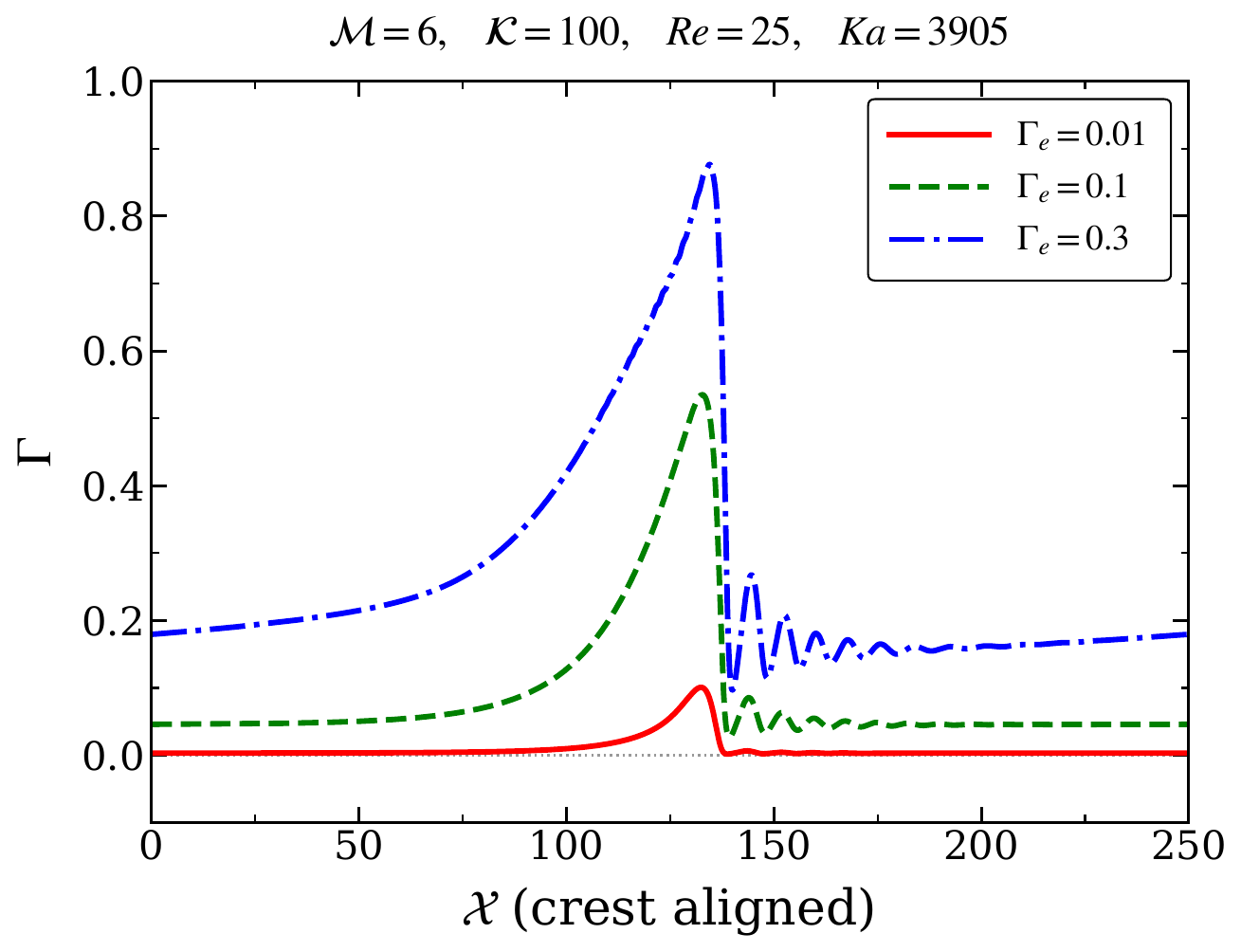}
    \par\smallskip
    {\small (d) Surface surfactant coverage $\Gamma$.}
\end{minipage}
\caption{
Computed nonlinear profiles for varying equilibrium
surfactant coverage at $\mathcal M=6$, $\mathcal K=100$,
$Re=25$ and $Ka=3905$.
Red solid, green dashed and blue dash-dotted curves
correspond to $\Gamma_e=0.01$, $0.1$ and $0.3$,
respectively.
Panels show (a) the free-surface elevation,
(b) the liquid thickness, (c) the substrate displacement
and (d) the surface surfactant coverage.
Each complete state is translated to align its principal
free-surface crest, preserving the relative phase of
all fields.
}
\label{fig:nl_coverage_profiles}
\end{figure*}

The influence of equilibrium surfactant coverage is examined
for $\Gamma_e\in\{0.01,0.1,0.3\}$ at
$Re=25$, $\mathcal{M}=6$, $\mathcal{K}=100$ and $Ka=3905$,
with the remaining parameters unchanged.
Figure~\ref{fig:nl_coverage_profiles} compares the nonlinear profiles at a common observation time.
A single spatial translation is applied jointly to all
fields of each case, aligning the principal free-surface
crests while preserving the relative positions of the
liquid, substrate and surfactant distributions.

The liquid profiles exhibit a principal crest, a steep
descending front and smaller capillary ripples.
At the sampled time, increasing $\Gamma_e$ reduces the
maximum free-surface elevation and maximum liquid thickness.
The distinction between these quantities follows from
$h=H+\zeta$: substrate deformation contributes to the
measured elevation, whereas $H$ determines the local
liquid thickness.
The accompanying reduction in membrane displacement
reflects the change in fluid traction produced by the
coupled liquid and surfactant dynamics.

The surface concentration is strongly non-uniform,
with enrichment near the descending wave front and
depletion over the adjacent region.
These gradients generate Marangoni stresses,
$\Sigma_x=-\mathcal{M}\Gamma_x$, which modify the surface
velocity and feed back on surfactant transport.
Increasing $\Gamma_e$ raises the absolute concentration
peak, while reducing its amplification relative to
the equilibrium loading.
Since the equilibrium bulk concentration and initial
total surfactant inventory also increase with $\Gamma_e$,
the profiles describe redistribution within different
surfactant loadings.
Absolute concentration peaks should therefore be
interpreted together with the reference coverage and
the spatial concentration gradients.

\subsection{Effect of the Marangoni parameter}
\label{sec:nl_marangoni}

\begin{figure*}
\centering
\begin{minipage}[t]{0.48\textwidth}
    \centering
    \includegraphics[width=\linewidth]
    {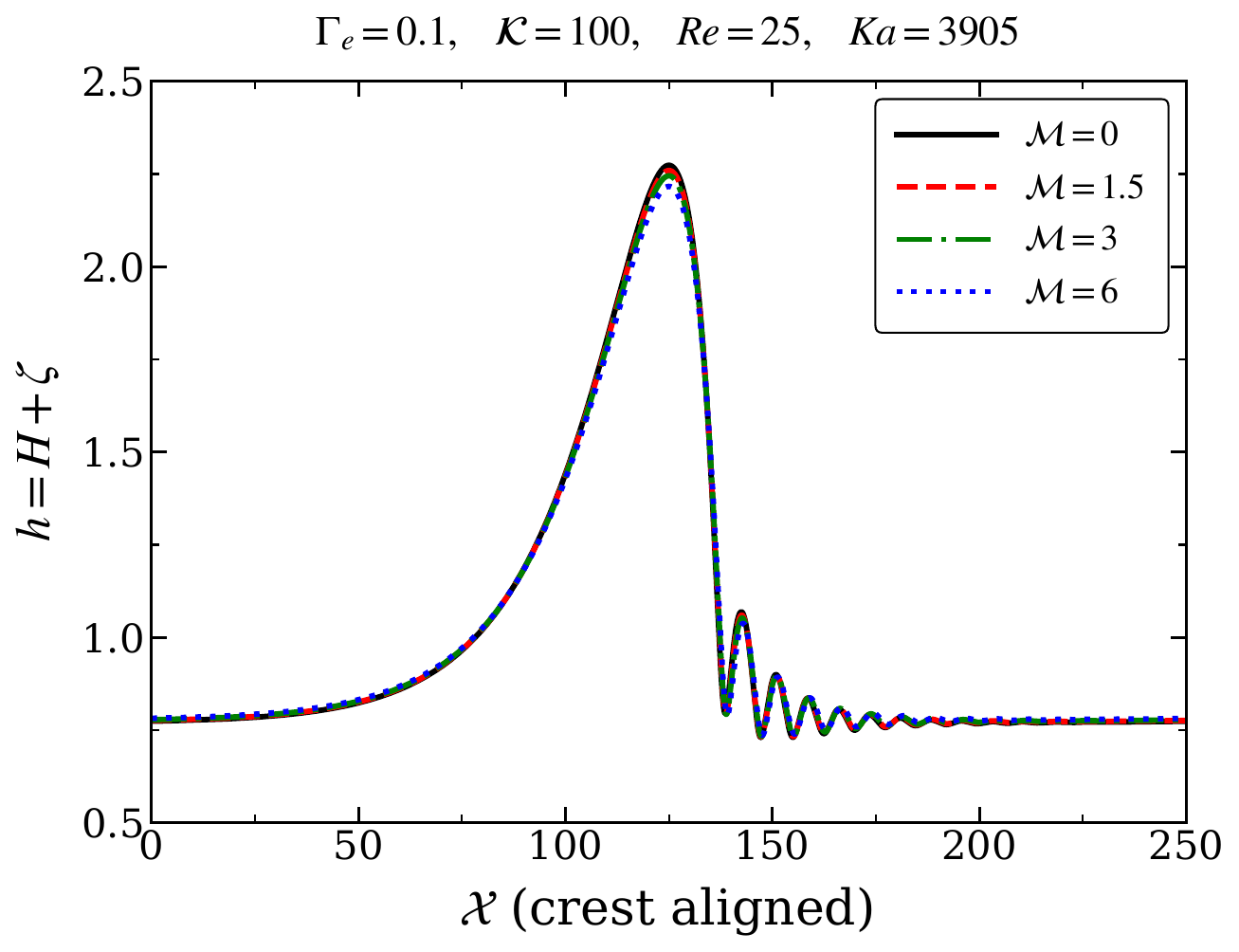}
    \par\smallskip
    {\small (a) Free-surface height $h$.}
\end{minipage}
\hfill
\begin{minipage}[t]{0.48\textwidth}
    \centering
    \includegraphics[width=\linewidth]
    {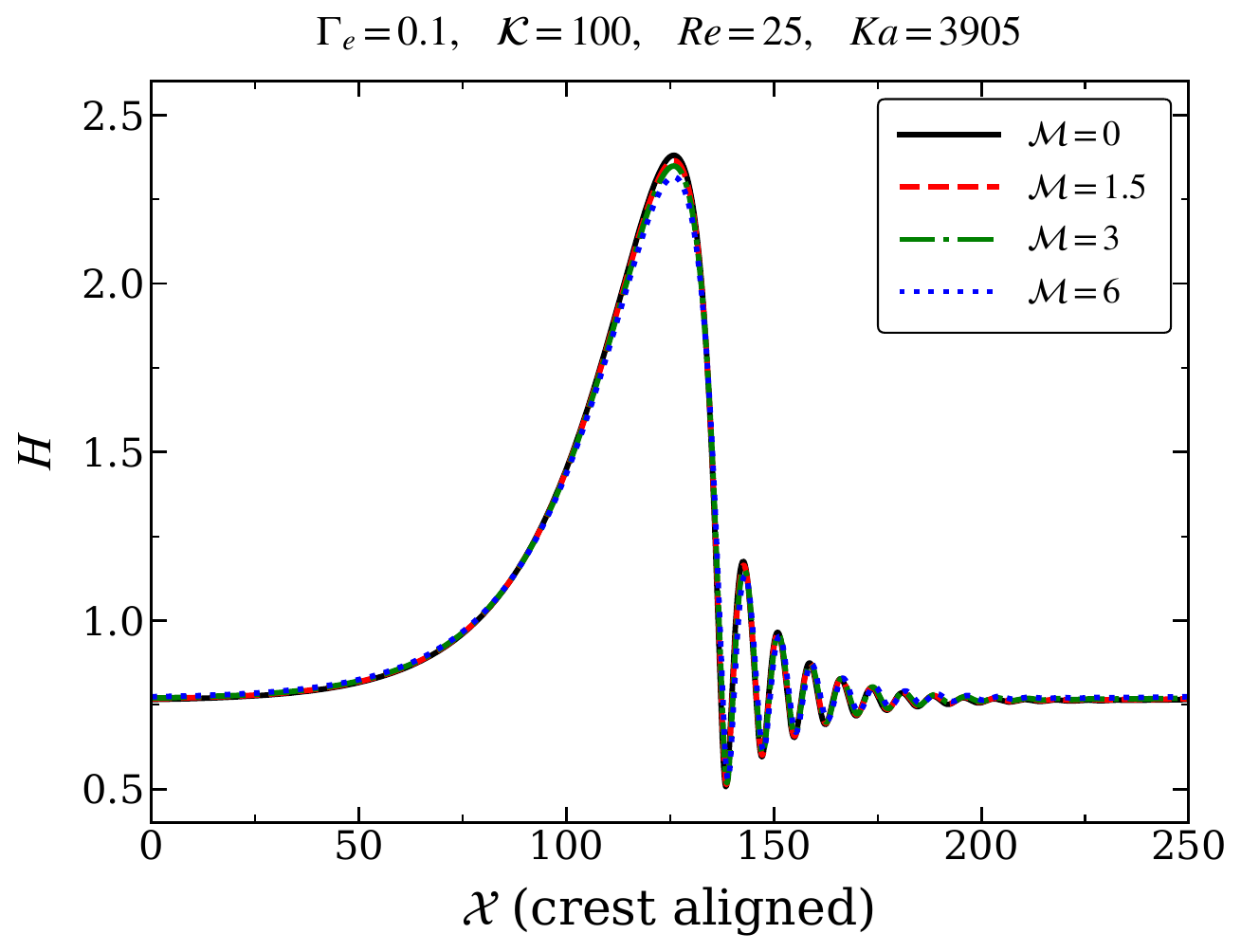}
    \par\smallskip
    {\small (b) Liquid thickness $H$.}
\end{minipage}

\medskip

\begin{minipage}[t]{0.48\textwidth}
    \centering
    \includegraphics[width=\linewidth]
    {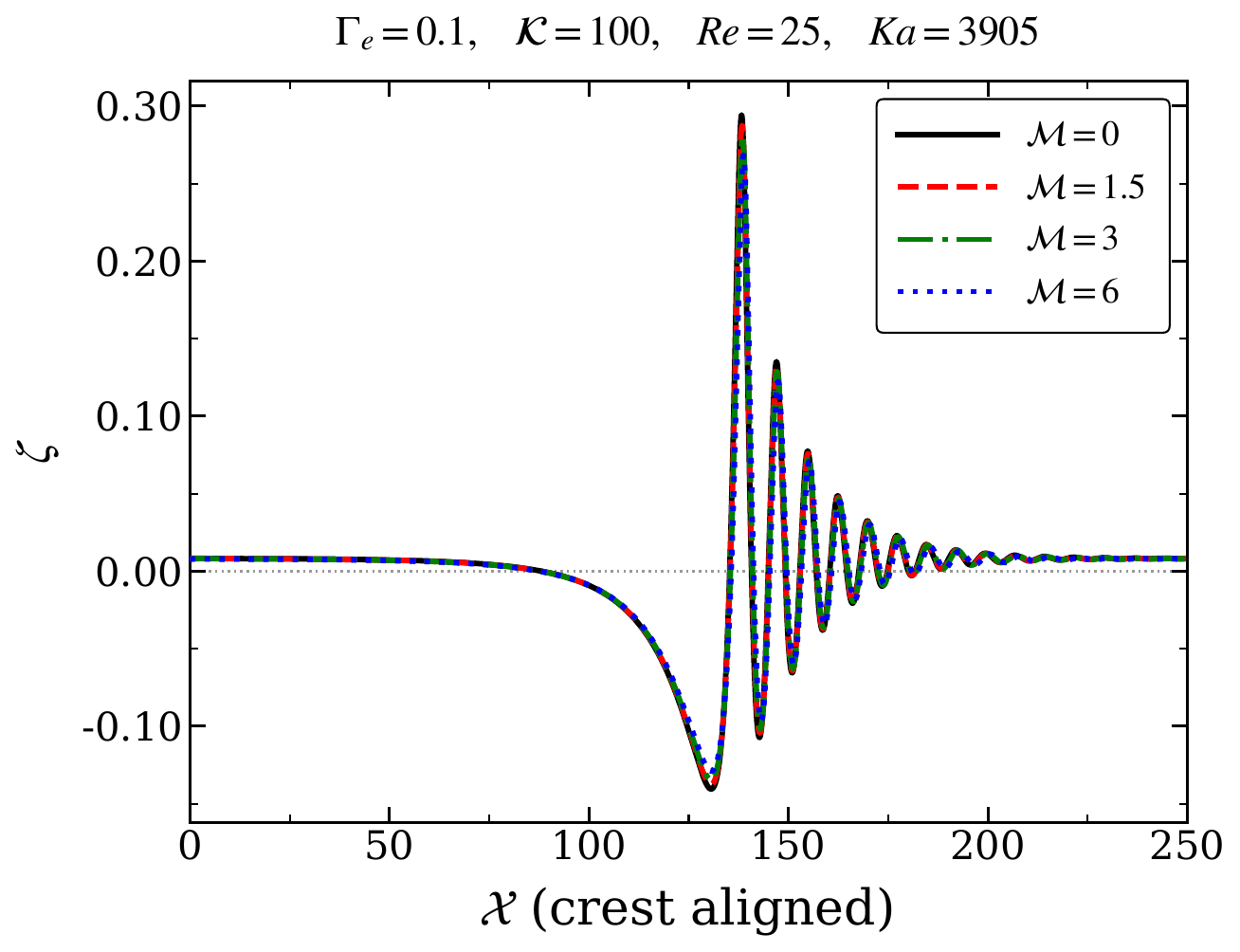}
    \par\smallskip
    {\small (c) Membrane displacement $\zeta$.}
\end{minipage}
\hfill
\begin{minipage}[t]{0.48\textwidth}
    \centering
    \includegraphics[width=\linewidth]
    {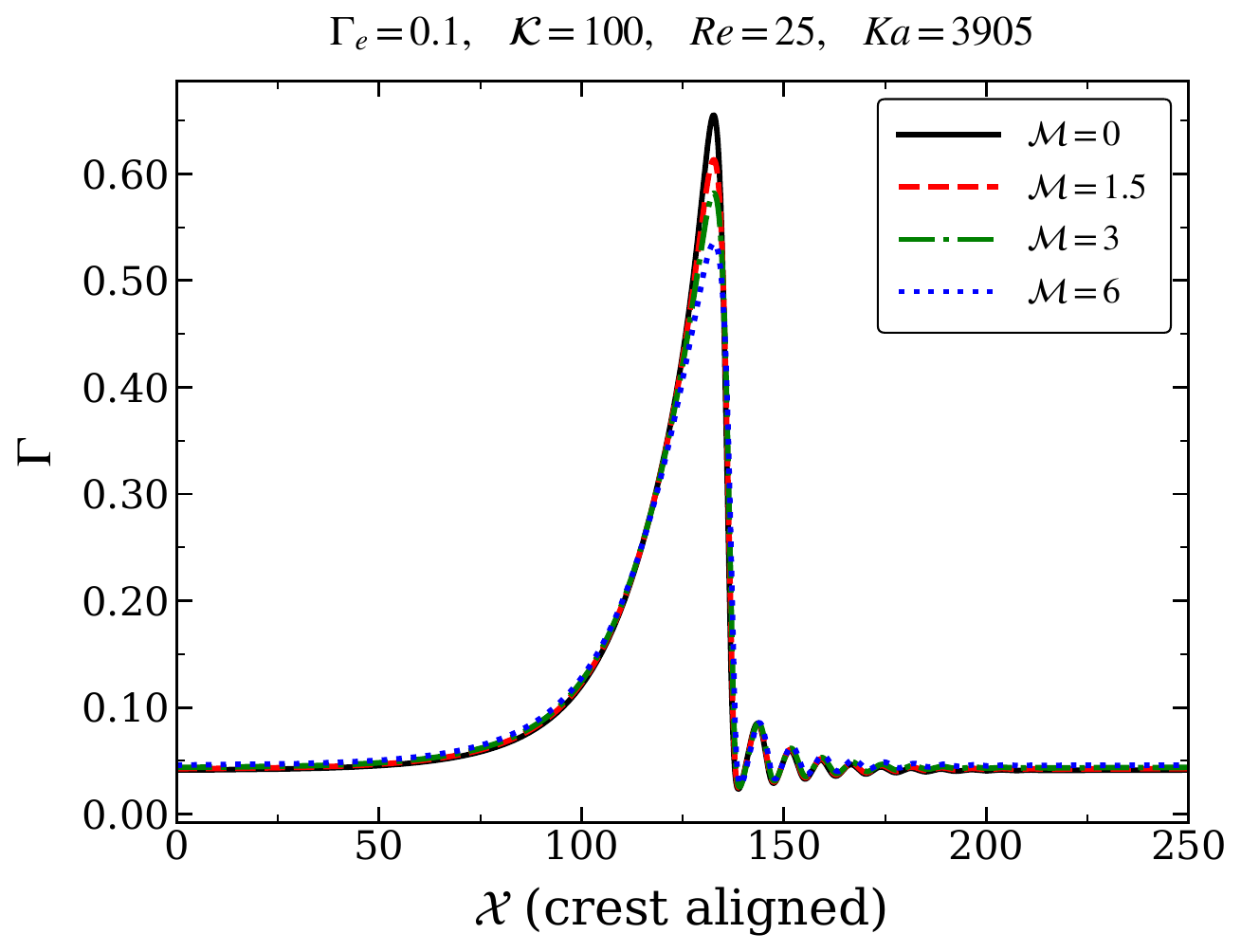}
    \par\smallskip
    {\small (d) Surface surfactant coverage $\Gamma$.}
\end{minipage}
\caption{
Nonlinear profiles for varying Marangoni parameter at
$\Gamma_e=0.1$, $\mathcal{K}=100$, $Re=25$ and $Ka=3905$.
Profiles are shown at $t=2000$, black solid, red dashed, green dash-dotted and blue dotted
curves correspond to $\mathcal{M}=0$, $1.5$, $3$ and $6$,
respectively.
Panels show (a) the laboratory free-surface height,
(b) the actual liquid thickness, (c) the membrane displacement
and (d) the surface surfactant coverage.
Each complete state is translated to place its principal
free-surface crest at $\mathcal{X}=L/2$, preserving the relative
phase between all fields.
}
\label{fig:nl_marangoni_profiles}
\end{figure*}

The influence of Marangoni coupling is examined for
$\mathcal{M}\in\{0,1.5,3,6\}$ at
$\Gamma_e=0.1$, $\mathcal{K}=100$, $Re=25$ and $Ka=3905$.
The remaining parameters and initial conditions are unchanged,
so all four cases have the same initial surfactant inventory.
Figure~\ref{fig:nl_marangoni_profiles} compares the resulting
liquid, membrane and surfactant profiles using the same
whole-state crest alignment.

The liquid profiles remain close across the sweep, with
moderate differences near the principal crest and the first
capillary ripples. The maximum free-surface height decreases
from approximately $2.273$ at $\mathcal{M}=0$ to $2.216$
at $\mathcal{M}=6$, while the maximum liquid thickness
decreases from approximately $2.380$ to $2.316$.
The minimum thickness increases slightly, from approximately
$0.507$ to $0.529$. The principal wave and capillary-ripple
train persist throughout the parameter range, while their
contrast is reduced as the Marangoni coupling strengthens.

The membrane response follows the changes in the liquid wave.
Its broad downward displacement and localized upward excursion
remain present, but the peak-to-peak displacement decreases
from approximately $0.435$ to $0.397$ between
$\mathcal{M}=0$ and $6$.
Because the mechanical properties of the membrane are fixed,
this reduction reflects the modified fluid forcing transmitted
to the substrate.

The surfactant concentration peak is more sensitive to
$\mathcal{M}$ than the overall liquid-wave geometry.
The maximum coverage decreases through approximately
$0.655$, $0.613$, $0.582$ and $0.535$ for
$\mathcal{M}=0$, $1.5$, $3$ and $6$, respectively.
The enrichment remains localized near the descending wave
front, but stronger Marangoni coupling reduces the
concentration contrast.

The surface surfactant flux in the weighted-residual model
can be written as
\begin{equation}
F_s
=
\frac{3q\Gamma}{2H}
-
\left(
\frac{1}{Pe_s}
+\frac{\mathcal{M}H\Gamma}{4}
\right)
\partial_x\Gamma.
\label{eq:nl_marangoni_surface_flux}
\end{equation}
For positive $H$ and $\Gamma$, the Marangoni contribution
acts opposite to the concentration gradient, transporting
surfactant away from locally enriched regions.
Increasing $\mathcal{M}$ strengthens this contribution,
providing a mechanism consistent with the reduction of
the concentration peaks. The molecular diffusivity remains
fixed; the additional redistribution results from the
coupling between surfactant gradients and surface motion.

At $\mathcal{M}=0$, surfactant transport and bulk-surface
exchange remain active, allowing the liquid wave to generate
a nonuniform concentration distribution while the Marangoni
stress vanishes. At finite $\mathcal{M}$, these concentration
gradients feed back onto the liquid motion. The resulting
changes in surface transport, liquid flux and fluid traction
account for the coupled variations of $\Gamma$, $H$ and
$\zeta$ in the figure.

\section{Consistency checks and validation}
\label{sec:verification}

This section connects the preceding Orr-Sommerfeld analysis,
weighted-residual formulation and nonlinear simulations through
independent consistency checks. The parameter definitions and
continuation paths are those established in
Sec.~III\, \ref{subsec:parameter_regime}.
The checks distinguish the algebraic and conservation properties
of the reduced model from its quantitative agreement with the
full linear problem and the accuracy of its numerical integration.

\subsection{Rigid straight-wall limit}
\label{subsec:rigid_limit}

The rigid straight-wall limit is obtained by setting
\begin{equation}
    \zeta=0,
    \qquad
    v=0,
    \qquad
    H=h,
    \label{eq:rigid_substitution}
\end{equation}
and suppressing the wall evolution equations.
All moving-substrate contributions vanish, including
$\mathcal{R}_\zeta$.

The retained no-slip momentum coefficients are
\begin{equation}
    \left(
        \frac{17}{7},
        \frac97,
        \frac52,
        \frac56,
        4,
        \frac92,
        6,
        \frac92,
        \frac1{48},
        \frac{15}{224},
        \frac{19}{336},
        \frac5{112}
    \right).
    \label{eq:coefficient_vector}
\end{equation}
For the adopted $G=2$ normalization, these agree with the
no-slip flow-rate coefficients of
\citet{mukhopadhyay2026}.
The surface transport equation likewise retains the conservative
form used in \citet{mukhopadhyay2026pof}.

The rigid-wall bulk diffusion operator follows from the
conservative bulk flux:
\begin{equation}
    \frac1{Pe_b}
    \partial_x(\chi_x+h\phi_x)
    =
    \frac1{Pe_b}
    \left(
        \chi_{xx}+h_x\phi_x+h\phi_{xx}
    \right).
    \label{eq:rigid_conservative_diffusion}
\end{equation}
In comparison, the printed bulk equation in \citet{mukhopadhyay2026} contains
\begin{equation}
    \frac1{Pe_b}
    \left(
        \chi_{xx}+h\phi_{xx}
        -\frac{3\chi}{h^2}h_x^2
    \right).
    \label{eq:printed_rigid_bulk_diffusion}
\end{equation}
The conservative boundary-flux completion is therefore
\begin{equation}
    \frac1{Pe_b}
    \left(
        h_x\phi_x+\frac{3\chi}{h^2}h_x^2
    \right).
    \label{eq:rigid_diffusion_correction}
\end{equation}
Both additional terms have zero first variation about a flat,
uniform equilibrium with $\chi=0$.
The completion consequently preserves the rigid-wall
linearization while modifying the nonlinear bulk transport
to retain its conservative flux structure
\citep{mukhopadhyay2026audit}.

Removing the surfactant subsystem and setting
$\mathcal{M}=0$ gives the clean compliant-film model.
Applying Eq.~\eqref{eq:rigid_substitution} additionally gives
the clean rigid-wall weighted-residual closure.
By contrast, the $\mathcal{M}=0$ case in the preceding
Marangoni sweep retains surfactant transport and exchange,
with vanishing Marangoni feedback on the liquid motion.

\subsection{Uniform equilibrium and fluid added mass}
\label{subsec:base_wall_checks}

For the prescribed backing pressure $P_e=0$, the uniform
reference equilibrium is
\begin{equation}
\begin{gathered}
    H=1,
    \qquad
    q=\frac G3,
    \qquad
    \zeta=v=0,
    \\
    \Gamma=\Gamma_e,
    \qquad
    \chi=0,
    \qquad
    \phi=\frac{\Gamma_e}{\kappa(1-\Gamma_e)}.
\end{gathered}
\label{eq:WR_base}
\end{equation}
Direct substitution gives zero residual in every reduced
equation. This verifies the gravitational forcing, viscous
relaxation, adsorption equilibrium and reference wall load
within a single base-state check.

A separate dynamical check follows from a spatially uniform
wall translation beneath a film of constant thickness $H_0$.
The free surface follows the substrate, and incompressibility
gives $w=v(t)$ throughout the liquid.
With zero free-surface curvature, the transverse momentum
equation yields
\begin{equation}
    \Pi(z)
    =
    \left(
        G\cot\theta+Re\,v_t
    \right)(h-z).
    \label{eq:uniform_pressure}
\end{equation}
The resulting wall equation is
\begin{equation}
\begin{aligned}
    (\mathcal{I}+ReH_0)v_t
    +\mathcal{D}v+\mathcal{K}\zeta
    &=
    P_e-G\cot\theta(H_0-1),
    \\
    \zeta_t&=v.
\end{aligned}
\label{eq:uniform_added_mass}
\end{equation}
For $H_0=1$, this becomes
\begin{equation}
    (\mathcal{I}+Re)\zeta_{tt}
    +\mathcal{D}\zeta_t+\mathcal{K}\zeta
    =
    P_e.
    \label{eq:uniform_wall_oscillator}
\end{equation}
The reduced wall equation,
Eq.~\eqref{eq:model_zeta}, reproduces this relation.

The term $ReH_0$ is the liquid added mass.
It supplements the structural inertia without redefining
$\mathcal{I}$, and explains why the nominal bare-foundation
damping ratio does not directly determine the damping of
the coupled fluid-membrane modes.

The restoring coefficient in the uniform translation problem
is $\mathcal{K}$.
The combination $\mathcal{K}-G\cot\theta$ appearing in a
pressure-eliminated Orr-Sommerfeld boundary condition must
be interpreted together with the associated fluid pressure
perturbation. It cannot independently replace
$\mathcal{K}$ in Eq.~\eqref{eq:uniform_wall_oscillator}.

\subsection{Recovery of the long-wave film branch}
\label{subsec:longwave_threshold}

The asymptotic Orr-Sommerfeld solution provides an independent
check on the linear limit of the weighted-residual model.
Consider $k\to0$ at fixed positive $Re$, with finite positive
$k_s$, $Pe_b$, $Pe_s$, $\kappa$ and $\mathcal{K}$.
Define
\[
    a_e=\kappa(1-\Gamma_e)^2.
\]
For the regular film branch, linearization of the reduced
equations gives the leading amplitude relations
\begin{equation}
    \widehat q=G\widehat H,
    \qquad
    \widehat\zeta
    =
    -\frac{G\cot\theta}{\mathcal{K}}\widehat H,
    \qquad
    \widehat\Gamma
    =
    \frac{6\Gamma_e a_e}{4+3a_e}\widehat H.
    \label{eq:longwave_amplitudes}
\end{equation}
These recover the leading liquid, membrane and surfactant
responses obtained from the full asymptotic problem.

The temporal eigenvalue is
\begin{equation}
\begin{aligned}
    \lambda
    ={}&
    -\mathrm{i}kG
    +k^2
    \left[
        \frac{2ReG^2}{15}
        -
        \frac{G\cot\theta}{3}
        \left(
            1-\frac{G\cot\theta}{\mathcal{K}}
        \right)
        -
        \frac{3\mathcal{M}\Gamma_e a_e}{4+3a_e}
    \right]
    +O(k^3).
\end{aligned}
\label{eq:longwave_lambda}
\end{equation}
Thus, the reduced model reproduces both the leading
propagation speed and the $O(k^2)$ temporal growth coefficient
of the full film branch.

For the fixed-$\mathcal{M}$ continuation adopted throughout
the parameter study, the long-wave neutral condition gives
\begin{equation}
    Re_c^{\mathrm{LW}}
    =
    \frac{5\cot\theta}{2G}
    \left(
        1-\frac{G\cot\theta}{\mathcal{K}}
    \right)
    +
    \frac{45\mathcal{M}\Gamma_e a_e}
    {2G^2(4+3a_e)}.
    \label{eq:Rec_fixed_M}
\end{equation}
The independence of this leading threshold from the transport
rates and membrane inertia and damping follows from the stated
regular long-wave limit. These coefficients remain relevant
to the finite-wavenumber spectrum and nonlinear evolution.

For completeness, substituting
$\mathcal{M}=Re\,Mn$ gives the corresponding implicit condition
for an alternative fixed-$Mn$ continuation:
\begin{equation}
    Re_c^{\mathrm{LW}}
    \left(
        1-\frac{45Mn\,\Gamma_e a_e}
        {2G^2(4+3a_e)}
    \right)
    =
    \frac{5\cot\theta}{2G}
    \left(
        1-\frac{G\cot\theta}{\mathcal{K}}
    \right).
    \label{eq:Rec_fixed_Mn}
\end{equation}
Division requires the factor multiplying
$Re_c^{\mathrm{LW}}$ to be nonzero.
Its sign determines how the leading growth coefficient
changes with $Re$. The numerical continuation presented
in this paper uses Eq.~\eqref{eq:Rec_fixed_M}.

For a clean film, $\mathcal{K}=G\cot\theta$ marks the
vanishing of the long-wave film threshold.
When $\mathcal{K}<G\cot\theta$, this branch is unstable
for arbitrarily small positive $Re$, with the long-wave
limit taken first at each fixed $Re$.
This is a property of the coupled film-wall dynamics,
rather than a universal restriction on mechanical
admissibility. Stability of the remaining wall, shear and
surfactant branches requires the full eigenvalue problem.

\subsection{Comparison with the full Orr-Sommerfeld solution}
\label{subsec:OS-WRL-linear-comparison}

The quantitative linear accuracy of the weighted-residual
model is assessed about the same flat-film equilibrium used
in the full Orr-Sommerfeld calculation.
Both formulations use the common scaling and parameter
paths of Sec.~III\, \ref{subsec:parameter_regime}.
Perturbations are proportional to
$e^{\lambda t+\mathrm{i}kx}$, with real $k$.

Linearizing the weighted-residual equations gives
\begin{equation}
    \mathsf{A}_{\mathrm{WR}}(Re,k)
    \boldsymbol{X}_{\mathrm{WR}}
    =
    \lambda
    \mathsf{B}_{\mathrm{WR}}(Re,k)
    \boldsymbol{X}_{\mathrm{WR}},
    \label{eq:wr-linear-comparison}
\end{equation}
where
\[
    \boldsymbol{X}_{\mathrm{WR}}
    =
    \left(
        \widehat H,\widehat q,\widehat\chi,
        \widehat\Gamma,\widehat\zeta,\widehat v
    \right)^{\mathsf T},
    \qquad
    \widehat v=\lambda\widehat\zeta.
\]
Here, $\widehat H=\widehat h-\widehat\zeta$.
The generalized pencil retains the mixed surfactant time
derivative in the momentum equation and the acceleration
terms in the substrate load.

The film branch is identified through its long-wave limit
and continued in wavenumber.
Its temporal growth rate is
\begin{equation}
    \sigma_f(Re,k)
    =
    \operatorname{Re}\lambda_f(Re,k)
    =
    k\alpha_{i,f}(Re,k),
    \qquad
    \lambda_f=-\mathrm{i}k\alpha_f.
    \label{eq:wr-os-growth-comparison}
\end{equation}
The corresponding neutral boundary satisfies
$\sigma_f(Re,k)=0$.
Positive-wavenumber roots are bracketed and refined using
Brent's method, while the analytical long-wave result supplies
the intercept at $k=0$.

\begin{figure}
\centering
\begin{subfigure}[t]{0.48\linewidth}
    \centering
    \includegraphics[width=\linewidth]
    {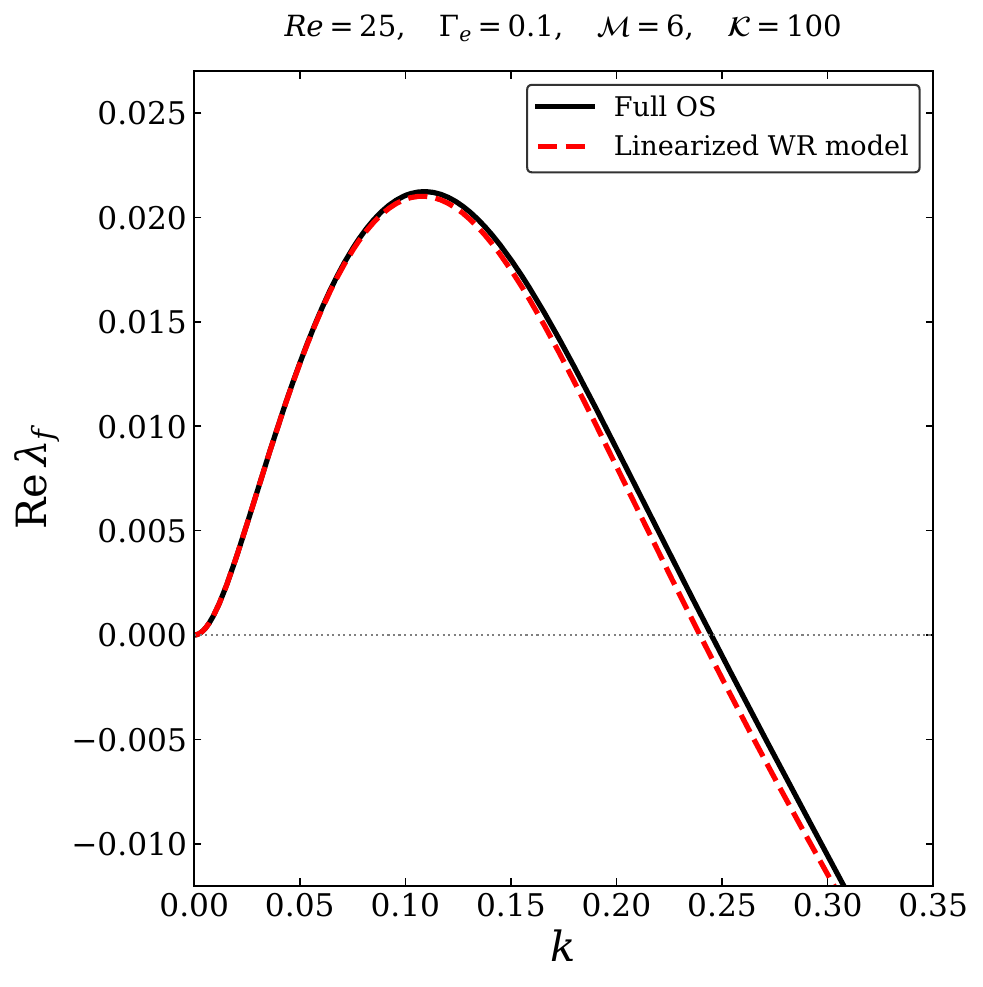}
    \caption{Film-branch growth at $Re=25$.}
    \label{fig:wr-os-growth}
\end{subfigure}
\hfill
\begin{subfigure}[t]{0.48\linewidth}
    \centering
    \includegraphics[width=\linewidth]
    {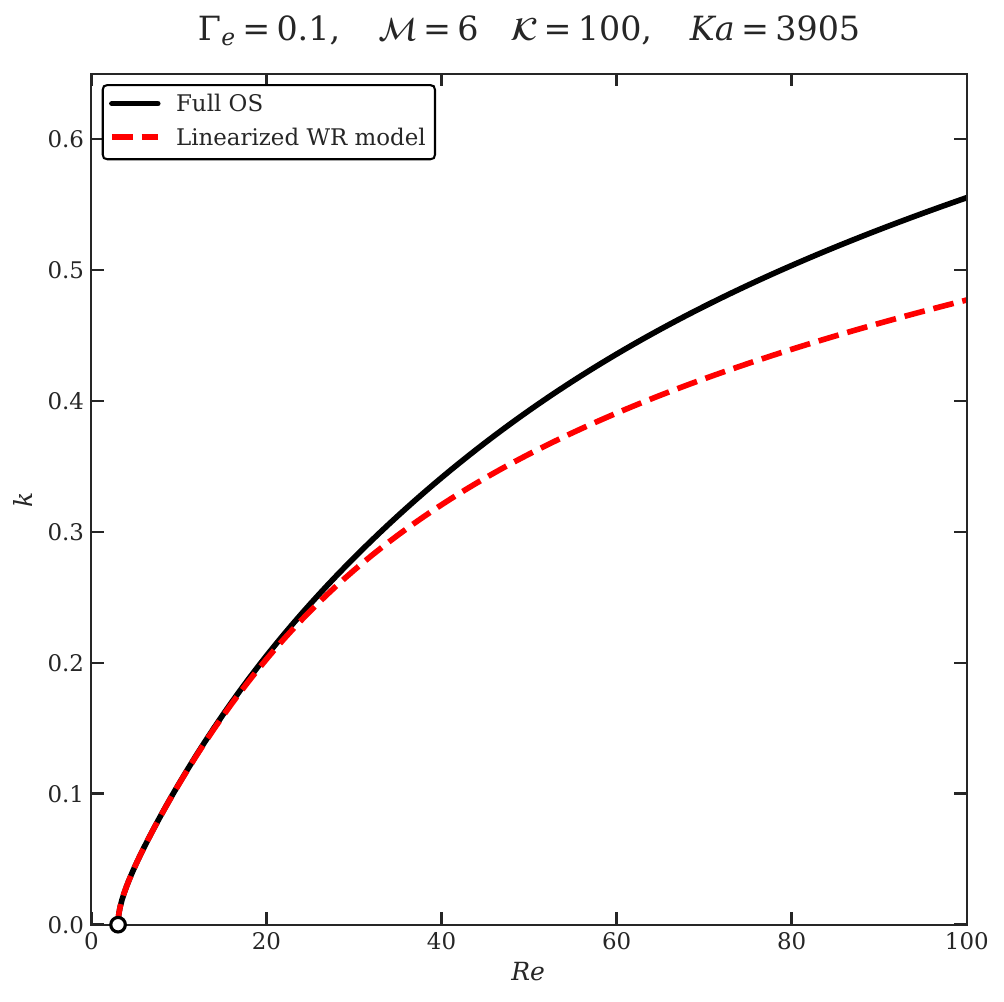}
    \caption{Film-branch marginal stability.}
    \label{fig:wr-os-neutral}
\end{subfigure}
\caption{
Comparison of the full Orr-Sommerfeld solution
(solid black) and the linearized weighted-residual model
(dashed red) for $\Gamma_e=0.1$, $\mathcal{M}=6$,
$\mathcal{K}=100$ and $Ka=3905$.
(a) Temporal film-growth rate
$\sigma_f=\operatorname{Re}\lambda_f=k\alpha_{i,f}$
as a function of real wavenumber at $Re=25$.
The dotted horizontal line denotes zero growth.
(b) Film-branch neutral boundaries
$\sigma_f(Re,k)=0$, with Reynolds number on the
horizontal axis.
The open circle marks the common analytical long-wave
threshold $Re_c^{\mathrm{LW}}\simeq3.056053$.
The remaining parameters and the fixed-$\mathcal{M}$,
fixed-$Ka$ continuation follow
Sec.~III\, \ref{subsec:parameter_regime}.
}
\label{fig:wr-os-comparison}
\end{figure}

Figure~\ref{fig:wr-os-comparison}(a) compares the dispersion
curves at the reference Reynolds number $Re=25$.
The peak locations and growth rates are
\begin{equation}
\begin{aligned}
    k_{\mathrm{peak}}^{\mathrm{OS}}
    &\simeq0.10906,
    &\qquad
    \sigma_{f,\mathrm{peak}}^{\mathrm{OS}}
    &\simeq0.021236,
    \\
    k_{\mathrm{peak}}^{\mathrm{WR}}
    &\simeq0.10792,
    &
    \sigma_{f,\mathrm{peak}}^{\mathrm{WR}}
    &\simeq0.021015.
\end{aligned}
\label{eq:wr_os_peak_comparison}
\end{equation}
The weighted-residual model underestimates the maximum
growth rate by approximately $1.04\%$.
The upper neutral wavenumber is approximately $0.244892$
for the full OS problem and $0.239507$ for WR,
corresponding to a $2.20\%$ underestimate.

Figure~\ref{fig:wr-os-comparison}(b) extends the comparison
to the film-branch neutral curves.
Both formulations recover
$Re_c^{\mathrm{LW}}\simeq3.056053$, and the curves nearly
coincide near this intercept.
Their separation increases with Reynolds number:
at $Re=50$, the neutral wavenumbers are approximately
$0.392744$ and $0.359438$ for OS and WR, respectively;
at $Re=100$, they are $0.555306$ and $0.477138$.
The corresponding WR underestimates are $8.48\%$ and
$14.08\%$.

These results quantify the accuracy of the reduced film
branch near onset and at the reference Reynolds number,
while demonstrating its increasing departure from the full
problem as the neutral boundary reaches larger wavenumbers.
They provide a linear benchmark for the nonlinear simulations
reported above. The finite-amplitude profiles and propagation
speeds are assessed through their nonlinear evolution and
travelling-frame diagnostics, rather than inferred from the
linear growth curves.

The comparison concerns the regular film branch over the
displayed range. Other spectral branches, including growing
short-wave branches of the WR closure at sufficiently large
wavenumber, require separate assessment against the full
OS spectrum.

The substrate-load closure is additionally assessed through
the two oscillating wall branches at the baseline parameters,
with $Re=25$ and $k=0.1$.
The full OS eigenvalues are approximately
\[
    \lambda_{\mathrm{w},1}^{\mathrm{OS}}
    =-0.189481+1.992143i,
    \qquad
    \lambda_{\mathrm{w},2}^{\mathrm{OS}}
    =-0.197657-2.120177i,
\]
whereas the WR model gives
\[
    \lambda_{\mathrm{w},1}^{\mathrm{WR}}
    =-0.188102+1.990403i,
    \qquad
    \lambda_{\mathrm{w},2}^{\mathrm{WR}}
    =-0.196372-2.118619i.
\]
The damping-rate discrepancies are approximately
$0.73\%$ and $0.65\%$, while the oscillation-frequency
discrepancies are below $0.09\%$.
The OS values agree across primitive collocation degrees
$N=32,40,48$.
This comparison tests the selected finite-wavenumber wall
response in addition to the uniform added-mass limit.

\subsection{Numerical consistency and scope of verification}
\label{subsec:verification_scope}

The projection coefficients can be checked independently
through the mapped derivative relations and polynomial moments.
For the velocity basis used in the reduced model,
\begin{equation}
    \int_0^1 f_0^2\,\mathrm{d}y=\frac2{15},
    \qquad
    \int_0^1 f_0(1-y^2)\,\mathrm{d}y=\frac{11}{60},
    \qquad
    \int_0^1 f_1(1-y^2)\,\mathrm{d}y=\frac1{10}.
    \label{eq:polynomial_moments}
\end{equation}
Together with the viscous projection and transverse-acceleration
moment, these identities verify the coefficients entering the
liquid and surfactant fluxes.

The conservative forms of equations~\eqref{eq:model_H}
and~\eqref{eq:total_surfactant_local} preserve the liquid
volume and projected total surfactant inventory on the
periodic domain. Figure~\ref{fig:nl_membrane_conservation}
verifies these invariants to floating-point accuracy,
without inventory renormalization or concentration clipping.

The linear limit of the nonlinear implementation is also
checked independently.
Its Fourier Jacobian is compared with the six-state WR
pencil, and a separate infinitesimal perturbation is seeded
with a film eigenmode.
At the reference parameters, using amplitude $10^{-6}$ and
$k=8\pi/L\simeq0.100531$, the measured growth rate is
approximately $0.0208874$, agreeing with the WR eigenvalue
prediction at the displayed precision.
The full OS growth rate at the same wavenumber is
approximately $0.0210699$, giving a WR discrepancy of about
$0.87\%$.
This test checks the time integrator independently of the
finite-amplitude disturbances used for wave formation.

The nonlinear calculations are interpreted within the stated
Fourier-Galerkin band.
Smaller time steps, increased padding and nearby spatial
cutoffs provide local accuracy checks on the computed
wave states. Conservation alone does not establish spatial
resolution, and local refinement within the tested bands
does not constitute unrestricted continuum convergence
of the reduced equations.

Travelling-wave assessment uses the complete evolved state
\[
    \boldsymbol{Y}
    =
    \left(
        H,q,C_{\mathrm{tot}},\Gamma,\zeta,v
    \right)^{\mathsf T}.
\]
The spatial grid and retained spectral band are distinct
numerical controls.
Increasing the number of grid points at fixed cutoff
improves the evaluation of the prescribed Galerkin system,
whereas increasing the cutoff changes the set of evolved
modes.
The flat-film WR spectrum contains an additional growing
branch at sufficiently large wavenumber, outside the
production band at the baseline parameters.
This does not by itself invalidate the retained long-wave
solutions, but it prevents unrestricted spectral refinement
from serving as an automatic convergence argument.
Accordingly, quantitative nonlinear conclusions are
restricted to states that satisfy the admissibility,
resolution and travelling-state checks.
Writing its evolution as
$\partial_t\boldsymbol{Y}=\boldsymbol{F}(\boldsymbol{Y})$,
a translating state satisfies
\begin{equation}
    \boldsymbol{\mathcal{R}}_{\mathrm{TW}}
    =
    \boldsymbol{F}(\boldsymbol{Y})
    +c\,\partial_x\boldsymbol{Y}
    \simeq0.
    \label{eq:verification_travelling_residual}
\end{equation}
The translating-state residual is normalized according to
\begin{equation}
    r_{\mathrm{TW}}(t)
    =
    \frac{
    \left\|\mathsf W\left[F(Y)+cY_x\right]\right\|_2
    }{
    \|\mathsf WF(Y)\|_2
    +|c|\|\mathsf WY_x\|_2
    },
\end{equation}
where the norm includes all six fields and the spatial grid.
The diagonal weights are $\mathsf W_{jj}=s_j^{-1}$, with
\begin{equation}
    s_j
    =
    \max\left[
    \left\langle(Y_j-Y_{j,e})^2\right\rangle_{x,t}^{1/2},
    10^{-8}\max(|Y_{j,e}|,10^{-3})
    \right].
\end{equation}
The averages are taken over the assessed late interval.
The speed used for classification is obtained from
amplitude-weighted phase fits of the significant
free-surface Fourier modes.

Amplitude stationarity is measured by
\begin{equation}
    d_{\mathrm{amp}}
    =
    \max_{j,n\in\mathcal A_j}
    \frac{
    \operatorname{std}_t|\widehat Y_{j,n}|
    }{
    \langle|\widehat Y_{j,n}|\rangle_t
    },
\end{equation}
where $\mathcal A_j$ contains positive-wavenumber modes
whose mean amplitudes exceed both $10^{-10}$ and
$10^{-3}$ times the largest modal amplitude of that field.
A state is classified as travelling only when
$\max_t r_{\mathrm{TW}}<10^{-3}$ and
$d_{\mathrm{amp}}<10^{-3}$.

The rigid-wall, equilibrium, added-mass and long-wave checks
establish the internal consistency of the stated closure.
The OS comparison quantifies its linear accuracy, while
the nonlinear diagnostics assess its computed finite-amplitude
solutions. The quadratic bulk-concentration approximation
does not resolve arbitrary transverse concentration structure,
and the hydrodynamic reduction assumes slow motion and small
slopes. In particular, a rapidly oscillating membrane mode
can exceed the temporal assumptions of the reduction even
when its streamwise wavelength is long.


\section{Discussion and conclusions}
\label{sec:conclusion}

Samanta \citeyearpar{samanta2025} established the linear
stability characteristics of a soluble-surfactant-laden film
flowing over a compliant substrate. The present study follows
the coupled flow into the finite-amplitude regime and examines
the resulting liquid thickness, surfactant redistribution and
mechanical work at the wall. These calculations show how
substrate motion changes the interpretation of the surface
profile and identify the load contributions that maintain
deformation beneath a propagating wave.

The nonlinear stiffness comparison shows that substrate
deformation and wave propagation respond quite differently
to changes in the support. Along the prescribed path
$\mathcal{D}=\sqrt{\mathcal{K}}$, increasing $\mathcal{K}$
from $100$ to $5\times10^6$ reduces the substrate
peak-to-peak displacement from approximately $0.397$ to
$2.23\times10^{-5}$. Over the same comparison, the measured
wave speed changes from approximately $2.792$ to $2.766$.
The displacement therefore decreases by more than four orders
of magnitude, while the propagation speed changes by less
than $1\%$. A finite-amplitude liquid wave persists as the
support approaches the rigid limit. Since stiffness and
damping vary together, this result describes the response
of the specified family of supports.

The distinction between free-surface elevation and liquid
thickness is particularly important in this comparison.
Figure~\ref{fig:nl_stiffness_profiles} shows that the minimum
laboratory elevation decreases from approximately $0.736$
to $0.557$, while the minimum liquid thickness increases
from approximately $0.529$ to $0.557$. Thus, a lower surface
trough can accompany a greater local liquid depth. The
free-surface crest changes only modestly, whereas the maximum
thickness decreases as the substrate approaches the rigid
limit. The nearly opposing surface and substrate
displacements, quantified by the phase diagnostic, are
consistent with these different trends. For a compliant
film, the extrema of $h$ cannot therefore be used directly
as measures of thinning or of the amplitude of the
liquid-thickness disturbance.

The signed work budget gives a mechanical explanation of
how the wave maintains substrate motion. For the two
compliant supports considered, capillary-pressure work
supplies approximately $97\%$ of the mean damping loss.
The viscous and acceleration contributions are individually
substantial, but largely cancel in the mean. The principal
net supply of work to these supports therefore comes from
capillary pressure. The additional control at
$\mathcal{K}=500$ shows that reducing damping from
$\mathcal{D}=\sqrt{500}$ to $10$ produces almost the same
peak-to-peak displacement at $t=2000$, despite a $53.6\%$
decrease in mean damping loss. This comparison shows that
a similar deformation amplitude can be maintained with
a substantially different mechanical transfer. The control
remains an unsteady propagating state, so this conclusion
concerns the selected finite-time responses rather than
their eventual travelling-wave limits.

The Marangoni comparison provides a clearer separation of
transport and mechanical feedback because the initial
surfactant inventory is held fixed. Increasing
$\mathcal{M}$ from $0$ to $6$ reduces the maximum thickness
from approximately $2.380$ to $2.316$, raises the minimum
thickness from approximately $0.507$ to $0.529$, and reduces
the substrate peak-to-peak displacement
(figure~\ref{fig:nl_marangoni_profiles}). The surface
concentration peak also decreases from approximately
$0.655$ to $0.535$. These changes arise from redistribution
of the same surfactant inventory and its coupling to the
velocity field. At $\mathcal{M}=0$, adsorption, desorption
and surfactant transport remain active, but the concentration
field exerts no Marangoni stress. This case therefore
provides a passive-surfactant reference for assessing
the nonlinear feedback.

Figure~\ref{fig:nl_coverage_profiles} compares the nonlinear
profiles at $t=2000$. The coverage variation produces
pronounced changes in the distribution of surfactant around
the wave, accompanied by more moderate changes in liquid
geometry. Enrichment near the wave front creates surface
concentration gradients that influence both the surface
velocity and subsequent transport. This comparison requires
a different interpretation from the Marangoni sweep:
changing $\Gamma_e$ changes the equilibrium bulk
concentration and the initial total inventory. Absolute
concentration peaks must therefore be considered relative
to the equilibrium loading and together with the spatial
gradients responsible for Marangoni stress.

The reconstructed internal flow provides further information
about the finite-amplitude states.
Figure~\ref{fig:nl_stiffness_streamlines} identifies closed
recirculating regions beneath the crest for all three
selected supports, including the nearly rigid case.
Recirculation therefore persists even when wall displacement
is very small. The fixed-origin snapshots in
figure~\ref{fig:nl_fixed_origin_snapshots} clarify the
associated substrate motion. The liquid wave and the
deformation pattern propagate downstream, while substrate
material points move locally in the transverse direction.
For a travelling profile,
$\zeta_t\simeq-c\zeta_x$, so the rising and falling portions
of the deformation have opposite vertical velocities.
The fluid motion relative to the wave is described by
$u-c$, which accounts for the upstream relative flow
around the propagating crest.

The spatial period of these nonlinear states must be
interpreted in relation to the computational domain and
initial disturbance. The baseline Orr-Sommerfeld spectrum
has its largest film-mode growth near $k=0.109$, corresponding
to a wavelength of approximately $58$. The nonlinear
calculations use a periodic domain of length $L=250$ and
an initial disturbance at its fundamental spatial mode.
Their subsequent evolution includes nonlinear generation
of harmonics. The fastest infinitesimal disturbance
therefore does not uniquely determine the period of the
computed finite-amplitude wave. Agreement with linear
growth is established separately through the small-amplitude
test.

The corrected nonlinear transport formulation is important
to these finite-amplitude comparisons. The surface balance
accounts consistently for the changing interfacial measure,
following the surface-transport correction examined by
Mukhopadhyay \emph{et al.}
\citeyearpar{mukhopadhyay2026pof}. The normal bulk-flux
conditions are used in deriving the depth-integrated balance
before the concentration profile is approximated. This
retains the moving-boundary diffusion contributions needed
to extend the conservative bulk-flux treatment
\citep{mukhopadhyay2026audit} to a deformable support.
These nonlinear corrections leave the equations linearized
about uniform equilibrium unchanged. Agreement with
linear stability predictions would consequently not,
by itself, detect their omission.

With this formulation, the bulk and surface exchange terms
cancel locally, giving a conservative balance for the
projected inventory $\chi+H\phi+\Gamma$. The periodic liquid
volume and total surfactant inventory remain constant to
floating-point round-off in the reported calculations,
without corrections to their means after integration
(figure~\ref{fig:nl_membrane_conservation}). The wall-pressure
closure supplies a separate mechanical check by recovering
fluid added mass in the uniform-translation limit.
Together with rigid-wall recovery, these results test the
transport and moving-wall closures. Conservation alone,
however, does not establish the accuracy of the transverse
profile approximations.

The linear results provide context for the nonlinear
comparisons. For the baseline compliant substrate,
increasing $\mathcal{M}$ from $0$ to $6$ raises the
long-wave critical Reynolds number from approximately
$2.090$ to $3.056$. Increasing foundation stiffness at
$\mathcal{M}=6$ and $\Gamma_e=0.1$ moves this threshold
towards the rigid-wall value of approximately $3.131$.
These trends agree qualitatively with the opposing
surface-mode effects reported by Samanta
\citeyearpar{samanta2025}. They describe the film branch;
the oscillating wall branches and the remaining
concentration and hydrodynamic modes require separate
assessment.

The linear coverage dependence is non-monotonic.
Intermediate coverage gives the greatest delay of film-mode
onset among the sampled cases, while enrichment towards
$\Gamma_e=0.9$ weakens the stabilization. The factor
$a_e=\kappa(1-\Gamma_e)^2$ in the long-wave concentration
response describes the changing relation between bulk
and surface concentration perturbations at equilibrium.
Greater equilibrium coverage therefore need not produce
a stronger restoring stress. This behaviour is consistent
with earlier studies of soluble-surfactant films
\citep{karapetsas2013,karapetsas2014,mukhopadhyay2026},
which identify transport and exchange kinetics as
important parts of the interfacial response.

The quantitative verification also sets the scope of the
nonlinear interpretation. At $Re=25$, the weighted-residual
model reproduces the peak film-mode growth rate and upper
neutral wavenumber within approximately $1\%$ and $2.2\%$,
respectively, and recovers the same long-wave onset as the
full Orr-Sommerfeld formulation
(figure~\ref{fig:wr-os-comparison}). The neutral-wavenumber
discrepancy increases to approximately $14.1\%$ at $Re=100$,
so this agreement should not be extrapolated to larger
Reynolds numbers. The nonlinear results are solutions of
the specified Fourier-Galerkin approximation, assessed
through time-step, padding and nearby spectral-band checks.
A separate short-wave growth branch of the closure limits
the interpretation of unrestricted spectral refinement
as convergence to the continuum reduced equations.
Likewise, linear Orr-Sommerfeld agreement does not validate
the full nonlinear Navier-Stokes problem. The physical
findings concern the computed states within the stated
long-wave, transverse-profile and substrate-load assumptions.

\par\medskip
\noindent\textbf{Acknowledgements.}
The authors thank their colleagues for valuable discussions.

\par\medskip
\noindent\textbf{Funding.}
This research received no specific grant from any funding
agency, commercial or not-for-profit sectors.

\par\medskip
\noindent\textbf{Competing interests.}
The authors declare none.

\par\medskip
\noindent\textbf{Data availability.}
The numerical codes and data supporting the findings of
this study are available from the corresponding author
on reasonable request.

\appendix
\section{Long-wave asymptotic solution for the regular film branch}
\label{app:longwave_film_branch}
For the regular film branch, the phase speed remains
finite as $k\to0$, and we write
\begin{equation}
\begin{aligned}
\alpha
=\alpha_0+k\alpha_1+O(k^2),\quad
\psi
=\psi_0+k\psi_1+O(k^2), \quad
\hat C
=C_0+kC_1+O(k^2),\\
\hat h
=h_0+kh_1+O(k^2),\quad
\hat\zeta
=\zeta_0+k\zeta_1+O(k^2),\quad
\hat\Gamma
=\gamma_0+k\gamma_1+O(k^2).
\end{aligned}
\label{eq:series}
\end{equation}
We normalize the liquid-thickness disturbance by
$\hat h-\hat\zeta=1$, so that
\begin{equation}
h_0-\zeta_0=1,
\qquad
h_1=\zeta_1.
\label{eq:thickness-normalization}
\end{equation}
This normalization remains regular at
$\mathcal{K}=G\cot\theta$, where the leading absolute
free-surface displacement vanishes.

At order $k^0$, the differential equations give
\begin{equation}
D^4\psi_0=0,
\qquad
D^2C_0=0.
\end{equation}
Thus, $C_0$ is independent of $z$. Using the base
adsorption-desorption equilibrium,
$C_B=\Gamma_B/[\kappa(1-\Gamma_B)]$, the leading
sorption condition becomes
\begin{equation}
\gamma_0=\kappa(1-\Gamma_B)^2C_0.
\label{eq:leading-sorption}
\end{equation}

Combining the upper and lower normal-stress
conditions at order $k$ gives
\begin{equation}
G\cot\theta\,h_0
+\left(\mathcal{K}-G\cot\theta\right)\zeta_0=0.
\end{equation}
Together with Eq.~\eqref{eq:thickness-normalization},
this yields
\begin{equation}
\zeta_0=-\frac{G\cot\theta}{\mathcal{K}},
\qquad
h_0=1-\frac{G\cot\theta}{\mathcal{K}}.
\label{eq:leading-wall}
\end{equation}
The no-slip, substrate-kinematic,
tangential-stress, and free-surface-kinematic
conditions then determine
\begin{equation}
\alpha_0=G,
\label{eq:alpha0}
\end{equation}
and
\begin{equation}
\psi_0(z)
=
G\left[
\frac12\left(1-\frac{G\cot\theta}{\mathcal{K}}\right)z^2
+\frac{G\cot\theta}{\mathcal{K}}z
-\frac{G\cot\theta}{\mathcal{K}}
\right],
\qquad
D\psi_0(1)=G.
\label{eq:leading-fluid}
\end{equation}
The leading propagation speed of the
liquid-thickness disturbance is therefore
independent of substrate compliance.

To determine the leading concentration amplitudes,
we integrate the order-$k$ bulk equation across
the liquid layer and apply the combined bulk-flux
and surface-surfactant condition. Since
$\int_0^1U(z)\,\dd z=G/3$, we obtain
\begin{equation}
\left[
\frac23+\frac{\kappa(1-\Gamma_B)^2}{2}
\right]C_0=\Gamma_B.
\end{equation}
Consequently,
\begin{equation}
\begin{aligned}
C_0
&=\frac{6\Gamma_B}
{4+3\kappa(1-\Gamma_B)^2},\\
\gamma_0
&=\frac{6\Gamma_B\kappa(1-\Gamma_B)^2}
{4+3\kappa(1-\Gamma_B)^2}.
\end{aligned}
\label{eq:leading-chemical}
\end{equation}
These amplitudes account for the exchange of
surfactant between the bulk and the interface.
Adsorption and desorption conserve the combined
surfactant inventory; the surface inventory is
not conserved separately.

Substitution of Eqs.~\eqref{eq:alpha0}
and \eqref{eq:leading-fluid} into the order-$k$
Orr-Sommerfeld equation gives
\begin{equation}
D^4\psi_1=-\ii ReG^2(1-z).
\label{eq:first-order-OS}
\end{equation}
Integrating this equation and applying the
upper normal-stress condition yields
\begin{equation}
D^3\psi_1(z)
=
\ii\left[
G\cot\theta
\left(1-\frac{G\cot\theta}{\mathcal{K}}\right)
-ReG^2\left(z-\frac{z^2}{2}\right)
\right].
\label{eq:first-order-third-derivative}
\end{equation}
The compatibility of the tangential-stress and
kinematic conditions gives
\begin{equation}
\alpha_1
=
-\frac{3\ii\Ma\Gamma_B\kappa(1-\Gamma_B)^2}
{4+3\kappa(1-\Gamma_B)^2}
-\int_0^1
\left(z-\frac{z^2}{2}\right)
D^3\psi_1(z)\,\dd z.
\label{eq:alpha1-solvability}
\end{equation}
Using
\begin{equation}
\int_0^1\left(z-\frac{z^2}{2}\right)\,\dd z
=\frac13,
\qquad
\int_0^1\left(z-\frac{z^2}{2}\right)^2\,\dd z
=\frac{2}{15},
\end{equation}
we obtain the explicit first-order phase-speed
correction
\begin{equation}
\alpha_1
=
\ii\left[
\frac{2ReG^2}{15}
-\frac{G\cot\theta}{3}
\left(1-\frac{G\cot\theta}{\mathcal{K}}\right)
-\frac{3\Ma\Gamma_B\kappa(1-\Gamma_B)^2}
{4+3\kappa(1-\Gamma_B)^2}
\right].
\label{eq:alpha1-explicit}
\end{equation}

The corresponding first-order streamfunction is
\begin{align}
\psi_1(z)={}&
G\zeta_1\left(1-z+\frac{z^2}{2}\right)
-\frac{G\cot\theta}{\mathcal{K}}\alpha_1
\nonumber\\
&-\frac{3\ii\Ma\Gamma_B\kappa(1-\Gamma_B)^2}
{4+3\kappa(1-\Gamma_B)^2}\,z^2
\nonumber\\
&-\ii G\cot\theta
\left(1-\frac{G\cot\theta}{\mathcal{K}}\right)
\left(\frac{z^2}{2}-\frac{z^3}{6}\right)
\nonumber\\
&+\ii ReG^2
\left(\frac{z^2}{6}
-\frac{z^4}{24}
+\frac{z^5}{120}\right).
\label{eq:psi1}
\end{align}
The first-order boundary displacement follows
from the order-$k^2$ normal-stress compatibility:
\begin{equation}
h_1=\zeta_1
=
\frac{\ii G}{\mathcal{K}}
\left[
1+(1-\mathcal{D})\frac{G\cot\theta}{\mathcal{K}}
\right].
\label{eq:zeta1}
\end{equation}
Although substrate damping affects $\zeta_1$,
its contribution cancels from the solvability
condition for $\alpha_1$.

\section{Leading long-wave film-mode calculation}
\label{app:longwave_WR}

Let $\hat H$ denote the thickness perturbation.
The leading local Nusselt relation and the
leading wall load give
\begin{equation}
    \hat q=G\hat H,
    \qquad
    \hat\zeta
    =
    -\frac{G\cot\theta}{\mathcal{K}}\hat H.
\end{equation}
The mass equation then gives
\begin{equation}
    \lambda=-\ii kG+O(k^2).
\end{equation}

For finite positive exchange rates, the leading
bulk concentration perturbation is uniform
across the film, with
\begin{equation}
    \hat\phi
    =
    \frac{\hat\Gamma}
    {\kappa(1-\Gamma_e)^2}.
\end{equation}

Linearizing the total bulk-surface inventory
and its leading advective flux gives
\begin{equation}
    \Gamma_e\hat H
    -
    \left(
        \frac{2}{3\kappa(1-\Gamma_e)^2}+\frac12
    \right)\hat\Gamma=0.
\end{equation}
Therefore,
\begin{equation}
    \hat\Gamma
    =
    \frac{6\Gamma_e\kappa(1-\Gamma_e)^2}
    {4+3\kappa(1-\Gamma_e)^2}\hat H.
\end{equation}

The first-gradient correction to the
depth-integrated streamwise velocity is
\begin{equation}
    \hat q
    =
    G\hat H
    +
    \ii k
    \left[
        \frac{2ReG^2}{15}\hat H
        -\frac{G\cot\theta}{3}\hat h
        -\frac{\Ma}{2}\hat\Gamma
    \right]
    +O(k^2).
\end{equation}
Using
\begin{equation}
    \hat h
    =
    \hat H+\hat\zeta
    =
    \left(
        1-\frac{G\cot\theta}{\mathcal{K}}
    \right)\hat H
\end{equation}
in $\lambda\hat H+\ii k\hat q=0$ gives
Eq.~\eqref{eq:longwave_lambda}.

This expansion holds with the kinetic and
mechanical parameters fixed as $k\to0$.
Simultaneous singular limits of exchange,
diffusion or wall parameters require a
separate ordering.

\section{Decomposition of the substrate load}
\label{app:jfm_red_load}

Using the bare structural inertia, the wall equation reads
$\mathcal I v_t+\mathcal Dv-\mathcal T\zeta_{xx}
+\mathcal B\zeta_{xxxx}+\mathcal K\zeta=\mathcal F_{\mathrm w}$.
Combining equations~\eqref{eq:pressure_at_wall},
\eqref{eq:wall_viscous_load} and~\eqref{eq:external_pressure} gives the
signed load components used in equation~\eqref{eq:jfm_red_work}:
\begin{equation}
\begin{aligned}
\mathcal F_e &= P_e,\\
\mathcal F_g &= -G\cot\theta\,(H-1),\\
\mathcal F_{\Sigma} &=
  \frac{\Sigma(\Gamma)h_{xx}}{(1+h_x^2)^{3/2}},\\
\mathcal F_{\nu} &= \frac{3q_x}{2H}
  -\frac{3qH_x}{2H^2}-\frac{3q\zeta_x}{H^2},\\
\mathcal F_M &= -\mathcal M\left[
  \left(\frac54H_x+\frac32\zeta_x\right)\Gamma_x
  +\frac H4\Gamma_{xx}\right],\\
\mathcal F_a &= -Re\left(Hv_t+\widetilde{\mathcal A}_{\mathrm w}\right).
\end{aligned}
\label{eq:jfm_red_load_components}
\end{equation}
The acceleration expression is the existing closure in
equation~\eqref{eq:Aw_explicit}. The work calculation retains the bare
structural energy defined in the main text and includes $-ReHv_t$ in
$\mathcal F_a$, keeping a single convention for fluid added mass.
In the numerical implementation, the acceleration closure uses its
padded-grid intermediate $q_t$ and $q_{tx}$ before the final Fourier
projection. The capillary term retains the local $\Sigma(\Gamma)$ even
though the explicit gradient-dependent Marangoni term is displayed
separately. The backing-pressure contribution is zero for $P_e=0$.
\par

\section{Pressure elimination and integral relations}
\label{app:pressure_integrals}

\subsection{Pressure elimination in the Orr-Sommerfeld problem}
The pressure amplitude obtained from the streamwise
momentum equation is
\begin{equation}
    \hat\Pi
    =
    -Re\left[(U-\alpha)D\psi-U'\psi\right]
    -\frac{\ii}{k}(D^2-k^2)D\psi.
    \label{eq:pressure_amplitude}
\end{equation}

The pressure amplitude at the substrate is
\begin{equation}
    \hat\Pi(0)
    =
    Re\left[\alpha D\psi(0)+G\psi(0)\right]
    -
    \frac{\ii}{k}
    (D^3-k^2D)\psi(0).
    \label{eq:wall_pressure_amplitude}
\end{equation}
The first term vanishes after applying
Eq.~\eqref{eq:wall_kinematic_OS}.

The linearized normal viscous traction is
\begin{equation}
    2\left[
        D\hat w(0)-\ii kG\hat\zeta
    \right]
    =
    -2\ii k
    \left[D\psi(0)+G\hat\zeta\right]
    =0.
    \label{eq:wall_viscous_cancellation}
\end{equation}

\subsection{Pressure reduction and integral relations}
\label{subsec:pressure_integral_relations}

We return to the long-wave coordinates of
Eq.~\eqref{eq:dimension}.
Integration of the transverse momentum equation
from $z$ to $h$ gives the exact identity
\begin{align}
    \Pi(z)
    ={}&
    G\cot\theta(h-z)+\Pi(h)
    +\varepsilon\left[w_z(z)-w_z(h)\right]
    \nonumber\\
    &+\varepsilon^2Re
    \int_z^h
    \left(w_t+u w_x+w w_z\right)\,\dd z'
    -\varepsilon^3
    \int_z^h w_{xx}\,\dd z'.
    \label{eq:pressure_exact_integral}
\end{align}

For the simplified streamwise momentum reduction,
the retained pressure terms are
\begin{equation}
    \Pi(z)
    \simeq
    G\cot\theta(h-z)
    -\frac{\varepsilon^2}{Ca}h_{xx}
    +\varepsilon
    \left[w_z(h)+w_z(z)\right].
    \label{eq:pressure_expression}
\end{equation}

Using continuity,
\begin{equation}
    \Pi_x
    \simeq
    G\cot\theta\,h_x
    -\frac{\varepsilon^2}{Ca}h_{xxx}
    -\varepsilon
    \frac{\partial}{\partial x}\left[u_x(h)\right]
    -\varepsilon u_{xx}.
    \label{eq:pressure_gradient}
\end{equation}

The resulting streamwise equation is
\begin{align}
    \varepsilon Re
    \left(u_t+u u_x+w u_z\right)
    ={}&
    G\left(1-\varepsilon\cot\theta\,h_x\right)
    +u_{zz}
    +2\varepsilon^2u_{xx}
    \nonumber\\
    &+\varepsilon^2
    \frac{\partial}{\partial x}\left[u_x(h)\right]
    +\frac{\varepsilon^3}{Ca}h_{xxx}.
    \label{eq:x_momentum_reduced}
\end{align}

The explicit factor $\varepsilon^3$ in the capillary
term follows from $-\varepsilon\Pi_x$.
A strong-capillarity ordering,
$Ca^{-1}=O(\varepsilon^{-2})$ for $Re=O(1)$,
retains this term in the long-wave momentum balance.

The boundary conditions required in the integral
calculation are
\begin{align}
    &z=\zeta:
    &&
    u=0,
    \qquad
    w=\zeta_t,
    \label{eq:lower_boundary_integral}\\
    &z=h:
    &&
    u_z
    =
    -\varepsilon\Ma\Gamma_x
    +4\varepsilon^2u_xh_x
    -\varepsilon^2w_x
    +O(\varepsilon^3),
    \label{eq:upper_tangential_integral}\\
    &z=h:
    &&
    w=h_t+u h_x.
    \label{eq:upper_kinematic_integral}
\end{align}

The pressure approximation used in
Eq.~\eqref{eq:x_momentum_reduced} does not retain
the transverse-acceleration integral.
That term enters the streamwise pressure gradient
at a higher order than the retained viscous terms.
However, it enters the substrate load without an
additional streamwise derivative. We therefore
retain its leading contribution when constructing
the wall-pressure closure in equation~\eqref{eq:pressure_at_wall}.

The integral identities for the inertial terms are
\begin{align}
    \int_\zeta^h u_t\,\dd z
    &=
    q_t-u(h)h_t,
    \label{eq:u_t_integral}\\
    \int_\zeta^h u u_x\,\dd z
    &=
    \frac12
    \left[
        \frac{\partial}{\partial x}
        \int_\zeta^h u^2\,\dd z
        -u^2(h)h_x
    \right],
    \label{eq:u_ux_integral}\\
    \int_\zeta^h w u_z\,\dd z
    &=
    u(h)\left[h_t+u(h)h_x\right]
    \nonumber\\
    &\quad+
    \frac12
    \left[
        \frac{\partial}{\partial x}
        \int_\zeta^h u^2\,\dd z
        -u^2(h)h_x
    \right].
    \label{eq:w_uz_integral}
\end{align}
Consequently,
\begin{equation}
    \int_\zeta^h
    \left(u_t+u u_x+w u_z\right)\,\dd z
    =
    q_t+
    \frac{\partial}{\partial x}
    \int_\zeta^h u^2\,\dd z.
    \label{eq:convective_integral}
\end{equation}

\section{Details of the conservative weighted-residual closure}
\label{app:wr_closure}

\subsection{Mapped operators and weighted-residual projection}
For a field expressed in the coordinates $(x,y,t)$,
the physical derivatives satisfy the matrix relation
\begingroup
\large
\setlength{\arraycolsep}{10pt}       
\renewcommand{\arraystretch}{1.4}   

\begin{equation}
\begin{pmatrix}
    \left.\partial_t\right|_{x,z}\\[2mm]
    \left.\partial_x\right|_{z,t}\\[2mm]
    \left.\partial_z\right|_{x,t}
\end{pmatrix}
=
\begin{pmatrix}
    1 & 0 &
    -\dfrac{\zeta_t+yH_t}{H}\\[3mm]
    0 & 1 &
    -\dfrac{\zeta_x+yH_x}{H}\\[3mm]
    0 & 0 &
    \dfrac{1}{H}
\end{pmatrix}
\begin{pmatrix}
    \left.\partial_t\right|_{x,y}\\[2mm]
    \left.\partial_x\right|_{y,t}\\[2mm]
    \left.\partial_y\right|_{x,t}
\end{pmatrix}.
\label{eq:operator_transformation}
\end{equation}

\endgroup

The corresponding contravariant transverse velocity is
\begin{equation}
    \mathcal{W}\equiv\frac{Dy}{Dt}
=
\frac{1}{H}
\left[
    w-\zeta_t-yH_t-u(\zeta_x+yH_x)
\right].
    \label{eq:mapped_velocity}
\end{equation}

Incompressibility and the kinematic conditions give
\begin{equation}
    \mathcal{W}_y
=
-\frac{1}{H}\left[H_t+(Hu)_x\right],
\qquad
\mathcal{W}(0)=\mathcal{W}(1)=0.
    \label{eq:mapped_velocity_identity}
\end{equation}
where the streamwise derivative on the right is
taken at fixed $y$.

Hence,
\begin{equation}
    \frac{Du}{Dt}
=
u_t+u\,u_x+\mathcal{W}\,u_y.
    \label{eq:mapped_material_derivative}
\end{equation}
with all derivatives on the right evaluated in
the mapped coordinates. The wall translation is
already included in the mapping and does not
produce an additional independent $\zeta_t$
contribution to the horizontal material derivative.

The basis functions in equation~\eqref{eq:basis_functions} satisfy
\begin{gather}
    f_0(0)=f_1(0)=0,
    \qquad
    f_0'(1)=0,
    \qquad
    f_1'(1)=-4,
    \nonumber\\
    \int_0^1f_0\,\dd y=\frac13,
    \qquad
    \int_0^1f_1\,\dd y=0.
    \label{eq:basis_properties}
\end{gather}

Using $f_0$ as the weight function, the normalized
parabolic inertia is
\begin{equation}
    \frac52H\int_0^1
    f_0\frac{Du_0}{Dt}\,\dd y
    =
    q_t+\frac{17}{7}\frac qHq_x
    -\frac97\frac{q^2}{H^2}H_x.
    \label{eq:inertial_projection}
\end{equation}

Including the terms linear in $u_1$ gives
\begin{align}
    \frac52H\int_0^1
    f_0\frac{Du}{Dt}\,\dd y
    ={}&
    q_t+\frac{17}{7}\frac qHq_x
    -\frac97\frac{q^2}{H^2}H_x
    \nonumber\\
    &-\Ma\left[
        \frac{H^2}{48}\Gamma_{tx}
        +\frac{15}{224}qH\Gamma_{xx}
        +\frac{19}{336}Hq_x\Gamma_x
        +\frac5{112}qH_x\Gamma_x
    \right]
    \nonumber\\
    &+\frac{\Ma^2H^2\Gamma_x}{1344}
    \left(
        38H_x\Gamma_x+25H\Gamma_{xx}
    \right).
    \label{eq:full_inertial_projection}
\end{align}
The last line is quadratic in the Marangoni
velocity and is not retained in the simplified
flow-rate equation.

The streamwise viscous terms use the physical operator
$X$ defined in equation~\eqref{eq:physical_x_operator}.

The transverse velocity associated with $u_0$ is
\begin{equation}
    w_0
    =
    v-H\int_0^yXu_0\,\dd y',
    \qquad
    v=\zeta_t.
    \label{eq:w0_definition}
\end{equation}
Equivalently,
\begin{align}
    w_0
    ={}&
    v+\frac{q_x}{2}(y^3-3y^2)
    +\frac{3qH_x}{2H}(2y^2-y^3)
    \nonumber\\
    &+\frac{3q\zeta_x}{2H}(2y-y^2).
    \label{eq:w0_expanded}
\end{align}

The complete projected viscous contribution is
\begin{align}
    \mathcal R_\nu
    ={}&
    5H\int_0^1f_0X^2u_0\,\dd y
    +
    \frac{5H}{6}
    \frac{\partial}{\partial x}\left[Xu_0(1)\right]
    \nonumber\\
    &+\frac54
    \left[
        4Xu_0(1)h_x-Xw_0(1)
    \right].
    \label{eq:viscous_projection}
\end{align}
The derivative outside the brackets differentiates
the boundary trace, whereas $X$ inside the brackets
is the derivative at fixed physical $z$.

Substitution of $u_0$ and $w_0$ gives
\begin{equation}
    \mathcal R_\nu
    =
    4\frac q{H^2}H_x^2
    -\frac9{2H}q_xH_x
    -6\frac qH H_{xx}
    +\frac92q_{xx}
    +\mathcal R_\zeta,
    \label{eq:viscous_projection_expanded}
\end{equation}
where
\begin{equation}
    \mathcal R_\zeta
    =
    -5\frac q{H^2}\zeta_x^2
    -\frac52\frac q{H^2}H_x\zeta_x
    -\frac{15}{4}\frac qH\zeta_{xx}
    -\frac54\zeta_{tx}.
    \label{eq:wall_geometry_correction}
\end{equation}
The term $\zeta_{tx}=v_x$ originates from the
wall-velocity contribution in $w_0$.
No $q_x\zeta_x/H$ term remains after the complete
projection.

\subsection{Normal traction and transverse-acceleration moment}
At the substrate, differentiation of the no-slip
condition along the wall gives
\begin{equation}
    u_x(\zeta)+u_z(\zeta)\zeta_x=0.
    \label{eq:wall_noslip_derivative}
\end{equation}
Together with incompressibility and
$w(\zeta)=\zeta_t$, this yields
\begin{equation}
    \tau_{nn,w}
    =
    -\frac{
        2\varepsilon^3\zeta_x\zeta_{tx}
    }{
        1+\varepsilon^2\zeta_x^2
    }.
    \label{eq:wall_traction_cancellation}
\end{equation}
Hence the explicit normal viscous traction has
zero linearization and is beyond the retained
pressure order for slow wall motion.
The viscous pressure contribution at the substrate
does not vanish.

Define the transverse-acceleration moment
\begin{equation}
    \mathcal A_w
=
H\int_0^1
\left[
    (w_0)_t
    +u_0(w_0)_x
    +\mathcal{W}_0(w_0)_y
\right]\,dy.
    \label{eq:Aw_definition}
\end{equation}
where
\begin{equation}
    \mathcal{W}_0
=
-\frac{1}{H}
\int_0^y
\left[
    H_t+(Hu_0)_x
\right]\,dy'.
    \label{eq:V0_definition}
\end{equation}
All derivatives inside these expressions are
taken at fixed $y$, and $H_t=-q_x$ is used.

Using the decomposition in equation~\eqref{eq:Aw_split},
polynomial integration gives

\begin{align}
    \widetilde{\mathcal A}_w
    ={}&
    -\frac{3H}{8}q_{tx}
    +\left(\frac58H_x+\zeta_x\right)q_t
    -\frac98qq_{xx}
    -\frac18q_x^2
    +2qv_x
    \nonumber\\
    &+\frac{33}{20H}H_xqq_x
    +\frac{33}{40H}H_{xx}q^2
    +\frac6{5H}q^2\zeta_{xx}
    +\frac{12}{5H}qq_x\zeta_x
    \nonumber\\
    &-\frac{33}{40H^2}H_x^2q^2
    -\frac6{5H^2}H_xq^2\zeta_x.
    \label{eq:Aw_explicit}
\end{align}

\subsection{Conservative completion of bulk diffusion}
For comparison, direct integration of the
streamwise second derivative alone gives
\begin{align}
    \int_\zeta^h C_{xx}\,\dd z
    ={}&
    \chi_{xx}+H\phi_{xx}
    -\frac{3\chi}{H^2}h_x^2
    +\frac3H\chi_x\zeta_x
    \nonumber\\
    &-\frac{3\chi}{H^2}H_x\zeta_x
    +\frac{3\chi}{2H}\zeta_{xx}.
    \label{eq:bulk_diffusion_operator}
\end{align}

Leibniz's rule requires
\begin{equation}
    \partial_x\int_\zeta^h C_x\,\dd z
    =
    \int_\zeta^h C_{xx}\,\dd z
    +C_x(h)h_x-C_x(\zeta)\zeta_x.
    \label{eq:bulk_diffusion_leibniz}
\end{equation}
For the adopted profile,
\begin{equation}
    \partial_x\int_\zeta^h C_x\,\dd z-\int_\zeta^h C_{xx}\,\dd z
    =
    H_x\phi_x+\frac{3\chi}{H^2}h_x^2
    -
    \frac32
    \left(
        \frac{\chi_x}{H}
        -\frac{\chi H_x}{H^2}
    \right)\zeta_x.
    \label{eq:bulk_diffusion_completion}
\end{equation}


\begin{thebibliography}{99}

\bibitem[Alexander et~al.(2020)]{alexander2020}
\textsc{Alexander, J.~P., Kirk, T.~L. \& Papageorgiou, D.~T.}
2020
Stability of falling liquid films on flexible substrates.
\emph{J. Fluid Mech.} \textbf{900}, A40.

\bibitem[Benjamin(1957)]{benjamin1957}
\textsc{Benjamin, T.~B.}
1957
Wave formation in laminar flow down an inclined plane.
\emph{J. Fluid Mech.} \textbf{2}, 554--573.
doi:10.1017/S0022112057000373.

\bibitem[Blyth \& Pozrikidis(2004)]{blyth2004}
\textsc{Blyth, M.~G. \& Pozrikidis, C.}
2004
Effect of surfactant on the stability of film flow down
an inclined plane.
\emph{J. Fluid Mech.} \textbf{521}, 241--250.
doi:10.1017/S0022112004001909.

\bibitem[Carpenter \& Garrad(1985)]{carpenter1985}
\textsc{Carpenter, P.~W. \& Garrad, A.~D.}
1985
The hydrodynamic stability of flow over Kramer-type
compliant surfaces. Part 1. Tollmien--Schlichting
instabilities.
\emph{J. Fluid Mech.} \textbf{155}, 465--510.

\bibitem[Carpenter \& Garrad(1986)]{carpenter1986}
\textsc{Carpenter, P.~W. \& Garrad, A.~D.}
1986
The hydrodynamic stability of flow over Kramer-type
compliant surfaces. Part 2. Flow-induced surface
instabilities.
\emph{J. Fluid Mech.} \textbf{170}, 199--232.

\bibitem[D'Alessio et~al.(2020)]{dalessio2020}
\textsc{D'Alessio, S.~J.~D., Pascal, J.~P.,
Ellaban, E. \& Ruyer-Quil, C.}
2020
Marangoni instabilities associated with heated
surfactant-laden falling films.
\emph{J. Fluid Mech.} \textbf{887}, A20.
doi:10.1017/jfm.2019.1058.

\bibitem[Huang et~al.(2008)]{huang2008}
\textsc{Huang, H., Lai, M.-C. \& Tseng, H.-C.}
2008
A parametric derivation of the surfactant transport
equation along a deforming fluid interface.
In \emph{Frontiers of Applied and Computational
Mathematics}, pp.~198--205. World Scientific.
doi:10.1142/9789812835291\_0021.

\bibitem[Karapetsas \& Bontozoglou(2013)]{karapetsas2013}
\textsc{Karapetsas, G. \& Bontozoglou, V.}
2013
The primary instability of falling films in the presence
of soluble surfactants.
\emph{J. Fluid Mech.} \textbf{729}, 123--150.
doi:10.1017/jfm.2013.291.

\bibitem[Karapetsas \& Bontozoglou(2014)]{karapetsas2014}
\textsc{Karapetsas, G. \& Bontozoglou, V.}
2014
The role of surfactants on the mechanism of the long-wave
instability in liquid film flows.
\emph{J. Fluid Mech.} \textbf{741}, 139--155.
doi:10.1017/jfm.2013.670.

\bibitem[Matar et~al.(2007)]{matar2007}
\textsc{Matar, O.~K., Craster, R.~V. \& Kumar, S.}
2007
Falling films on flexible inclines.
\emph{Phys. Rev. E} \textbf{76}, 056301.
doi:10.1103/PhysRevE.76.056301.

\bibitem[Mukhopadhyay et~al.(2026a)]{mukhopadhyay2026}
\textsc{Mukhopadhyay, S., Millet, S., Di Pierro, B.
\& Mukhopadhyay, A.}
2026a
Solutocapillary instability in slipping falling films.
\emph{Phys. Rev. Fluids} \textbf{11}, 084004.
doi:10.1103/5xrv-cjwg.

\bibitem[Mukhopadhyay et~al.(2026b)]{mukhopadhyay2026pof}
\textsc{Mukhopadhyay, S., Millet, S., Di Pierro, B.
\& Mukhopadhyay, A.}
2026b
Eliminating artificial surfactant-mass drift in nonlinear
weighted-residual models of soluble-surfactant-laden
falling films.
\emph{Phys. Fluids} \textbf{38}, 092118.
doi:10.1063/5.0344560.

\bibitem[Mukhopadhyay \& Mukhopadhyay(2026)]{mukhopadhyay2026audit}
\textsc{Mukhopadhyay, S. \& Mukhopadhyay, A.}
2026
A conservative bulk-surface closure for
soluble-surfactant falling films: an analytical and
numerical mass-balance audit.
\href{https://arxiv.org/abs/2609.37869}
{arXiv:2609.37869} [physics.flu-dyn].

\bibitem[Orszag(1971)]{Orszag1971}
\textsc{Orszag, S.~A.}
1971
Accurate solution of the Orr--Sommerfeld stability
equation.
\emph{J. Fluid Mech.} \textbf{50} (4), 689--703.
doi:10.1017/S0022112071002842.

\bibitem[Pascal et~al.(2019)]{pascal2019}
\textsc{Pascal, J.~P., D'Alessio, S.~J.~D.
\& Ellaban, E.}
2019
Stability of inclined flow of a liquid film with soluble
surfactants and variable mass density.
\emph{Phys. Rev. Fluids} \textbf{4}, 054004.
doi:10.1103/PhysRevFluids.4.054004.

\bibitem[Ruyer-Quil \& Manneville(2000)]{ruyerquil2000}
\textsc{Ruyer-Quil, C. \& Manneville, P.}
2000
Improved modeling of flows down inclined planes.
\emph{Eur. Phys. J. B} \textbf{15}, 357--369.

\bibitem[Ruyer-Quil \& Manneville(2002)]{ruyerquil2002}
\textsc{Ruyer-Quil, C. \& Manneville, P.}
2002
Further accuracy and convergence results on the modeling
of flows down inclined planes by weighted-residual
approximations.
\emph{Phys. Fluids} \textbf{14}, 170--183.
doi:10.1063/1.1426103.

\bibitem[Samanta(2025)]{samanta2025}
\textsc{Samanta, A.}
2025
Role of soluble surfactant in linear stability of a
liquid film flowing down a compliant substrate.
\emph{J. Fluid Mech.} \textbf{1011}, A6.

\bibitem[Sisoev et~al.(2010)]{sisoev2010}
\textsc{Sisoev, G.~M., Matar, O.~K., Craster, R.~V.
\& Kumar, S.}
2010
Coherent wave structures on falling fluid films flowing
down a flexible wall.
\emph{Chem. Eng. Sci.} \textbf{65}, 950--961.
doi:10.1016/j.ces.2009.09.046.

\bibitem[Stone(1990)]{stone1990}
\textsc{Stone, H.~A.}
1990
A simple derivation of the time-dependent
convective-diffusion equation for surfactant transport
along a deforming interface.
\emph{Phys. Fluids A} \textbf{2}, 111--112.

\bibitem[Yih(1963)]{yih1963}
\textsc{Yih, C.-S.}
1963
Stability of liquid flow down an inclined plane.
\emph{Phys. Fluids} \textbf{6}, 321--334.
doi:10.1063/1.1706737.

\end{thebibliography}
\end{document}